%% file: main_R1_EJ_0625.tex
\documentclass[11pt, a4paper, pdftex, twoside, dvipsnames]{article} 
\usepackage{amsmath,amssymb,amsthm,amsfonts}
\usepackage{bm}  

\everymath{\displaystyle}
\allowdisplaybreaks
\usepackage{mathtools}
\usepackage[makeroom]{cancel}
\usepackage[title]{appendix}
\usepackage{epigraph}
\usepackage{marginnote}
\usepackage{soul}
\usepackage[labelfont=bf,font=small]{caption}
\usepackage{rotating}
\usepackage{float}
\usepackage{graphicx}  

\usepackage[section]{placeins}
\usepackage{siunitx}

\usepackage{tabularx}
\usepackage{numprint}
\usepackage[dvipsnames]{xcolor}

\usepackage{xstring}
\usepackage{flafter}

\usepackage{threeparttable}
\usepackage{multirow}
\usepackage{multicol}
\usepackage{blkarray}
\usepackage{booktabs}
\usepackage{arydshln}
\usepackage{tabularx}
\usepackage{longtable, array}
\newcolumntype{P}[1]{>{\raggedright\arraybackslash}p{#1}}
\usepackage{tikz}
\usetikzlibrary{arrows.meta}

\usepackage[bookmarks=false, bookmarksnumbered=true,
colorlinks=true,setpagesize=false,
pdftitle={},pdfauthor={SimonScheidegger},pdfsubject={},
pdfkeywords={}, bookmarkstype=toc, urlcolor=black, citecolor=blue,
linkcolor=black, filecolor=black]{hyperref}

\hypersetup{
    colorlinks=true,
    linkcolor=red,
    filecolor=magenta,      
    urlcolor=cyan
    }
\usepackage{float}
\usepackage{graphicx} 
\usepackage{subcaption}
\usepackage{indentfirst}
\usepackage{verbatim}
\usepackage{enumitem}
\usepackage{cases}
\usepackage{url}
\usepackage{setspace}
\usepackage{ulem}
\usepackage{lscape}
\usepackage{here}
\usepackage{authblk}
\usepackage{framed}
\usepackage{adjustbox}
\usepackage{pdflscape}
\usepackage{listings}
\usepackage{cleveref}
\crefname{equation}{Eq.}{Eqs.}
\crefname{figure}{Figure}{Figures}
\crefname{table}{Table}{Tables}
\crefname{line}{Algorithm}{Algorithms}
\crefname{asmp}{Assumption}{Assumptions}
\crefname{section}{Section}{Sections}
\crefname{chapter}{Chapter}{Chapters}
\crefname{appsec}{Appendix}{Appendixes}

\renewcommand{\ref}{\cref}

\usepackage{lipsum}

\usepackage[bottom]{footmisc}

\DeclareFontFamily{U}{mathx}{}
\DeclareFontShape{U}{mathx}{m}{n}{<-> mathx10}{}
\DeclareSymbolFont{mathx}{U}{mathx}{m}{n}
\DeclareMathAccent{\widehat}{0}{mathx}{"70}
\DeclareMathAccent{\widecheck}{0}{mathx}{"71}
\newcommand{\too}[2]{\text{#1} \! \to \! \text{#2}}
\newcommand{\bb}[1]{\boldsymbol{\bold{#1}}}

\newcommand{\bbt}[1]{\tilde{\boldsymbol{\mathrm{#1}}}}

\newcommand{\argmin}[1]{\underset{#1}{\text{argmin}} }

\newcommand{\ameq}[0]{\bb{a},\bbt{m}}
\newcommand{\expnum}[2]{
\ifnum#1=1 
  10^{#2} 
\else 
  #1 \! \cdot \! 10^{#2}
\fi
}

\renewcommand{\emph}[1]{\textit{#1}}

\newcommand{\fixmeAlex}[1]{{{\textcolor{red}{/* TBD Alex: #1 */}}}}

\newcommand{\fixmeDoris}[1]{{{\textcolor{red}{/* TBD Doris: #1 */}}}}

\usepackage[a4paper,left=2cm,right=2cm,top=2.5cm,bottom=2.5cm]{geometry}
\usepackage{fancyhdr}
\fancypagestyle{firstpagestyle}{

\fancyhf{} 
\fancyfoot[C]{\thepage}
}
\fancypagestyle{CVfirstpagestyle}{

\fancyhf{} 
\fancyfoot[C]{\thepage}
}
\usepackage{natbib}
\title{Building Interpretable Climate Emulators for Economics\thanks{We thank Christian Traeger, as well as seminar participants at the University of Lausanne, the University of Zurich, and the EGU General Assembly Conference, as well as two anonymous referees, for their valuable discussions and comments. The authors express their sincere gratitude to Pratyuksh Bansal for his dedication in initiating this project. This work was supported by the Swiss National Science Foundation (SNSF) under project ID ``Can Economic Policy Mitigate Climate Change?''}}
\author{
Aryan Eftekhari\thanks{Institute of Computing, Universit\`a della Svizzera italiana; Email: \href{aryan.eftekhari@usi.ch}{aryan.eftekhari@usi.ch}.}, \;
Doris Folini\thanks{Institute for Atmospheric and Climate Science, ETHZ; Email:
    \href{mailto:doris.folini@env.ethz.ch}{doris.folini@env.ethz.ch}.}, \;
Aleksandra Friedl\thanks{ifo Institute, Ludwig-Maximilian University of Munich; Email: \href{mailto:friedl.aleksandra@gmail.com}{friedl@ifo.de}.}, \; 
Felix K\"ubler\thanks{Department for Banking and Finance, University of Z\"urich; Swiss Finance Institute (SFI); Email: \href{mailto:fkubler@gmail.com}{felix.kuebler@bf.uzh.ch}.}, \\ \;
Simon Scheidegger\thanks{Department of Economics, University of Lausanne; Grantham Research Institute, London School of Economics; Email: \href{mailto:simon.scheidegger@unil.ch}{simon.scheidegger@unil.ch}}, \;
Olaf Schenk\thanks{Institute of Computing, Universit\`a della Svizzera italiana; Email: \href{mailto:olaf.schenk@usi.ch}{olaf.schenk@usi.ch}.}
  }
\date{\today}

\begin{document}
\maketitle

\begin{abstract}

We introduce a framework for developing efficient and interpretable climate emulators (CEs) for economic models of climate change. The paper makes two main contributions. First, we propose a general framework for constructing carbon-cycle emulators (CCEs) for macroeconomic models. The framework is implemented as a generalized linear multi-reservoir (box) model that conserves key physical quantities and can be customized for specific applications. We consider three versions of the CCE, which we evaluate within a simple representative agent economic model: (i) a three-box setting comparable to DICE-2016, (ii) a four-box extension, and (iii) a four-box version that explicitly captures land-use change. While the three-box model reproduces benchmark results well and the fourth reservoir adds little, incorporating the impact of land-use change on the carbon storage capacity of the terrestrial biosphere substantially alters atmospheric carbon stocks, temperature trajectories, and the optimal mitigation path. Second, we investigate pattern-scaling techniques that transform global-mean temperature projections from CEs into spatially heterogeneous warming fields. We show how regional baseline climates, non-uniform warming, and the associated uncertainties propagate into economic damages. \\



\end{abstract}

\newpage

\newpage

%

\section{Introduction}
\label{sec:introduction}

\paragraph{Motivation, Research Questions, and Key Contributions}
There is a rapidly developing literature on the macroeconomics of climate change (see, e.g., \citealp{DIETZ20241} and~\citealp{annurev:/content/journals/10.1146/annurev-economics-091124-045357} for recent reviews) that uses so-called \lq\lq climate emulators\rq\rq \ (CEs) within macroeconomic models
 to explore the economic consequences of industrial carbon emissions.
%
%
These emulators are reduced‐form models that map emissions into atmospheric $\mathrm{CO}_{2}$, global‐mean temperature, and regional warming, while reproducing the salient behavior of Earth System Models (ESMs; see, e.g., \citealp{Geoffroy2013a}).

\citet{folini2024climate} scrutinize the three‐box climate emulator embedded in DICE-2016 \citep{nordhausRevisitingSocialCost2017} and develop a calibration-validation protocol for that specific structure. Their contribution, however, stops short of asking whether the three-box architecture itself is adequate and how emulator uncertainty propagates when regional impacts are required.
In this paper, we address these two questions that are pivotal for contemporary models in climate economics.
\begin{enumerate}[label=(\roman*)]
  \item \textbf{Carbon-cycle resolution.}  Does replacing the DICE-2016 three‐box representation with richer multi-reservoir models materially alter carbon stocks, temperatures, and optimal policies?
  \item \textbf{Spatial downscaling.} How robust is standard pattern-scaling when the economic model demands grid‐level or regional climate inputs (cf.\ \citealp{Desmet2024})? Specifically, how much additional uncertainty does the downscaling step inject beyond the already considerable uncertainty in global $\mathrm{CO}_{2}$ concentrations and temperatures \citep{Santer1990,Tebaldi2014,Lynch2017}?
\end{enumerate}


To answer these questions, we introduce an open-source toolbox that lets economists design physically consistent linear box-model carbon-cycle emulators (CCEs),  then auto-calibrate and rigorously test them. 
The same framework links to pattern-scaling libraries, keeping emulator, pattern, and baseline uncertainties in a transparent manner.

%
\begin{figure}[ht!]
\centering
\begin{tikzpicture}[
    >=Stealth,
    main node/.style    = {circle, draw, minimum size=1cm, font=\sffamily\Large\bfseries},
    climate node/.style = {circle, draw, minimum size=7cm},
    reservoir/.style    = {draw, rectangle, minimum width=1.8cm, minimum height=0.8cm, font=\sffamily\small},
    arrow/.style        = {->, thick},
    bidir/.style        = {<->, thick},
  ]

  \node[main node] (economy) at (-2,0)   {economy};
  \node[main node] (damages) at (4,-6)   {damages};

  \node[climate node] (climate) at (6,0) {};

  \node[font=\sffamily\Large\bfseries] 
        at ([yshift= 2.8cm]climate.center) {climate};

  \draw[arrow] (economy) to[bend left=20] node[above]      {CO$_2$} (climate);
  \draw[arrow] (climate) to[bend left=20] node[right]      {$T^{AT}$} (damages);
  \draw[arrow] (damages) to[bend left=20] node[left]       {US\$~$$} (economy);

  \begin{scope}[shift={(climate.center)}, yshift=-0.5cm]
    \node[reservoir] (atm)  at (0,  1.5)  {Atmosphere};
    \node[reservoir] (land) at (-1.8,0)   {Land Biosphere};
    \node[reservoir] (uo)   at (+1.8,0)   {Upper Ocean};
    \node[reservoir] (deep) at (0,-1.8)   {Deep Ocean};
    \draw[bidir] (atm)  -- (land);
    \draw[bidir] (atm)  -- (uo);
    \draw[bidir] (uo)   -- (deep);
  \end{scope}

\end{tikzpicture}
\caption{Stylized schematic of an IAM. In their structural form, an IAM in macroeconomics typically links three components: (i) an economic module that generates fossil-fuel \(\mathrm{CO}_{2}\) emissions;  
(ii) a CE, that is, a reduced-form physical model that maps those emissions into atmospheric \(\mathrm{CO}_{2}\) concentrations and global-mean temperature \(T^{\mathrm{AT}}\); and (iii) a damage module that transforms the resulting temperature trajectory into economic losses (measured, e.g., in US\$). 
The inset within the CE visualizes one particular CCE: a four-reservoir box model that tracks \(\mathrm{CO}_{2}\) exchanges among the atmosphere, land biosphere, upper ocean, and deep ocean. Double-headed arrows indicate the bidirectional carbon fluxes between these reservoirs. For clarity, the separate climate sub-component that links radiative forcing to temperature is not depicted.}
\label{fig:DICE-Box1}
\end{figure}

\paragraph{Carbon Cycle}

Integrated assessment models (IAMs) often employ a single linear rule that maps cumulative emissions to global temperature change (see, e.g.,~\citealp{Dietz2019}, and references therein).  
This cumulative-emissions climate emulator is computationally attractive because it collapses climate dynamics into a single stock of cumulative $\mathrm{CO}_{2}$, but this parsimony masks four critical weaknesses: temperature change in response to changing atmospheric $\mathrm{CO}_{2}$ is not instantaneous but takes about 10 years~\citep{stjern-et-al:23}; forcings from non-$\mathrm{CO}_{2}$ greenhouse gases or aerosols are relevant but are ignored~\citep{jenkins-et-al:21}; the non-distinction between land and ocean carbon reservoirs prevents any study addressing their relative role in climate or economic terms; the assumed linear relation negates a priori the existence of any non-linearities as expected, for example, in the context of negative emissions or climate tipping points~\citep{Lenton2008, zickfeld-et-al:21}. 
Consequently, relying on this linear relationship can lead to misleading temperature projections and, as a direct consequence, flawed policy guidance.

A second strand of literature employs two separate modules to predict the link between emissions and temperature: the former maps emissions into atmospheric concentration by means of a CCE, and the latter 
maps concentrations to temperature. 
The present paper focuses on the link between emissions and concentration. For this, there exist two alternative modeling frameworks. The literature adopts either multi-reservoir box models \citep{nordhausRevisitingSocialCost2017,dorheim-et-al:24} or impulse-response formulations (IRFs; see, e.g.,~\citealp{leach-et-al:21,gasser-et-al:17,meinshausen-et-al:11}). Comparison studies document the strengths and weaknesses of each family \citep{nicholls-et-al:21,melnikova-et-al:23}. The two approaches can be shown to be formally equivalent when modeling the decay of a single pulse of carbon emitted to the atmosphere~\citep{li-jarvis-leedal-09, raupach-et-al:11, raupach-13}. Both approaches have to be calibrated against properly chosen benchmark data. 
 
In this article, we focus on a box-model CCE, where the global system is partitioned into a handful of well-mixed ``boxes” (reservoirs) between which carbon diffuses.\footnote{In most IAMs, a two‐reservoir energy‐balance module handles the evolution of the temperature of the atmosphere and the ocean. Since this aspect of climate emulators is extensively discussed in \cite{folini2024climate}, we do not address it in this paper.} In a box model, the atmosphere, the ocean layers, the land biosphere, and occasionally additional reservoirs such as permafrost or the shallow ocean are treated as homogeneous pools connected by fluxes that obey mass balance, as illustrated in a stylized fashion in the “climate” bubble of Figure~\ref{fig:DICE-Box1} for one concrete, illustrative example. Along with the number of reservoirs, the number of free parameters and characteristic time scales increases. We argue that the box-model formulation of the CCE is intuitive in its interpretation and adaptation to use-case-specific modifications, including accommodation of non-linearities~\citep{hooss-et-al:01, glotter-et-al:14, zhang-et-al:21}. The box approach naturally captures feedbacks that arise because carbon emissions to the atmosphere perturb the equilibrium among the different carbon reservoirs, yet remains computationally light.
 
The canonical DICE-2016 emulator (\citealp{nordhausRevisitingSocialCost2017} and \citealp{folini2024climate}), for instance, considers a three-reservoir carbon-cycle box model.
The model contains no explicit land-biosphere reservoir, even though terrestrial carbon stocks and biological carbon dioxide removal (CDR) dominate many low-cost mitigation pathways. Our central question in this paper is how modeling choices, such as how many carbon reservoirs to include, which data sets to fit, and which physical constraints to enforce, shape projected concentrations (and ultimately economic damages).

We present an open-source toolbox that lets economists design, calibrate, and deploy bespoke CCEs that are physically sound and policy-specific. 
A key advantage of this modular approach is that it makes modeling choices, how many reservoirs to include, which data sets to fit, and which physical constraints to impose, both transparent and testable. 

We construct three emulators, \textbf{3SR}, \textbf{4PR}, and \textbf{4PR-X}, of increasing complexity, tracing how each additional layer sharpens economic insight:\footnote{Acronyms indicate reservoir count and routing of carbon: \textbf{3SR} has three serial reservoirs (atmosphere $\rightarrow$ upper ocean $\rightarrow$ deep ocean); \textbf{4PR} adds a land-biosphere reservoir in parallel with the upper ocean but keeps its storage capacity fixed; \textbf{4PR-X} builds on 4PR by allowing the land-biosphere capacity to vary with land-use, thus capturing afforestation and deforestation feedbacks.}

\begin{itemize}
  \item \textbf{3SR}: a re-implementation of the DICE-2016 three-reservoir carbon cycle, used as our validation anchor;
  \item \textbf{4PR}: a variant that adds a land-biosphere box with \emph{static} carbon holding capacity; and 
  \item \textbf{4PR-X}: a new model in which the land reservoir holding capacity adjusts endogenously to land-use change, capturing feedbacks from deforestation or afforestation.\footnote{Afforestation, reforestation, and soil-carbon enhancement could store hundreds of petagrams of carbon \citep{minx-et-al:18,fuss-et-al:18,nemet-et-al:18}.  
Sophisticated IAMs such as GCAM \citep{Calvin_2014} and IMAGE \citep{Doelman_2020} model these processes but at a high computational and transparency cost.  
Earlier attempts to link DICE with forestry models \citep{Sohngen_2006} lacked a land box in the climate block.  
By embedding both static and dynamic land reservoirs inside a lightweight CE, we bridge this gap while retaining the tractability of cost–benefit IAMs.}
\end{itemize}



Three- and four-box emulators (3SR and 4PR), calibrated to both pre-industrial (PI) and present-day (PD) conditions, replicate the historical and long-run evolution of atmospheric $\mathrm{CO}_{2}$ and global temperature to within about $5\%$ and $3\%$, respectively, over a 500-year horizon, an accuracy widely regarded as ``fit for purpose'' in policy analysis. Including a static land box, therefore, leaves the headline results essentially unchanged.
To illustrate the extent to which the choice of a CCE, the associated calibration targets, and the selected hyperparameters can influence quantitative results, we evaluate the three carbon-cycle models within the standard DICE economy–climate framework under PI conditions and a range of extreme scenarios.
While 3SR and 4PR produce broadly similar policy-relevant trajectories, the 4PR-X variant, which permits the land biosphere to evolve dynamically under deforestation or urbanization, yields markedly higher atmospheric carbon stocks and temperatures: by~2100, atmospheric carbon is almost $80\,\mathrm{GtC}$ (about~$6\%$) higher, translating into an additional $0.2^{\circ}\mathrm{C}$ (also~$\sim6\%$) of warming relative to the static models. Consequently, the optimal social cost of carbon rises by as much as 17 \$US  per tCO$_2$, necessitating stronger mitigation.
These results underscore the importance of explicitly representing the land-biosphere reservoir when modeling land-use change. Deforestation and urbanization materially alter optimal carbon policy, highlighting the need for coordinated measures that address both industrial emissions and land management. Under an extreme warming scenario, the incremental impact of deforestation is comparable in magnitude, reinforcing its relevance. Overall, our framework provides a robust, interpretable basis for climate-economic analysis, overcoming the limitations of reduced-reservoir models and enabling the study of extreme cases.

\paragraph{Pattern Scaling}

Pattern scaling \citep{Santer1990,Tebaldi2014,Lynch2017} remains the workhorse for down-scaling global-mean projections to gridded fields, recently entering economic analyses of spatial damages and insurance \citep{Krusell2022,Desmet2024,Cruz2024,Kotlikoff2024}.  
Using the pattern library of \citet{Lynch2017}, our module turns the global mean temperature path from any emulator into a gridded warming map and, when absolute temperatures matter, anchors that map to an observation-based climatology such as ERA5.\footnote{ERA5 is the European Center for Medium-Range Weather Forecasts (ECMWF) fifth-generation reanalysis, an hourly, global data set that merges observations with a modern forecast model; see \url{https://www.ecmwf.int/en/forecasts/dataset/ecmwf-reanalysis-v5}.}  
Because the module keeps the three main uncertainty sources, emulator calibration, ESM pattern choice, and observational baseline, explicit and separate, analysts can see immediately how each factor propagates to regional damages.
Pattern scaling, our computationally efficient bridge from global to local climate, unveils substantial geographic and methodological uncertainty.  
Using the pattern library of \citet{Lynch2017}, which is derived from the 41 global climate models in the Coupled Model Intercomparison Project, Phase~5 (CMIP5), shows that land areas typically heat about 50\% faster than the planet as a whole, although the exact amplification varies markedly between models and regions.
Anchoring a given pattern to different present day temperature maps, for example, the ERA5 reanalysis versus the climatology of an individual model, can further shift the projected 2100 regional means by up to 3 °C.\footnote{CMIP5 is the international multi-model ensemble that underpinned the IPCC’s Fifth Assessment Report; it provides internally consistent past-to-future climate simulations for dozens of Earth-system models.  
See \url{https://wcrp-cmip.org/cmip-phases/cmip5} for CMIP5 and \url{https://www.ipcc.ch} for the Intergovernmental Panel on Climate Change (IPCC).}
Passing these alternative temperature fields through a standard hump-shaped damage function can therefore flip sub-regions from “relative winners’’ to “relative losers’’ and vice-versa.  

\paragraph{Organization of the Article}
Section~\ref{sec:2} defines the class of carbon-cycle emulators we study, and Section~\ref{sec:3} details the constrained-optimization routine that calibrates any CCE for use in a full climate emulator.  
Section~\ref{sec:4} examines how the emulator design and fitting choices propagate through a dynamic economic model and shape the resulting optimal climate policies.
Section~\ref{sec:pattern_scaling} formalizes a plug-and-play pattern-scaling module, quantifies how different ESMs and observational baselines affect regional warming and absolute temperatures, and demonstrates, through a spatially resolved damage function, how these uncertainties translate into heterogeneous economic impacts.  
Section~\ref{sec:conclusion} concludes.  
All source code is openly available at \url{https://github.com/ClimateChangeEcon/Building_Interpretable_Climate_Emulators_forEconomics}.

\section{Carbon Cycle}\label{sec:2}
In this section, we present the most general formulation of the linear-box models for the carbon cycle used in this paper.
The carbon cycle governs atmospheric CO$_2$ concentrations and, through interactions among several carbon reservoirs, modulates global temperature, as illustrated in Figure~\ref{fig:DICE-Box1}.
Our objective is to demonstrate to quantitative economists how transparent CCEs can be selected, or even custom-designed, to address specific research questions within a given IAM, such as those concerning deforestation or reforestation. To that end, we analyze three representative CCE specifications.

We begin in Section~\ref{sec:2.1} with a formal description of a multi-reservoir, linear carbon cycle comprising three reservoir classes: 
atmosphere ($\text{A}$), ocean ($\text{O}$), and land biosphere ($\text{L}$);
We first analyze a three-reservoir, serial configuration (denoted as \(3\text{SR}\)), in which carbon moves sequentially from the atmosphere through two vertically stacked ocean reservoirs, that is, the upper ocean  ($\text{O}_1$) and deep ocean ($\text{O}_2$). Next, we introduce a novel four-reservoir, parallel configuration (denoted as \(4\text{PR}\)), where atmospheric carbon is partitioned into concurrent flows toward the land biosphere and the upper ocean. We then extend the \(4\text{PR}\) configuration to a dynamic variant (denoted as \(4\text{PR}\text{-X}\)) by incorporating a time-dependent operator that captures shifts in the equilibrium state of the carbon cycle, most notably changes in the land-biosphere storage capacity, thereby enabling the simulation of scenarios with diminished carbon uptake, such as those induced by deforestation. Section~\ref{sec:2.3} outlines potential challenges in calibrating these models and motivates the fitting methodology detailed in Section~\ref{sec:3}. 
After calibration, we evaluate each CCE in two settings:  
(i) within a dynamic economic model analyzed in Sections~\ref{sec:4} of the main text, and  
(ii) in a complementary suite of standalone climate-science experiments reported in Appendix~\ref{sec:appendix:additional_experiments}.


\subsection{Multi-Reservoir Model}\label{sec:2.1}

Let $\bb{m}_{t}\!\in\!\mathbb{R}^{n}$ be the vector of carbon masses held in $n$ reservoirs at discrete times $t=0,\dots ,T-1$ and let
$\bb{A}\!\in\!\mathbb{R}^{n\times n}$ denote the linear operator whose entry $\bb{A}_{ij}$ quantifies the flux \emph{from} reservoir $j$ \emph{to} reservoir $i$.\footnote{Throughout, subscripts denote time in time-dependent variables, whereas superscripts index vector or matrix components. Thus, $\bb{m}_{t}^{1}$ and $\bb{m}_{t}^{\text{A}}$ represent the reservoir mass of component 1 and reservoir A, respectively, at time $t$. For equilibrium quantities, only subscripts are used to label the components; for example, $\bbt{m}_{\text{A}}$ is the equilibrium carbon mass in the atmosphere.
    %
}
By choosing which entries of $\bb{A}$ are non-zero, the modeler stipulates the routes along which carbon is allowed to circulate.
Figures~\ref{fig:DICE-Box1} depict the permissible  CO$_2$ flow patterns of the four-reservoir carbon-cycle emulator ($4$PR, $n=4$), comprising the atmosphere, upper ocean, deep ocean, and land biosphere.
In this model, CO$_2$ is exchanged bidirectionally between the atmosphere and the upper ocean, between the atmosphere and the land biosphere, and between the upper and deep ocean layers.\footnote{
    The three-reservoir, serial scheme ($3$SR, $n=3$) is obtained by suppressing the A–L exchange pathway.
}
%
%
\begin{figure}[ht!] 
    \input{fig/capartment_model}
    \caption{
    The left panel depicts the non-zero pattern of the operator matrix $\bb{A}$ for the four-reservoir carbon-cycle emulator ($4$PR), whose reservoirs are the atmosphere (A), upper ocean (O$_1$), deep ocean (O$_2$), and land biosphere (L).
    Black squares indicate fixed, non-zero coefficients, while colored squares mark free parameters that are calibrated (cf.~\ Equation~\eqref{eq:4} below).
    Matrix rows (``To'') correspond to recipient reservoirs and columns (``From'') to donor reservoirs.
    In the right panel, we show the directed graph of the same model, which corresponds to the possible transfer pathways of carbon.
    Nodes denote reservoirs, and arrows represent admissible bidirectional carbon fluxes; colored arrows map one-to-one onto the free parameters highlighted in the left panel.
    For example, the flux $\text{O}_1 \to \text{O}_2$ is entry $\bb{A}_{3,2}$.
    Eliminating the land–atmosphere exchange by setting $\bb{A}_{4,1}=0$ reduces the system to the three-reservoir ($3$SR) model.
    }
    \label{fig:1}
\end{figure}
%
%

Denote $\mathbf{e}_t \in \mathbb{R}^{n}$ as the external carbon emissions at discrete time $t$ (positive values correspond to inputs to the system).
The evolution of the reservoir-mass vector $\bb{m}_t \in \mathbb{R}^{n}$ is governed by the linear first-order difference equation
\begin{align}\label{eq:1}
  \mathbf{m}_{t+1} - \mathbf{m}_{t} \;=\; \mathbf{A}\,\mathbf{m}_{t} \;+\; \mathbf{e}_{t},
  \qquad t = 0,1,\dots
\end{align}
with known initial condition $\bb{m}_0$.
Unless stated otherwise, we adopt an annual time step, so $t$ indexes calendar years. 
Emissions may arise from both anthropogenic activity and natural processes.\footnote{
    Examples include fossil-fuel combustion, land-use change, volcanic eruptions, and wildfires.
}
Below we summarize the structural properties of the operator $\mathbf{A}$ that are required for the calibration procedure in Section~\ref{sec:3.1}.

\begin{itemize}
    \item The operator $\mathbf{A}$ possesses real eigenvalues $\lambda_i(\mathbf{A})$ confined to the interval $(-1,0]$. For each non-zero eigenvalue, we define the associated dynamic timescale as $\tau_i = 1/|\lambda_i|$.\footnote{An explicit form of the operator is provided in Appendix~\ref{APX:D}.} Accordingly, the admissible timescales for the linear carbon-cycle model satisfy $\tau_i \in (1,\infty)$.\footnote{The eigenvalue $\lambda_i = 0$ represents the equilibrium mode and implies an infinite timescale; this mode is omitted from the reported set.}
    
    \item The operator $\mathbf{A}$ is subject to the equilibrium (steady-state) constraint  \begin{equation}\label{eq:2}
  \mathbf{A}\,\tilde{\mathbf{m}} = \mathbf{0},
  \end{equation}
  where $\tilde{\mathbf{m}}$ denotes the vector of equilibrium carbon masses and is, up to normalization, the eigenvector associated with the zero eigenvalue of $\mathbf{A}$.

    \item The principle of mass conservation for  $\bb{A}$ is upheld by ensuring that for all $t$ the following holds:\footnote{The principle of mass conservation states that the total mass in a closed system remains constant over time, meaning that mass can only be re-distributed, but is neither created nor destroyed. In the context of the carbon cycle, this principle ensures that the total carbon mass across all reservoirs remains constant, regardless of internal exchanges among reservoirs.}
    \begin{align}\label{eq:3}
    \bb{1}^\top (\bb{m}_{t+1}-\bb{m}_{t}) = \bb{1}^\top \bb{e}_{t} \quad \iff \bb{1}^\top \bb{A} =\bb{0}.
    \end{align}

    \item For each pair of reservoirs $i$ and $j$, a carbon transfer pathway exists if $\mathbf{A}_{ij} \neq 0$, and no pathway exists if $\mathbf{A}_{ij} = 0$ (cf. Figure~\ref{fig:1}). 
    
    \item We denote $\mathcal{I}:=\{(i,j):\bb{A}_{ij}\neq 0,i>j\}$ as the strictly lower triangular nonzero indices of $\bb{A}$. For all model configurations, the upper triangular values of the operator satisfying both Equations~\eqref{eq:2} and \eqref{eq:3} are given by:
    \begin{align} \label{eq:4}
    	\bb{A}_{ji} = \bb{A}_{ij} \cdot \frac{\bbt{m}_j}{\bbt{m}_i} \; \forall \; (i,j) \in \mathcal{I}, \; \text{and} \;  \bb{A}_{ii}= -\!\!\!\sum_{j=1,j\neq i}^p \bb{A}_{ji}.
    \end{align}
    The parameters defining the operator, denoted $\bb{A}[\ameq]$, are the strictly lower triangular nonzero elements $\bb{a} := \{ \bb{A}_{ij} : (i,j) \in \mathcal{I} \}$, along with the equilibrium masses $\bbt{m}$.

    \item The set of admissible parameters includes those that satisfy $-1 < \bb{\lambda}_i(\bb{A}[\ameq]) \leqslant 0$. For a predefined sequence of emissions $\bb{E} := (\bb{e}_{1}, \bb{e}_{2}, \ldots, \bb{e}_T)$ of length $T$, the resulting carbon cycle simulation is denoted as
    \begin{align} \label{eq:5}
    	\bb{M}[\ameq] := (\bb{m}_{1}, \bb{m}_{2}, \ldots, \bb{m}_{T}),
    \end{align}
    where each $\bb{m}_{t}$ is defined according to expression Equation~\eqref{eq:1}.
    For example, the carbon mass in the deep-ocean reservoir ($\text{O}_2$) at time $t$ is denoted as $\bb{M}[\ameq]_{t}^{\text{O}_2} = \bb{m}_{t}^{\text{O}_2}$.
    
\end{itemize} 

Having introduced the general framework, we now focus on the four-reservoir (4PR) model. 
%
The system of equations described in expression Equation~\eqref{eq:1} for the four-reservoir model can be explicitly expressed as follows in Equation~\eqref{eq:NEW_ALL}:
\begin{align}\label{eq:NEW_ALL}
\bb{m}_{t+1}^{\text{A}} &= (1 - \bb{A}_{2,1}-\bb{A}_{4,1}) \cdot \bb{m}_{t}^{\text{A}} + \bb{A}_{2,1} \cdot \frac{\bbt{m}_{\text{A}}}{\bbt{m}_{\text{O}_1}} \cdot \bb{m}_{t}^{\text{O}_1} + \bb{A}_{4,1} \cdot \frac{\bbt{m}_{\text{A}}}{\bbt{m}_{\text{L}}} \cdot \bb{m}^{\text{L}}_{t} \\
\bb{m}_{t+1}^{\text{O}_1} &= \bb{A}_{2,1} \cdot \bb{m}_{t}^{\text{A}} + \left(1-\bb{A}_{2,1} \cdot \frac{\bbt{m}_{\text{A}}}{\bbt{m}_{\text{O}_1}} - \bb{A}_{3,2} \right) \cdot \bb{m}_{t}^{\text{O}_1} + \bb{A}_{3,2}    \cdot \frac{\bbt{m}_{\text{O}_1}}{\bbt{m}_{\text{O}_2}} \cdot \bb{m}^{\text{O}_2}_{t} \nonumber\\ 
\bb{m}_{t+1}^{\text{O}_2} &= \bb{A}_{3,2} \cdot \bb{m}_{t}^{\text{O}_1} + \left(1-\bb{A}_{3,2} \cdot \frac{\bbt{m}_{\text{O}_1}}{\bbt{m}_{\text{O}_2}} \right) \cdot \bb{m}_{t}^{\text{O}_2} \nonumber\\
\bb{m}_{t+1}^{\text{L}} &= \bb{A}_{4,1} \cdot \bb{m}_{t}^{\text{A}} + \left(1-\bb{A}_{4,1} \cdot \frac{\bbt{m}_{\text{A}}}{\bbt{m}_{\text{L}}} \right) \cdot \bb{m}_{t}^{\text{L}}. \nonumber
\end{align}
Note that the $3$SR carbon cycle can be derived from the same set of equations by setting $\bb{A}_{4,1}=0$. 
%

\paragraph{Time-dependent Land Capacity}\label{sec:2.2}
In principle, the operator $\bb{A}$ may be either time-invariant (i.e., constant)  or time-dependent.
As defined in Equation~\eqref{eq:4}, it is time-invariant under the assumptions of (i) a fixed equilibrium partitioning of the total carbon mass across reservoirs, expressed by the ratios $\bbt{m}_j/\bbt{m}_i$, and (ii) constant inter-reservoir exchange coefficients $\bb{A}_{ij}$.
However, land-use-related emissions, primarily permanent deforestation and agricultural expansion, alter the storage capacity of the land biosphere~\citep{land_use_emissions, meinshausen-et-al:11}.
Within the box-model framework, the equilibrium carbon mass of the land reservoir, therefore, cannot remain constant.

When changes in the equilibrium mass of the land biosphere $\bbt{m}_{\text{L}}$ are taken into account over time, the operator $\bb{A}$ becomes time-dependent and we denote it as $\bb{A}_t$.
Given land-use-change emissions $\bb{e}^{t}_{\text{L}}$ at time $t$, and assuming that a fraction $r$ of these emissions results from deforestation, the equilibrium mass of the land biosphere in the next time step is
\begin{align}\label{eq:lb_m_stp}
  \bbt{m}_{t+1}^{\text{L}} \;=\; \bbt{m}_{t}^{\text{L}} \;-\; r \cdot \bb{e}_{t}^{\text{L}} .
\end{align}
This modification yields the time-dependent version of the operator in Equation~\eqref{eq:4}.
In our simplified setting we adopt a one-to-one correspondence ($r=1$): every unit of land-use emission decreases the equilibrium mass $\bbt{m}_{\text{L}}$ by the same amount, so that all land-use emissions directly reduce the land-biosphere reservoir.
The equilibrium mass in the box model nonetheless only approximates the true carbon stored on land.\footnote{By contrast, the atmospheric carbon stock in the model matches the historical value of $589$ Gt C in $1750$; cf. Table~\ref{tab:1}.} 
%
Therefore, a one-to-one correspondence between real-world carbon emissions from land use change and the change in equilibrium land-biosphere mass in the emulator may not be guaranteed. The proportionality factor $r$ could be different from $1$.

The sensitivities of atmospheric and land-biosphere reservoir masses to changes in land-biosphere equilibrium mass can be expressed as follows
\begin{align}
%
  \frac{\partial \bb{m}_{t+1}^{\text{A}}}{
\partial \bbt{m}_{t}^{\text{L}} 
  } \;\propto\; -\frac{1}{(\bbt{m}_{t}^{\text{L}})^{2}} 
  ,\qquad
 \frac{\partial \bb{m}_{t+1}^{\text{L}}}{
\partial \bbt{m}_{t}^{\text{L}}
 } \;\propto\; \frac{1}{(\bbt{m}_{t}^{\text{L}})^{2}}.
\end{align}
Hence, a decrease in the land-biosphere equilibrium mass 
will (i) increase the carbon remaining in the atmosphere, and (ii) reduce the carbon retained in the land biosphere.
Simulation results confirming this behavior are presented in Appendix~\ref{sec:4.2}.
Although the time-stepping formulation in Equation~\eqref{eq:1} remains linear, the above analysis indicates a potentially nonlinear sensitivity of carbon mass exchanges with respect to changes in the land-biosphere equilibrium mass.
\subsection{Challenges of Degrees of Freedom}\label{sec:2.3}
 
A central challenge in constructing a climate emulator is to determine a parameter set that allows the reduced model to replicate the carbon-cycle behavior of more detailed Earth-system models.
The usual remedy is to solve an optimization problem that minimizes the discrepancy between the emulator's output and a benchmark simulation.
In practice, however, this calibration task is often ill-posed: different parameter vectors can generate virtually identical carbon-flux trajectories, even though many of those vectors are physically implausible.

The indeterminacy originates partly from Equation~\eqref{eq:4}, which shows that only the ratios of the equilibrium masses enter the operator~$\bb{A}$.
Hence, for any scalar $c>0$, the replacement $\bbt{m}\!\mapsto\! c\,\bbt{m}$ leaves the model response unchanged.
In Section~\ref{sec:3} we resolve this degeneracy by introducing a physics-informed calibration scheme that (i) enforces additional physical constraints and (ii) exploits the emulator's linear structure to construct a weighted operator capable of capturing the extreme dynamical regimes present across the admissible model configurations.

\section{A Calibrated Climate Emulator}\label{sec:3}

In this section, we introduce a systematic calibration procedure that turns a box-type CCE, configured for a specific integrated-assessment purpose (cf.~Section~\ref{sec:2}), into a climate-data-constrained component ready for quantitative modeling. Furthermore, we show how the calibrated CCE can be coupled with a temperature module to form a full climate emulator that can be embedded in quantitative IAMs (Figure~\ref{fig:DICE-Box1}). Recall that throughout this paper, the term emulator refers to this combined carbon-cycle and temperature model.

Section~\ref{sec:3.1} presents a generic fitting framework for the linear box model with $n$ reservoirs (cf. Section~\ref {sec:2}), encompassing data selection, parameter estimation, and hyperparameter tuning. As an example, we utilize the pulse-decay database of~\cite{joos2013carbon}, which compiles simulations of instantaneous carbon release from a suite of state-of-the-art climate models worldwide; these trajectories serve as our fitting target. 
Individual simulations differ, among others, in the climate model used, the amplitude of the pulse, or the background conditions into which the pulse is released. In our work below, we use the so-called multi-model mean (average over more than ten individual models) of the simulated decay of a 100 GtC carbon pulse under pre-industrial (PI) conditions, corresponding to the year 1750. To illustrate that other choices are possible, we provide in Appendix~\ref{APX:C} a corresponding illustration of our method for present-day (PD) background conditions, representative of the year 2010. Any pulse-decay dataset, not only that of \cite{joos2013carbon}, can be used with our toolbox to calibrate the CCE. The selection should match the intended application: for example, whether average or extreme decay behavior is required \citep{folini2024climate}, and which background state (PI, PD, or another) is most relevant.

Next, in Section~\ref{sec:3.2}, we outline the temperature model and its parameterization, which is primarily based on the work of~\cite{Geoffroy2013, Geoffroy2013a}, with slight adjustments to account for external radiative forcing factors.\footnote{Radiative forcing factors are constituents or processes that disturb Earth’s energy balance by altering the net flux of incoming solar or outgoing long-wave radiation. Typical examples include long-lived greenhouse gases (e.g.\ CO$_2$, CH$_4$, N$_2$O), short-lived species such as aerosols or tropospheric ozone, surface-albedo changes from land-use, and variations in solar irradiance or volcanic emissions.}

To build a bridge to practical applications, Appendix~\ref{APX:D} provides a concise guide for climate economists, explaining how to deploy each emulator presented in this paper to various targets, including PI, PD, multi-model mean, and extreme scenarios.

\subsection{Carbon Cycle Fitting Procedure} \label{sec:3.1}
%

To fit the CCE’s free parameters to the data shown in the left panel of Figure~\ref{fig:1}, we apply a constrained optimization procedure based on the pulse-decay dynamics of \citet{joos2013carbon}.\footnote{A detailed discussion of why we chose this data set to fit the CCE is provided in~\cite{folini2024climate}.} The calibration proceeds in two steps, detailed in Sections~\ref{sec:3.1.1} and~\ref{sec:3.1.2}. First, we adjust the parameters so that the model reproduces the mean atmospheric pulse-decay trajectory derived from benchmark simulations. Second, we rescale this calibrated model to capture extreme carbon-cycle responses.

\begin{figure}[ht!]
\centering
    \includegraphics[width=\textwidth]{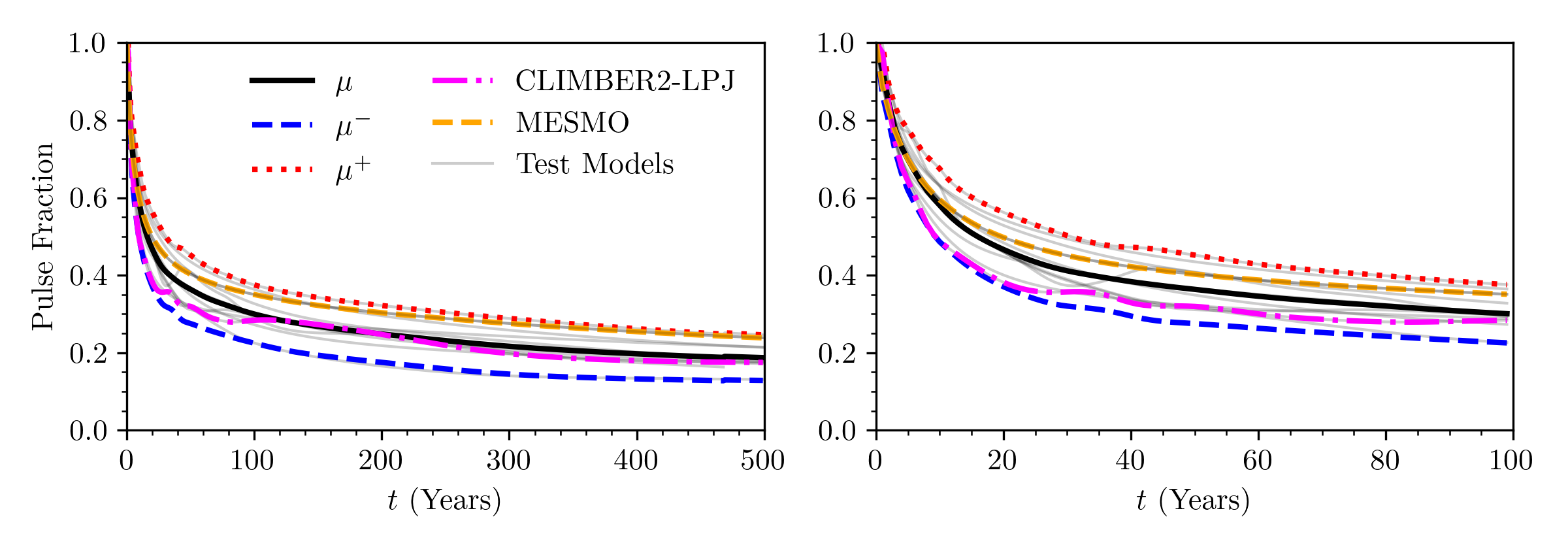}
    \caption{
    %
    Simulated atmospheric CO$_2$ decay after a $100$ GtC pulse under PI equilibrium conditions with experimental data from \cite{joos2013carbon}.
    In the left and right panels, we show the decay trajectories of different models for $0–500$ and $0–100$ years, respectively.
    Shown is the simulated pulse decay across various test models (thin gray lines), including ESMs and EMICs.
    Shown also is the multi-model mean ($\mu$, thick black line), with $\mu^{+}$ (thick dashed blue) and $\mu^{-}$ (thick dotted red) representing two standard deviations above and below the multi-model mean, respectively.
    CLIMBER2-LPJ~\citep{sitch2005impacts} and MESMO~\citep{gmd-14-2265-2021} illustrate two models with fast and slow decay trajectories, respectively, representing extremes relative to the multi-model mean.
    }
    \label{fig:2}
\end{figure}

The simulation benchmarks are the atmospheric \(\mathrm{CO}_{2}\) decay trajectories generated by several carbon–cycle models after a \(100\;\text{GtC}\) pulse is injected into an initially equilibrated system (i.e., no net carbon flux between reservoirs).\footnote{See \url{https://climatehomes.unibe.ch/~joos/IRF_Intercomparison/results.html} for further details.} Figure~\ref{fig:2} shows the benchmark data: the multi-model mean, denoted $\mu$, together with $\mu^+$ and $\mu^-$, which lie two standard deviations above and below $\mu$, respectively, as reported by \citet{joos2013carbon}. For reference, two individual models, CLIMBER2-LPJ and MESMO, illustrate very fast and slow pulse-decay responses. 
%
%

\subsubsection{Multi-Model Mean Calibration} \label{sec:3.1.1}

Let $\bb{y}^{\mu} \in \mathbb{R}^{T}$ denote the atmospheric decay trajectory for the $\mu$-benchmark, that is, the multi-model mean, for $T$ years after the introduction of the $100$ GtC pulse.\footnote{Throughout we use the superscript \(\mu\) for quantities derived from the benchmark data set, here the multi-model mean.  Subsequent sections introduce analogous notation for other data sets.}
The fit error, which is measured here based on the deviation between the emulated atmospheric masses and the multi-model mean, that is, our so-called $\mu$-benchmark, is defined as follows:\footnote{Note that, whereas \citet{folini2024climate} employed the maximum~($\ell_\infty$) norm, we use the Euclidean~($\ell_2$) norm in the present work; both choices are equally defensible. Their choice of the maximum norm was motivated by obtaining a better fit to the early part of the pulse decay, whereas the $\ell_2$-norm places relatively greater weight on the tail of the decay. 
Our fitting period spans 250~years, compared with the 100~years used in that study.}
 \begin{align}\label{eq:7}
 	\mathcal{L}(\bb{a}^{\mu},\bbt{m}^{\mu} ):=\frac{1}{T} \Big \| \bb{M}[\ameq]^\text{A} - \bb{y}^\mu \Big \|_2,
 \end{align}
For example, in the $4$PR model, $\bb{a}^{\mu}$ is a three-element vector representing carbon mass transfer rates between different reservoirs (see Figure~\ref{fig:1} for details).
Likewise, $\bbt{m}^{\mu}$ is a four-element vector containing the equilibrium carbon mass of each reservoir.
The atmospheric equilibrium mass is fixed to the estimated preindustrial value $589$ GtC (\citealp{IPCC_carbon_cycle}; see Appendix~\ref{APX:C} for the present-day calibration).

Since the benchmark data only contains atmospheric carbon masses, the models will be over-parameterized relative to the objective.
Consider the $4$PR model with atmospheric equilibrium fixed: six parameters are fitted to a single output $\bb{m}_t^{\text{A}}$, while the other carbon reservoirs remain free to take on any value.
This is problematic because it results in a highly under-constrained model, where different parameter sets can yield the same atmospheric reservoir mass, but vastly different carbon distributions across the other reservoirs.
To address this, we introduce three penalty terms, $q_1$, $q_2$, and $q_3$, which penalize deviations from physically motivated behavior observed in comprehensive carbon cycle simulations. 
%
For the models considered in this work, we have selected penalty functions to target deviations in dynamic timescales ($q_1$), equilibrium mass variability ($q_2$), and reservoir absorption ratios ($q_3$).\footnote{
Additional penalty terms may be introduced to capture further physical aspects of the carbon cycle, but they must be chosen carefully to avoid redundancy.
Each penalty should target a specific feature, such as a statistical regularity or a dynamical constraint, identified from observations or large-scale simulations.
}
Given the non-negative scalar hyperparameters $\rho_1$, $\rho_2$, and $\rho_3$, each corresponding to the respective penalty functions, the $\mu$-benchmark fitted parameters are
 \begin{align}\label{eq:8}
	\Big \{\bb{a}^{\mu},\bbt{m}^{\mu} \Big \} = \argmin{\{ \bb{a},\bbt{m} \}}  
	\Bigg\{  
		\mathcal{L}(\bb{a},\bbt{m})+
		\rho_1 q_1(\ameq)   +  
		\rho_2 q_2(\bbt{m}) +   
		\rho_3 q_3(\ameq)  
	\Bigg\}.
 \end{align}
In our tests, we use $T=250$, aligning with the time scales associated with the carbon exchange between the atmosphere and the Earth's surface, ranging from decades to centuries.
Below, we provide a detailed description of the penalty functions.

 
\paragraph{Dynamic Timescales ($q_1$):}
%
%

In addition to surface–atmosphere carbon exchange, we also account for the deep-ocean carbon cycle, which operates on centennial to millennial timescales~\citep{IPCC_carbon_cycle}.
As such, we penalize model parameters $\bb{a}$ and $\bbt{m}$ that yield an operator $\bb{A}$ with short dynamic timescales.
Each eigenvalue $\lambda_i \leqslant 0$ of $\bb{A}$ defines an exponential decay mode with dynamic timescale $\tau_i = 1/|\lambda_i|$ (see, e.g., \citealp[Chap.~5]{aastrom2021feedback}); smaller $|\lambda_i|$ therefore implies slower decay and thus larger values of $\tau_i$.  
To enforce this characteristic in the model, we propose the following penalty function
\begin{align}\label{eq:9}
  q_1(\ameq) := -\frac{1}{n}\,\mathrm{tr}\bigl(\bb{A}[\ameq]\bigr)
              = -\frac{1}{n}\sum_{i=1}^{n}\lambda_i=  \frac{1}{n}\sum_{i=1}^{n} | \lambda_i|.
\end{align}
Notice this penalty function is strictly non-negative because every $\lambda_i\leqslant 0$.
When included in the objective, this penalty function suppresses the average eigenvalue magnitude of the operator, promoting slower system dynamics and longer dynamic timescales.

\paragraph{Equilibrium Mass Variability ($q_2$):}

The optimization problem stated in Equation~\eqref{eq:8} is non-convex, as is evident from expression Equation~\eqref{eq:4}, which depends only on the ratios of $\bbt{m}$, making any positive scalar multiple of $\bbt{m}$ equally valid in determining $\bb{A}$.
Consequently, the fitted parameters $\bbt{m}$ (and thus $\bb{a}$) may vary substantially across equally valid solutions.
%
%
To enforce consistency with observational constraints, we add a penalty term for squared deviations between the model’s equilibrium reservoir masses and their target values, where the targets here are the published estimates of PI (or, alternatively, PD) carbon stocks.
Let $\bbt{m}^*$ denote these estimated equilibrium masses.
We penalize the relative difference between $\bbt{m}$ and a reference equilibrium vector $\bbt{m}^*$ using the penalty function:
\begin{align}\label{eq:10}
	q_2(\bbt{m}) := \frac{1}{n} \Big\| (\bbt{m} - \bbt{m}^*) \oslash \bbt{m}^* \Big\|_2,
\end{align}
where $\oslash$ denotes element-wise division.
In our tests, $\bbt{m}^*$ is defined based on the PI estimates from~\cite{IPCC_carbon_cycle}, where the equilibrium masses of the reservoirs for the atmosphere, upper ocean, lower ocean, and land biosphere are $589$, $900$, $37100$, and $550$ GtC, respectively.
We emphasize that the values of $\bbt{m}^*$ do not represent the absolute total equilibrium masses of carbon in each reservoir.
Instead, these values estimate the carbon masses most actively involved in the carbon cycle dynamics.
In contrast, carbon stored in fossil fuel reserves, permafrost, and deep soil organic matter is generally sequestered on timescales ranging from centuries to millennia, and is therefore excluded from equilibrium carbon mass estimates.\footnote{However, carbon stored in permafrost can rapidly re-enter the active carbon cycle when disturbed, particularly through climate-induced thawing. While this represents a potentially significant perturbation to the carbon cycle, it is not accounted for in the present analysis and warrants further investigation. }

\paragraph{Reservoir Absorption Ratios ($q_3$):}

Under a $100$ GtC pulse, the $4$PR model may achieve a low fit error while exhibiting negligible flux to the land biosphere, effectively replicating the behavior of the $3$SR model (cf. Section~\ref{sec:3} for further details).
This observation is specific to the $4$PR configuration, where parallel carbon transfer paths---from the atmosphere to either the ocean or the land biosphere---allow for an arbitrary partitioning of carbon.
To better reflect the behavior of more complex Earth System Models (ESMs), we will aim to mimic the ratio of cumulative ocean to land biosphere fluxes resulting from pulse emissions.
Let $\eta$ denote the target ratio of cumulative ocean to land biosphere carbon masses at time $t^{\text{ref}}$; we penalize deviations from this ratio using the following penalty function
\begin{align}\label{eq:11}
	q_3(\ameq) := \left \| \frac{  
	\bb{M}[ \ameq ]^{\text{O}}_{t^{\text{ref}}}
	}{
	\bb{M}[ \ameq]^{\text{L}}_{t^{\text{ref}}}} - \eta \right \|_2,
\end{align}
where $\bb{M}[\bb{a},\bbt{m}]_{t^{\text{ref}}}^{\text{O}}$ is the total mass of all ($\text{O}_1$ and $\text{O}_2$) ocean reservoirs, and $\bb{M}[\bb{a},\bbt{m}]_{t^{\text{ref}}}^{\text{L}}$ denotes the total mass of all land biosphere reservoirs.
Following the findings of~\cite{joos2013carbon}, we set $\eta = 1$, corresponding to equal distribution of $30$ GtC between ocean and land biosphere reservoirs at $t^{\text{ref}} = 20$ years after a $100$ GtC atmospheric pulse.

%

\subsubsection{Calibration For Extreme Scenarios} \label{sec:3.1.2}
%

As demonstrated in Figure~\ref{fig:2}, different ESMs exhibit varying $100$ GtC pulse decay trajectories.
In this section, we aim to capture the extrema, defined as two standard deviations above and below the mean atmospheric response across the trajectories of different ESMs, labeled as the $\mu^+$-benchmark and $\mu^-$-benchmark, respectively. 
%
%
We emphasize that the benchmarks used here (and in the previous section, too) do not correspond to any specific ESM but represent plausible upper and lower bounds for atmospheric carbon masses.

A simple way to calibrate the model for the different extrema is to rescale the parameters $\bb{a}^\mu$ obtained in Equation~\eqref{eq:8}.
%
%
This formulation allows IAM researchers to interpolate smoothly between the mean response and its extremes, eliminating the need to re-calibrate the CCE each time a more or less extreme scenario is examined.

Given the pulse decay trajectories $\bb{y}^{\mu^+}\!\!\!, \bb{y}^{\mu^-}\!\!\! \in \mathbb{R}^T$, we define extreme scaling factors as
\begin{align}\label{eq:12}
  c^{\mu^+} \!= \argmin{c} \Bigg\{ \frac{1}{T} \Big \| \bb{M}[c \cdot \bb{a}^\mu, \bbt{m}^{\mu}]_\text{A} - \bb{y}^{\mu^+} \Big \|_2 \Bigg\},\; \text{and} \;
  c^{\mu^-} \!= \argmin{c} \Bigg\{ \frac{1}{T} \Big \| \bb{M}[c \cdot \bb{a}^\mu, \bbt{m}^{\mu}]_\text{A} - \bb{y}^{\mu^-} \Big \|_2 \Bigg\}.
\end{align}
For $\bb{A}^{\mu}\!\! := \bb{A}[\bb{a}^\mu, \bbt{m}^\mu]$, we define the respective extrema operators as $\bb{A}^{\mu^+}\!\!:= c^{\mu^+}\!\! \cdot \bb{A}^{\mu}$ and $\bb{A}^{\mu^-}\!\!:= c^{\mu^-}\!\!\cdot \bb{A}^{\mu}$.
In our test, we keep $T$ consistent with the value used in Equation~\eqref{eq:8}.
Note that $\bb{A}^{\mu^+}$ and $\bb{A}^{\mu^-}$ are scaled versions of $\bb{A}^{\mu}$, with all eigenvalues multiplied by $c^{\mu^+}$ and $c^{\mu^-}$, respectively.\footnote{
Specifically, $\bb{A}[c \cdot \bb{a}, \bbt{m}]$ is equivalent to $c \cdot \bb{A}_{ij}$; see Equation~\eqref{eq:4} for details.
} 
Scaling the eigenvalues also shifts the range of dynamic timescales. 
The solutions in Equation~\eqref{eq:12} will satisfy $c^{\mu^+} <1<c^{\mu^-}$.
Due to the linearity of the carbon cycle model, we can formulate a parameterized weighted operator that satisfies the conditions outlined in Section~\ref{sec:2.1}, including the equilibrium condition and mass conservation.
Given $\alpha \in [-1,1]$, we write the weighted operator as 
\begin{align}\label{eq:13}
	\bb{A}^{\alpha} := 
	\begin{cases}
		(1-\alpha)\bb{A}^{\!\mu}  +  \alpha \bb{A}^{\mu^+}   & \text{if $\alpha > 0$,} \\
		(1+\alpha)\bb{A}^{\!\mu}  -  \alpha \bb{A}^{\mu^-}   & \text{if $\alpha \leqslant 0$,}
	\end{cases}
\end{align}
where $\alpha=0$ represents the mean atmospheric carbon content across various ESMs.
Setting $\alpha=1$ or $\alpha=-1$ emulates ESMs with higher or lower atmospheric carbon content, respectively, corresponding to slower or faster carbon absorption from the atmosphere.

\subsubsection{The Choice of Hyperparameters} \label{sec:3.1.3}
\begin{figure}[t]
\noindent 
\begin{minipage}{0.5\textwidth}
	\noindent
	\centering
	\hspace{2em}\small{\textsc{3SR}}
\end{minipage}%
\begin{minipage}{0.5\textwidth}
	\noindent
	\centering
	\small{\textsc{4PR}}
\end{minipage}
\vspace{-1em}
\includegraphics[width=\textwidth]{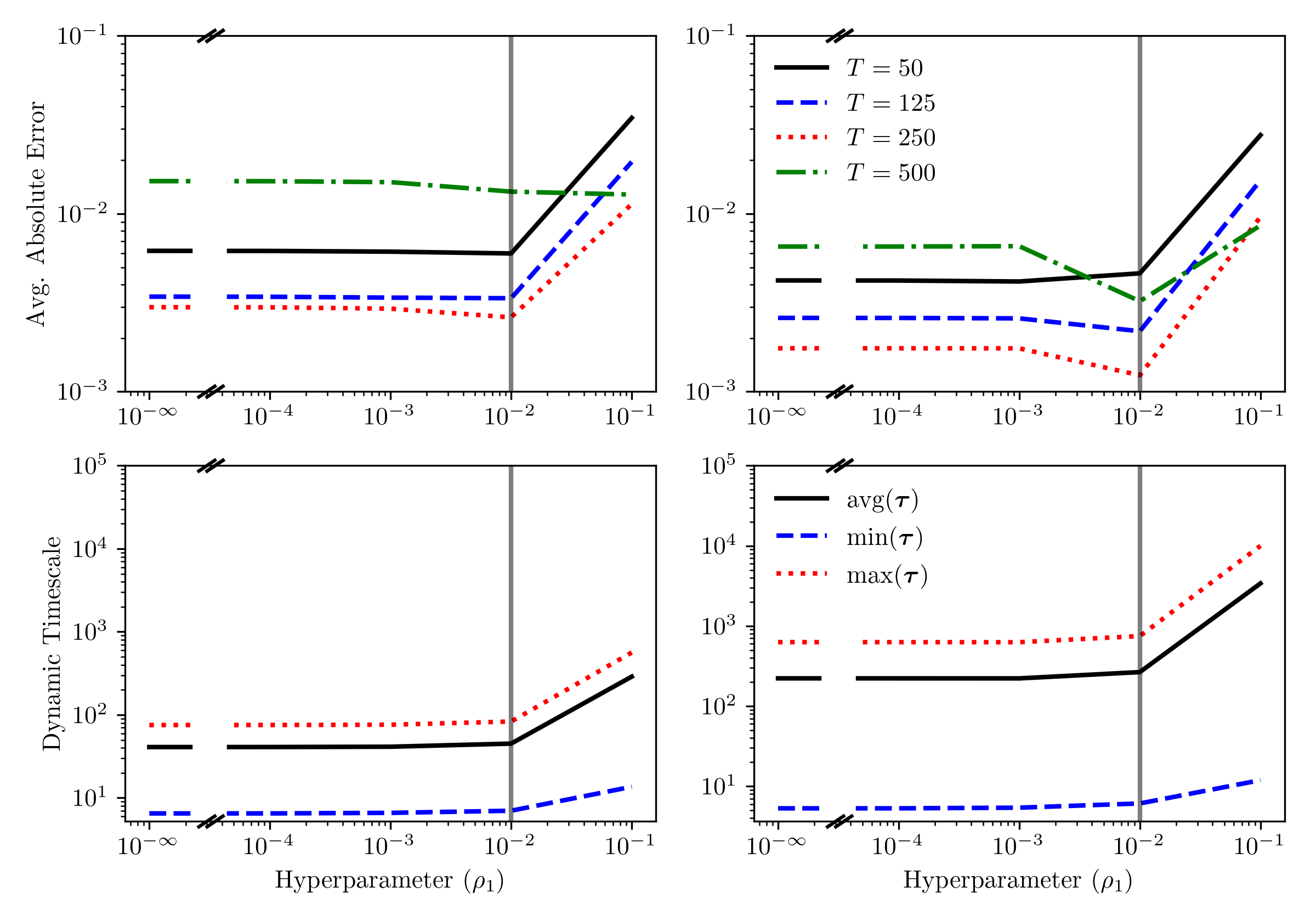}
\caption{
    The top panel shows the average absolute error $(\|\mathbf{m}_{\text{A}} - \bb{y}^{\mu}\|_1 / T)$, and the bottom panel shows the dynamic timescales $\bb{\tau}$ for the $3$SR (left) and $4$PR (right) models across varying $\rho_1$, with $\rho_2 = \rho_3 = \expnum{1}{-4}$ fixed.
    The average absolute error represents the mean absolute difference between the emulator simulation and the $\mu$-benchmark over a time span of $T$ years following a $100$ GtC carbon pulse.
    The vertical line indicates the selected value of $\rho_1 = \expnum{1}{-2}$.
}
\label{fig:3}
\end{figure}
%
In this section, we describe the methodology and criteria used to select the hyperparameters $\rho_1$, $\rho_2$, and $\rho_3$ in Equation~\eqref{eq:10}, and report the chosen values for each model.
As a general principle, smaller values for these hyperparameters are preferred to avoid the penalty terms overwhelming the model fit error in the optimization objective Equation~\eqref{eq:7}.

We begin with $\rho_2$ and $\rho_3$, which are associated with the penalty terms for equilibrium mass variability and reservoir absorption ratios, respectively (cf. Section~\ref{sec:3.1.1}).
For the $4$PR model configuration, our numerical results indicate that setting $\rho_2 = \rho_3 = \expnum{1}{-4}$ minimizes the fit error while achieving desirable low equilibrium mass variations and reservoir absorption ratios. 
The results are similar for the $3$SR configuration, except that $q_3$ has no effect due to the absence of a land biosphere; thus, we select $\rho_2 = \expnum{1}{-4}$ and $\rho_3=0$.
For a detailed analysis and numerical results on model behavior related to $\rho_2$ and $\rho_3$ see Appendix~\ref{sec:appendix:additional_experiments}.

With $\rho_2$ and $\rho_3$ fixed, we now focus on selecting an appropriate value for $\rho_1$.
To justify our choice, we evaluate the model under varying values of $\rho_1$, analyzing (i) the minimum, average, and maximum dynamical timescales, and (ii) the average absolute error in atmospheric carbon content relative to the $\mu$-benchmark over the time horizon $50 \leqslant T \leqslant 500$ (cf. Figure~\ref{fig:2}).
This analysis is visualized in Figure~\ref{fig:3}.
For sufficiently small values of $\rho_1$ (those for which the penalty function does not overwhelm the fit error), the $3$SR model exhibits minimal sensitivity in terms of both average absolute error and dynamical timescales, primarily due to its reduced number of parameters compared to the $4$PR model.
In contrast, for the $4$PR model, increasing $\rho_1$ leads to a notable reduction in average  absolute error, particularly at longer simulation times $T$. 
In both model configurations, increasing $\rho_1$ results in a relatively small increase in the dynamic timescales.
Importantly, selecting relatively small hyperparameter values, even when they have little effect on the estimated parameters, can significantly improve the numerical stability and time to solution of the optimization problem in Equation~\eqref{eq:8}.
This observation is well established in the numerical optimization literature (see, for example, \cite{GVK502988711} for a detailed discussion of conditioning and the effects of regularization).
We proceed with selecting $\rho_1=\expnum{1}{-2}$ for both the $3$SR and $4$PR model  configurations.
  
\subsubsection{Fitted Model Parameter}
\begin{table}[t]
\setlength{\tabcolsep}{5pt}
\centering
\small
\begin{tabular}{cccccccccc||cc}
& 
\multicolumn{9}{c}{\textsc{Fitted parameter values}} &  \\
&
	$\bb{a}^{\mu}_{\text{A}   \to \text{O}_1}$ & 
	$\bb{a}^{\mu}_{\text{O}_1 \to \text{O}_2}$ & 
	$\bb{a}^{\mu}_{\text{A}   \to \text{L}  }$ &
	$\bbt{m}^{\mu}_{\text{A}}$    &
	$\bbt{m}^{\mu}_{\text{O}_1}$  &
	$\bbt{m}^{\mu}_{\text{O}_2}$  &
	$\bbt{m}^{\mu}_{\text{L}}$   &
	$c^{\mu^+}$ &
	$c^{\mu^-}$ &
$\bb{m}^{\text{O}}_{t^{\text{ref}}}/\bb{m}^{\text{L}}_{t^{\text{ref}}} $ & 
$\tau_1/\tau_2/\tau_3$
\\[4pt]
\toprule
\toprule
\textit{Lower}& ${\expnum{1}{-6}}$ & ${\expnum{1}{-6}}$  & ${\expnum{1}{-6}}$ & ${589}$ &  ${\expnum{1}{-6}}$    & ${\expnum{1}{-6}}$  & ${\expnum{1}{-6}}$ & ${\expnum{1}{-6}}$  & ${1}$ &  &\\
\textbf{3SR} & $\num{0.0769419}$ & $\num{0.0109353}$ & - & $589$ & $752 $    & $1,\!289 $  & - & $\num{0.47464535096875327}$ & $\num{2.4558563750016473}$ & - & $7/83/-$ \\
\textbf{4PR} & $\num{0.0208104}$ & $\num{0.0025498}$ & $\num{0.0613352}$ & $589$ & $1,\!078$ & $37,\!220$ & $387$ & $\num{0.47006381598196945}$ & $\num{2.407426003806048}$ & $\num{0.71500}$&$6/42/748$\\
\textit{Upper}& ${0.3}$ & ${0.3}$ & ${0.3}$  & ${589}$ & ${1,800}$ & ${74,\!200}$ & ${1,\!100}$ & ${1}$ & ${5}$ &  &\\
\end{tabular}
\caption{The table presents the calibrated parameter values obtained using the $\mu$-benchmark, based on the PI pulse decay experiments reported by~\cite{joos2013carbon}.
    For calibration using PD pulse decay data, see Appendix~\ref{APX:C}.
    The table also lists the \emph{Lower} and \emph{Upper} search bounds for the optimization problems in Equation~\eqref{eq:8} and Equation~\eqref{eq:12}, specifying the allowable range for each parameter (see Sections~\ref{sec:3.1} and~\ref{sec:3.2} for details on model calibration and the fitting procedure).
    In addition, it includes the ratio of ocean to land biosphere pulse absorption $(\bb{m}^{\text{O}}_{t^{\text{ref}}}/\bb{m}^{\text{L}}_{t^{\text{ref}}})$ and the model's dynamic timescales $(\tau_1, \tau_2, \tau_3)$.
    The ocean-to-land pulse absorption ratio is evaluated at $t^{\text{ref}}=20$, see Section~\ref{sec:3.1.1} for penalty functions.
    The reported values are based on the hyperparameters $\rho_1 = \expnum{1}{-2}$, $\rho_2 = \expnum{1}{-4}$, and $\rho_3 = 0$ for the $3$SR model, and $\rho_1 = \expnum{1}{-2}$, $\rho_2 = \expnum{1}{-4}$, and $\rho_3 = \expnum{1}{-4}$ for the $4$PR model.
}
\label{tab:1}
\end{table}
%
Given the hyperparameters outlined in the previous section, we now perform the multi-model mean and extrema calibrations, as discussed in Sections~\ref{sec:3.1.1} and~\ref{sec:3.1.2}, respectively.
Table~\ref{tab:1} presents the resulting calibrated model parameters $\bb{a}^{\mu}$ and $\bbt{m}^{\mu}$, obtained as solutions to Equation~\eqref{eq:10}, along with the scaling factors $c^{\mu^+}$ and $c^{\mu^-}$, which are the solutions to Equation~\eqref{eq:12}.\footnote{
    The notation $\bb{a}^\mu$ refers to carbon transfer coefficients between the respective reservoirs, which can also be expressed in matrix form.
    For example, in the $4$PR model: $\bb{A}_{2,1}:=\bb{a}^{\mu}_{\text{A} \to \text{O}_1}$, $\bb{A}_{3,2}:=\bb{a}^{\mu}_{\text{O}_1 \to \text{O}_2}$, and $\bb{A}_{4,1}:= \bb{a}^{\mu}_{\text{A} \to \text{L}}$.
    In the $3$SR model, the same notation applies, except that the corresponding rows and columns of $\bb{A}$ related to the land biosphere are omitted.
}
We provide a high-level discussion of these results and their implications for model structure and dynamics below. When calibrated to either pre-industrial or present-day conditions, the three- and four-box emulators reproduce the historical and 500-year atmospheric CO$_2$ and global-mean temperature trajectories with fitting errors below 5\% and 3\%, respectively (see the lower panels of Figures~\ref{fig:5} and~\ref{fig:6_pd} in the Appendix). This level of accuracy is widely regarded as ``fit for purpose'' in policy analysis.

The parameters are all estimated within prescribed lower and upper bounds, and provided in Table~\ref{tab:1}.
The first observed difference between the $3$SR and $4$PR model parameters lies in the mass transfer coefficients $\bb{a}^\mu$, which are consistently smaller in the $4$PR model compared to the $3$SR model.
This follows directly from the $4$PR model structure, where atmospheric carbon is split between parallel fluxes to the ocean and land biosphere, allowing smaller individual transfer coefficients to yield the same net outflow.
A second key difference between the $3$SR and $4$PR models concerns the range of dynamic timescales each can represent.
The $3$SR model, with only two nonzero eigenvalues, captures two distinct timescales, whereas the $4$PR model, with three eigenvalues, resolves three timescales, including a much larger dynamic timescale associated with slower processes.
A noted limitation of the $3$SR model is its tendency to average the medium- and long-term dynamics, as reflected in the estimated dynamic timescale in Table~\ref{tab:1} (also observable in the lower panel of Figure~\ref{fig:3}).
For the equilibrium masses $\bbt{m}^\mu$, the $4$PR configuration yields significantly larger values than the $3$SR configuration and closely aligns with those reported in~\cite{IPCC_carbon_cycle} (see discussion related to Equation~\eqref{eq:10}).

Using the calibrated parameters for the multi-model mean (i.e., calibration based on the $\mu$-benchmark), we now turn to the extrema parameters $c^{\mu^+}$ and $c^{\mu^-}$, which are defined according to Equation~\eqref{eq:12}, and which are calibrated using the $\mu^+$ and $\mu^-$ benchmarks, respectively.
As shown in Table~\ref{tab:1}, these values are approximately the same for both the $3$SR and $4$PR configurations.
This is consistent with our results in Appendix~\ref{sec:4.1}, as both the $3$SR and $4$PR models exhibit similar atmospheric pulse decay for the $\mu$-benchmark (since both are calibrated to this benchmark); consequently, the extrema parameters, which are based solely on atmospheric concentrations, are expected to coincide.
However, this does not imply that the two models exhibit similar dynamics under general conditions.
As shown in the extended results in Appendix~\ref{sec:appendix:additional_experiments} and further discussed in Appendix~\ref{sec:4.1}, the $3$SR and $4$PR models exhibit substantial differences in carbon distribution across individual reservoirs and in their responses to general atmospheric carbon perturbations (i.e., those not used in the calibration procedure).

%

%
%
%

\subsection{Temperature Model}\label{sec:3.2}
\begin{table}[t]
\setlength{\tabcolsep}{5pt}
\centering
\small
  \begin{tabular}{cccccc}
    \multicolumn{6}{c}{\textsc{Parameters for Two-layer energy-balance model}}   \\
    & $C$ & $C_0$ & $\gamma$ & $\lambda$ & $\mathcal{F}_{\mathrm{4 \times \text{CO}_2}}$ \\
    & $\scriptstyle(\mathrm{W\,yr\,m^{-2}\,K^{-1}})$ 
    & $\scriptstyle(\mathrm{W\,yr\,m^{-2}\,K^{-1}})$ 
    & $\scriptstyle(\mathrm{W\,m^{-2}\,K^{-1}})$ 
    & $\scriptstyle(\mathrm{W\, m^{-2}\, K^{-1}})$ 
    & $\scriptstyle(\mathrm{W \,m^{-2}})$\\
    \toprule
    \toprule
    Multi-model mean   & $7.3$ & $106$ & $0.73$ & $1.13$ & $6.9$ \\
    Standard deviation& $1.1$  & $62$ & $0.18$ & $0.31$ & $0.9$ \\
  \end{tabular}
\caption{
    The parameters used for the two-layered energy balance model outlined by~\cite{Geoffroy2013}, representing the global thermal characteristics of coupled atmosphere-ocean general circulation models within a two-layer energy balance framework.
    For our purposes, we assume $\mathcal{F}_{\mathrm{4 \times \text{CO}_2}} = 2\mathcal{F}_{\mathrm{2 \times \text{CO}_2}}$.
}
\label{tab:2}
\end{table}

%
In this section, for completeness, we present the equations governing the evolution of temperature.\footnote{As discussed in the introduction, this aspect of climate emulators is extensively treated in \cite{folini2024climate} and is therefore beyond the scope of this paper.}
The temperature change is given by a time-dependent radiative forcing amplitude parameter $\mathcal{F}_t$ at discrete time $t$, along with the following set of free parameters: the combined heat capacity of the atmosphere, upper ocean (and land biosphere) $C$; the deep-ocean heat capacity $C_0$; the heat exchange coefficient $\gamma$; and the radiative feedback parameter $\lambda$ (not to be confused with the eigenvalues discussed earlier).
The summary of these parameters is presented in Table~\ref{tab:2}.
%
Let $\mathbf{T}_t = (\bb{T}_t^{\text{A}},\bb{T}_t^{\text{O}})^{\top} \in \mathbb{R}^2$ denote global temperature at time $t$ in a two-layer model.  The first component, $\bb{T}_t^{\text{A}}$, is the temperature of the atmosphere, upper ocean, and land biosphere (hereafter simply ``atmospheric''), while the second, $\bb{T}_t^{\text{O}}$, is the temperature of the deep ocean (hereafter ``oceanic'').
%
%
The temperature dynamics are modeled as a first-order system of difference equations, expressed as
\begin{align}\label{eq:14}
   \bb{T}_{t+1} - \bb{T}_{t} = \Delta t \cdot ( \bb{B}\bb{T}_{t} +\bb{b}_{t}),\; \text{where} \;
    \bb{B}:= 
    \begin{bmatrix}  
    -(\lambda + \gamma)/C & \gamma /C \\
    \gamma/C_0             & -\gamma / C_0 
    \end{bmatrix} ,\; \text{and}\;
    \bb{b}_{t}:=
    \begin{bmatrix}  
    \mathcal{F}_{t}/C \\ 0
    \end{bmatrix}.
\end{align}

The total radiative forcing amplitude is modeled as a function of the relative CO$_2$ concentration, following the formulation of~\citet{Geoffroy2013} and \cite{houghton1990climate}.
To account for additional factors, we scale the CO$_2$ relative forcing by a factor of $\kappa$, giving
\begin{align}\label{eq:15}
\mathcal{F}_{t} := \kappa \cdot \frac{\mathcal{F}_{\mathrm{2 \times \text{CO}_2}}}{\ln(2)} \cdot \ln\left( \frac{\bb{m}_{t}^{\text{A}}}{\bb{m}_0^{\text{A}}}\right),
\end{align}
where  $\bb{m}_{t}^{\text{A}}/\bb{m}_{0}^{\text{A}}$ is the ratio of atmospheric CO$_2$ masses (or concentrations) at time $t$ relative to pre-industrial (PI) levels, and $\mathcal{F}_{\mathrm{2 \times \text{CO}_2}}$ denotes the net radiative forcing associated with a doubling of atmospheric CO$_2$ concentration.
%
Alternatively, explicit definitions for the temperatures of the atmosphere ($\bb{T}_{t}^{\text{A}}$), as well as the ocean ($\bb{T}_{t}^{\text{O}}$) are provided below: 
\begin{align}
 \bb{T}_{t+1}^{\text{A}} &= \bb{T}_{t}^{\text{A}} + \frac{1}{C} \left(
 \kappa \cdot \frac{\mathcal{F}_{\mathrm{2 \times \text{CO}_2}}}{\ln(2)} \cdot \ln\left( \frac{\bb{m}_{t}^{\text{A}}}{\bb{m}_0^{\text{A}}}\right)  
 -\gamma \cdot (\bb{T}_{t}^{\text{A}} - \bb{T}_{t}^{\text{O}} ) 
 -\lambda \cdot \bb{T}_{t}^{\text{A}} \right), \\ 
 \bb{T}_{t+1}^{\text{O}} &= \bb{T}_{t}^{\text{O}} + \frac{\gamma}{C_0} \left( \bb{T}_{t}^{\text{A}} - \bb{T}_{t}^{\text{O}} \right).  \nonumber
\end{align}
%
%
We fix $\kappa = 1.2$ as the scaling factor for CO$_2$ forcing across all Representative Concentration Pathway (RCP) scenarios~\citep{folini2024climate}; this is discussed in more detail in Appendix~\ref{sec:4.2}. 
For the controlled atmospheric perturbation tests in Appendix~\ref{sec:4.1}, which are based solely on synthetic CO$_2$ forcing, we use $\kappa = 1$.

 \section{Numerical Experiments} \label{sec:4}

This section systematically investigates how alternative emulator designs influence outcomes once they are embedded in a dynamic economic model. After presenting the core model in Section~\ref{sec:4.3}, we analyze (i) a business-as-usual (BAU; no-policy) case in Section~\ref{sec:BAU_econ}, (ii) an optimal-mitigation case in Section~\ref{sec:optimal_mitigatoin}, and (iii) a hypothetical carbon-capture scenario in Section~\ref{sec:carbon_storage}.\footnote{Additional numerical experiments drawn from the climate-science literature, which validate the full emulator framework, that is, combining the carbon-cycle and temperature modules discussed in Sections~\ref{sec:3.1} and~\ref{sec:3.2}, are reported in Appendix~\ref{sec:appendix:additional_experiments}.}
For clarity, we briefly recall the three carbon-cycle emulator variants introduced above, which are compared in the experiments to follow:
\begin{itemize}
    \item[(i)] \textbf{$3$SR}: a three-reservoir emulator that sequentially links the atmosphere, upper ocean, and deep ocean;
    \item[(ii)] \textbf{$4$PR}: an extension of $3$SR that adds the land biosphere as a parallel carbon sink from the atmosphere;
    \item[(iii)] \textbf{$4$PR-X}: a $4$PR variant that introduces a time-dependent operator to adjust the land-biosphere equilibrium for land-use-related emissions.
\end{itemize}

%

%

\subsection{A Baseline Model} \label{sec:4.3}

We embed the three CCEs ($3$SR, $4$PR and $4$PR-X) introduced in \cref{sec:2} and validated in standalone tests from climate science (cf.\ \cref{sec:4.1,sec:4.2}) within the DICE-2016 framework~\citep{nordhausRevisitingSocialCost2017}.\footnote{All IAMs were solved using “Deep Equilibrium Nets,” a deep-learning method for dynamic stochastic models; see \cite{Azinovic2022} for the general approach and \cite{Friedl2023} for its IAM application.} Models are calibrated to the multi-model mean and to extreme scenarios (CLIMBER2-LPJ, MESMO), with $3$SR as the baseline and $4$PR/$4$PR-X as the cases of interest. Our objective is to assess whether alternative emulators materially alter macroeconomic outcomes, ensuring a transparent foundation for comparison.

In what follows, we will demonstrate that emulator design, especially the addition of carbon reservoirs or time-dependent feedback, can significantly influence IAM projections. The following analyses provide an illustrative first look at richer climate modules and offer preliminary recommendations on when and how to incorporate additional reservoirs or dynamic operators into climate-economy models.

Our economic set-up consists of a single, infinitely lived, representative consumer and a single firm. We describe the equilibrium allocation as the solution to a social planner problem where the planner maximizes a constant relative risk aversion (CRRA) utility function of per capita consumption, ${C_t}/{L_t}$. Here, $C_t$ represents consumption, $L_t$ denotes labor, with a constant intertemporal elasticity of substitution (IES), $ \psi >1 $, and a discount factor, $0 < \beta < 1$.

The value of the lifetime utility, $ V_0 $ is given  by the following expression:
\begin{align}
\label{eq:obj_2016}
V_0=  & \max_{\left\{C_t, \mu_t\right\}_{t=0}^{\infty}} \sum_{t=0}^{\infty} \beta^{t}
  \frac{\left(\frac{C_t}{L_t}\right)^{1-1/\psi} -1}{1-1/\psi} L_t\\
  \text{s.t.} \quad \label{eq:kplus_2016}
  & K_{t+1} = \left(1  -\Theta\left(\mu_{t}\right) - \Omega\left(\bb{T}^{\text{A}}_{t}\right) \right)
     K_t^{\alpha} \left(A_{t}L_t \right)^{1-\alpha} + (1-\delta) K_t - C_{t}  \\
    \nonumber & \text{Climate \cref{eq:1,eq:14,eq:15}}\\
  & \label{eq:Kplus_nonnegative} 0 \leq K_{t+1} \\
  & \label{eq:mu_range} 0 \leq \mu_t \leq 1
\end{align}
%
where emissions that enter Equation~\eqref{eq:1} are defined by:
\begin{align}
    & \label{eq:emissions}  e_t=\sigma_t Y^{\text{Gross}}_{t}(A_t,K_t,L_t) (1-\mu_t) + E^{\text{Land}}_t,
\end{align}
and where $E^{\text{Land}}_t$ are exogenous land emissions.
Output $Y^{\text{Gross}}_{t}(A_t,K_t,L_t)$ is produced using a Cobb-Douglas technology, with capital $K_t$, total factor productivity (TFP) $A_t$, and labor $L_t$, where $\alpha$ represents capital elasticity. The capital stock depreciates at rate $\delta$. Mitigation efforts, denoted by rate $\mu_t$, are costly and reduce output by a factor $ \Theta(\mu_t)$. Additionally, higher temperatures decrease output through the damage function $ \Omega(\bb{T}^{\text{A}}_{t})$ where $\bb{T}^{\text{A}}_{t}$ denotes atmospheric temperature.\footnote{For the sake of brevity, we do not explicitly state the functional forms of some of the model equations, as well as parametrization, unless it is necessary for presenting the results. Appendix~\ref{APX:D} contains all the relevant information about the complete model specification, detailed parameterization of all equations, including exogenous variables, and the procedure for integrating the climate emulator up to present-day conditions.} 

We first consider the BAU case, where the social planner does not invest in mitigation $\mu_t$. In this scenario, the planner only chooses the investment sequence, setting the mitigation sequence to zero at all times. Next, we consider the optimal mitigation case, where the planner simultaneously chooses investment in both capital and mitigation. In the optimal case, 
we examine the social cost of carbon, defined as the marginal cost of atmospheric carbon in terms of the numeraire good. Following the literature (see, e.g.,~\citealp{Traeger2014} and~\citealp{Cai2019}), we define the SCC as the planner's marginal rate of substitution between atmospheric carbon concentration and the capital stock:
\begin{align}
  \label{eq:SCCdef}
  SCC_{t} = -\frac{\partial V_{t}/\partial
  \bb{m}_{t}^{\text{A}}}{\partial V_{t}/\partial
  K_{t}}.
\end{align}
The SCC equals the optimal carbon tax when $\mu_{t}<1$.

In Sections~\ref{sec:BAU_econ} and \ref{sec:optimal_mitigatoin}, we present benchmark solutions for the BAU and optimal mitigation cases, respectively; Section~\ref{sec:carbon_storage} evaluates emulator performance under a hypothetical carbon capture and storage technology. We also analyze sensitivity to damage functions and discount rates (Appendix~\ref{APX:E}) and report present‐day simulation results in Appendix~\ref{APX:C}.


\subsection{Business-As-Usual}
\label{sec:BAU_econ}

We begin by examining the BAU trajectory across the carbon-cycle emulators introduced above. In this and all subsequent figures, the $3$SR emulator is shown as a solid blue line, the $4$PR emulator as a dashed green line, and the $4$PR-X emulator as a dotted black line.\footnote{Initial land-biosphere stocks differ slightly because each emulator is integrated forward to present-day conditions with its own numerical solver; see Appendix~\ref{APX:D} for details.} The extreme calibrations, MESMO and CLIMBER2-LPJ, are plotted in orange and red, respectively: solid lines denote their $3$SR variants, whereas dashed lines indicate their $4$PR variants.
%
\begin{figure}[H]
    \centering
    \includegraphics[width=\textwidth]{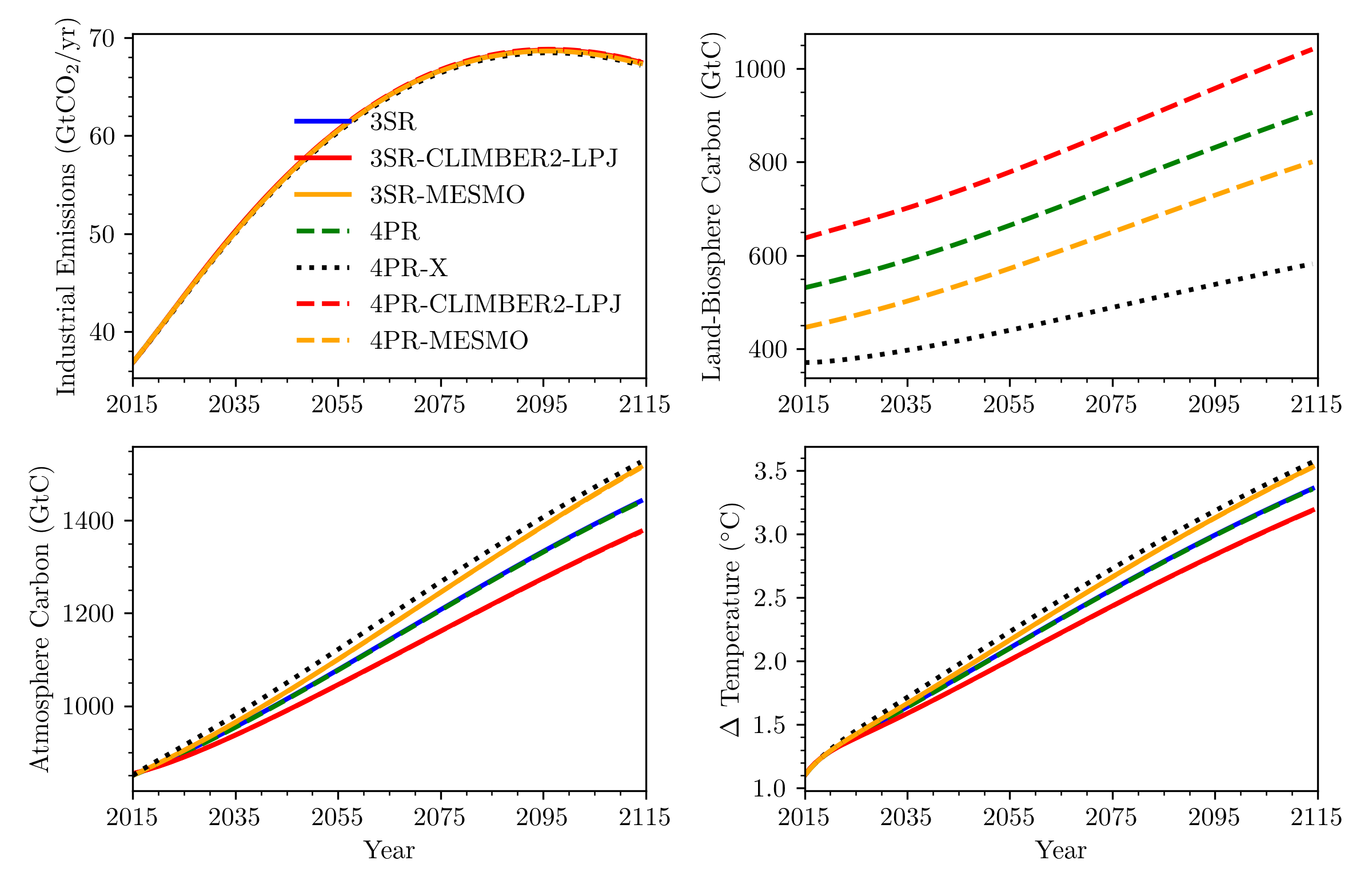}
    \caption{The simulation results for the BAU case ($\mu_t=0$).}
    \label{fig:bau}
\end{figure}
In \cref{fig:bau} (top left panel), we observe that industrial emissions are identical across all carbon cycle models. 
However, the carbon mass in the land biosphere (\cref{fig:bau}, top right panel) and atmosphere (\cref{fig:bau}, bottom left panel) differ between the $4$PR and $4$PR-X models. This variation arises because the dynamic $4$PR-X model simulates a decreasing land biosphere capacity, which leads to increased carbon uptake by other reservoirs, thereby raising atmospheric carbon mass and, consequently, temperature (\cref{fig:bau}, bottom right panel). The rather low carbon content of the land reservoir of the $4$PR-X model in the year 2015 (below 400 GtC and thus slightly outside the estimated range in~\citealp{IPCC_carbon_cycle}) is linked to the fact that we chose to simply remove all land-use change emissions from the land biosphere equilibrium mass instead of applying some scaling as discussed in Section~\ref{sec:2.2}. 
Notably, the extreme variants $3$SR–MESMO and $4$PR–MESMO yield virtually the same atmospheric-carbon burden and temperature rise as the $4$PR-X emulator (Figure~\ref{fig:bau}, bottom panels). This similarity highlights that a diminished land-biosphere sink can produce climate outcomes comparable to the most pessimistic warming scenarios.

\cref{tab:bau} presents the exact values for atmospheric carbon mass and temperatures in 2020, 2050, and 2100, along with the percentage differences between the $3$SR and $4$PR-X models.
\begin{table}[ht]
    \centering
    \begin{tabular}{c|c|c|c|c|c}
        \toprule
        Year & Variable & 3SR (1) & 4PR (2) & 4PR-X (3) & diff (1) \& (3) \\
        \midrule
        2020 & ${\bb{m}_{t}^{\text{A}}}_{2020}$ (GtC) & 874.37 & 873.52 & 883.14 & 8.77 GtC (1.00\%) \\
        2050 & ${\bb{m}_{t}^{\text{A}}}_{2050}$ (GtC) & 1046.07 & 1045.01 & 1085.54 & 39.47 GtC (3.77\%) \\
        2100 & ${\bb{m}_{t}^{\text{A}}}_{2100}$ (GtC) & 1363.19 & 1361.66 & 1439.97 & 76.78 GtC (5.63\%) \\
        \hline
        2020 & $\bb{T}^{\text{A}}_{2020}$ ($^\circ$C) & 1.29 & 1.29 & 1.30 & 0.01 $^\circ$C (0.99\%) \\
        2050 & $\bb{T}^{\text{A}}_{2050}$ ($^\circ$C) & 1.99 & 1.98 & 2.10 & 0.11 $^\circ$C (5.84\%) \\
        2100 & $\bb{T}^{\text{A}}_{2100}$ ($^\circ$C) & 3.09 & 3.09 & 3.29 & 0.20 $^\circ$C (6.34\%) \\
        \bottomrule
    \end{tabular}
     \caption{Data for atmospheric carbon mass (${\bb{m}_{t}^{\text{A}}}$) and atmospheric temperature ($\bb{T}^{\text{A}}_{t}$) across selected years. Absolute differences are $4$PR-X relative to $3$SR, with percentage differences in parentheses.}
    \label{tab:bau}
\end{table}
%
In the BAU case, the carbon masses and temperatures in 2020 show no significant variation across the different carbon cycle models. However, by 2050, differences become substantial: the dynamic $4$PR-X model shows nearly 4\% higher atmospheric carbon and approximately 6\% higher atmospheric temperature than the $3$SR model. 

\begin{figure}[H]
    \centering
    \includegraphics[width=\textwidth]{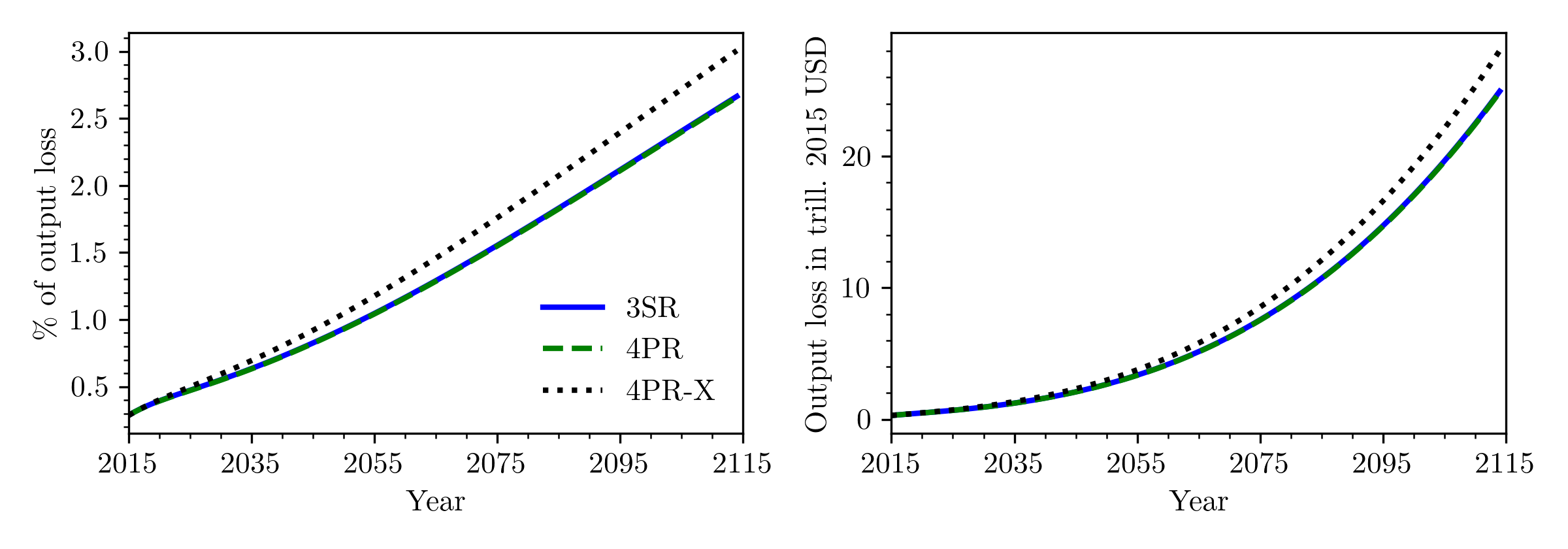}
    \caption{Share of output lost to temperature damages (left) and the resulting monetary loss (right).}
    \label{fig:dam}
\end{figure}

In 2100, the dynamic scenario experiences an additional \(0.2\,^\circ\mathrm{C}\) of warming relative to the static baseline (cf.\ \cref{tab:bau}).  
To translate this temperature increment into economic terms, we measure the fraction of gross output lost to climate damages with the function
\(\Omega\bigl(\mathbf{T}^{\mathrm{A}}_{t}\bigr)\), parameterized by the coefficients \(\psi_1=0\) and \(\psi_2=0.00236\).  
Specifically, 
\begin{align} 
    \label{eq:damages2016}
    \Omega\bigl(\mathbf{T}^{\mathrm{A}}_{t}\bigr)
    &= \psi_1 \mathbf{T}^{\mathrm{A}}_{t} \;+\; \psi_2\bigl(\mathbf{T}^{\mathrm{A}}_{t}\bigr)^2 ,
\end{align}
so that higher atmospheric temperatures \(T^{\mathrm{AT}}_{t}\) lead to an increasingly nonlinear share of output lost.  
The associated absolute loss is obtained by multiplying this share with gross output,
\begin{align}
    \label{eq:damagesGDP}
    D_t \;=\;\Omega\bigl(\mathbf{T}^{\mathrm{A}}_{t}\bigr)\,
    K_t^{\alpha}\bigl(A_t L_t\bigr)^{1-\alpha},
\end{align}
and is reported in trillion 2015 USD. 
Figure~\ref{fig:dam} contrasts the evolution of the damage share \(\Omega\) (left panel) with the corresponding monetary loss \(D_t\) (right panel).
Figure~\ref{fig:dam} shows that both the damage share and the associated monetary loss track the temperature trajectory across all models; consequently, losses are largest in the dynamic $4$PR-X case and smaller in the static $3$SR and $4$PR cases.
\begin{table}[ht]
    \centering
    \begin{tabular}{c|c|c|c|c|c}
        \toprule
        Year & Variable & 3SR (1) & 4PR (2) & 4PR-X (3) & diff (1) \& (3) \\
        \midrule
        2020 & $\Omega\left(\bb{T}^{\text{A}}_{t}\right)$ p.p. & 
        0.3938 &	0.3922 &	0.4016	& 0.0078 (1.98\%)  \\
        2050 & $\Omega\left(\bb{T}^{\text{A}}_{t}\right)$ p.p. & 0.9321 &	0.9288 &	1.0442	& 0.1121 (12.02\%)\\
        2100 & $\Omega\left(\bb{T}^{\text{A}}_{t}\right)$ p.p. & 2.2603 &	2.2542	& 2.5562	& 0.2958 (13.08\%) \\
        \hline
        2020 & $D_t$ (Trill 2015 USD) & 0.4881 &	0.4858  &	0.4965 &	0.0084 (1.73\%) \\
        2050 & $D_t$ (Trill 2015 USD) & 2.6684	& 2.6578 &	2.9802 &	0.3118 (11.68\%) \\
        2100 & $D_t$ (Trill 2015 USD) & 17.0611	& 17.0127 &	19.2261	& 2.165 (12.68\%) \\
        \bottomrule
    \end{tabular}
     \caption{Data for share of output loss and equivalent monetary loss  across selected years. Absolute differences are $4$PR-X relative to $3$SR, with percentage differences in parentheses.}
    \label{tab:dam}
\end{table}
%
In the dynamic $4$PR-X model, the additional 0.2 °C of warming relative to the static emulators increases damages in 2050  by 11.7 percent and in 2100 by 12.7 percent (see \cref{tab:dam}).  This underscores that deforestation-induced warming imposes substantial additional economic damages.




\subsection{Optimal Mitigation}
\label{sec:optimal_mitigatoin}

\begin{figure}[ht]
    \centering
    \includegraphics[width=\textwidth]{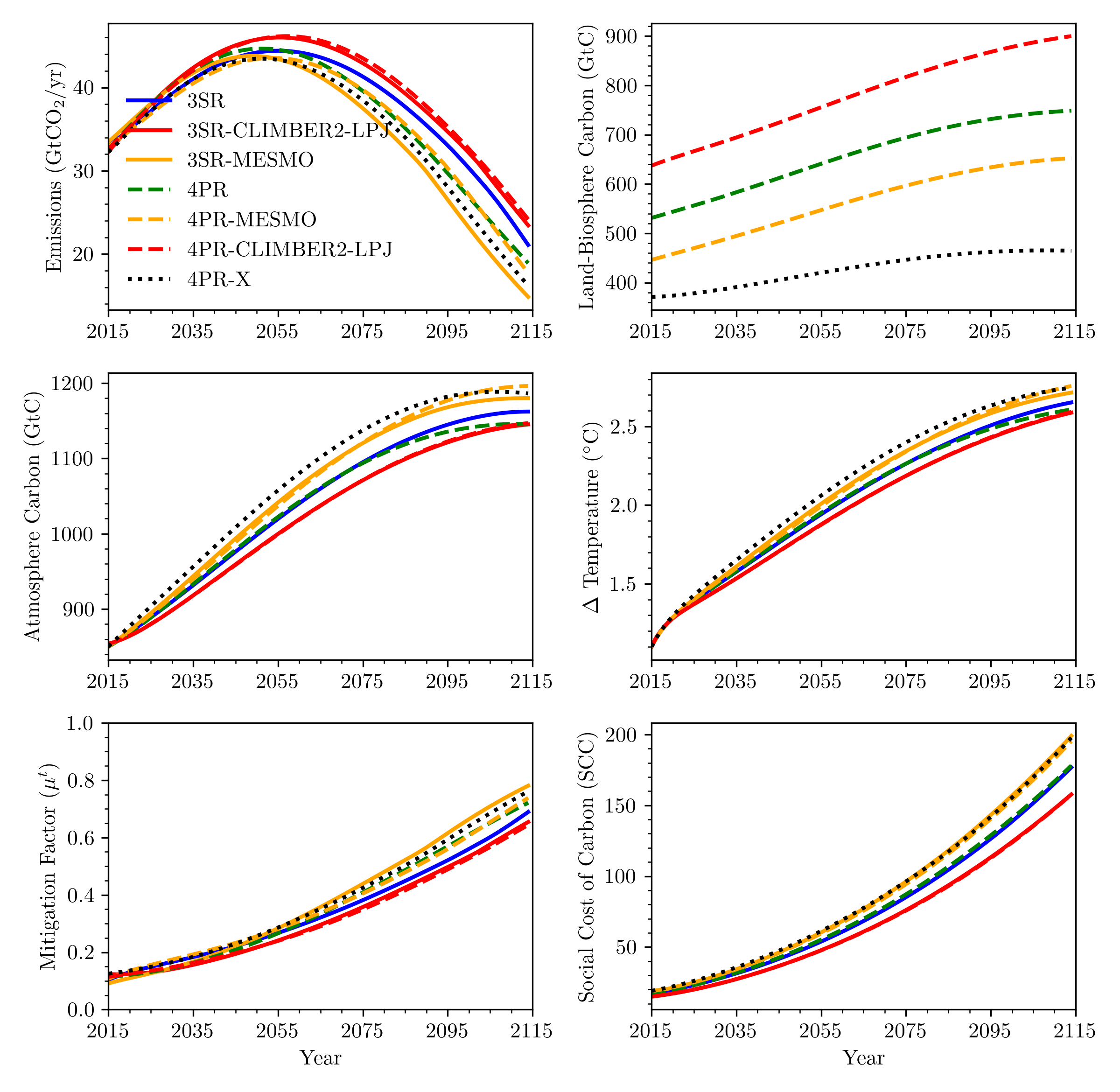}
    \caption{Simulation results for the optimal abatement scenario.}
    \label{fig:opt}
\end{figure}

Figure~\ref{fig:opt} traces the same variables under the optimal-mitigation policy. The ordering of outcomes, $4$PR-X above $4$PR and $3$SR, mirrors the BAU experiment. The optimal run additionally displays the mitigation rate and the SCC (bottom-right panel). Both series are consistently higher in the dynamic $4$PR-X emulator than in the static $3$SR and $4$PR counterparts, reflecting its warmer temperature path. Table~\ref{tab:scc_opt} quantifies these differences: the $4$PR-X SCC exceeds the $3$SR value by 11.9 \% in 2020 and by 13.9 \% in 2050. As in the BAU case, the extreme calibrations $3$SR–MESMO and $4$PR–MESMO closely track $4$PR-X, confirming that a depleted land-biosphere sink, e.g., extensive deforestation, can induce effects comparable to an extreme-warming scenario.
%
%
\begin{table}[h!]
    \centering
    \begin{tabular}{c|c|c|c|c|c}
        \toprule
        Year & Variable & 3SR (1) & 4PR (2) & 4PR-X (3) & diff (1) \& (3) \\
        \midrule
        2020 & SCC & 19.85 & 19.72 & 22.20 & 2.35 (11.87\%) \\
        2050 & SCC & 47.47 & 48.57 & 54.07 & 6.60 (13.90\%) \\
        2100 & SCC & 138.58 & 140.82 & 155.45 & 16.87 (12.18\%) \\
        \bottomrule
    \end{tabular}
    \caption{Social Cost of Carbon (2015 USD per ton of CO$_2$) across different years. Absolute differences are $4$PR-X relative to $3$SR, with percentage differences in parentheses.}
    \label{tab:scc_opt}
\end{table}
Our benchmark comparison reveals no substantive difference between the static three- and four-reservoir emulators: adding a land-biosphere reservoir has no impact on either BAU or optimal-policy outcomes. By contrast, allowing the land-biosphere stock in the four-reservoir model to decline over time raises BAU temperatures relative to the static cases. Explicitly modeling deforestation dynamics further increases the SCC, implying that land-use change materially shifts the optimal mitigation strategy and requires a higher carbon price to curb damages.

The mechanism is straightforward. Deforestation diminishes the land-biosphere sink’s uptake capacity, so carbon that would otherwise have been sequestered remains in the atmosphere, raising temperatures and amplifying damages. This feedback translates into a higher optimal carbon price; omitting land-use change, therefore, leads to a systematic underpricing of carbon.

%
 \subsection{Carbon Capture \& Storage}
\label{sec:carbon_storage}
To illustrate how emulator design can significantly impact model outcomes, we examine a carbon capture and storage (CCS) scenario. CCS removes CO$_2$ directly from the atmosphere, thereby targeting the primary driver of warming. If deforestation is ignored, however, the apparent effectiveness of CCS can be overstated: deforestation weakens the land-biosphere sink, leaving more carbon in the atmosphere, so CCS must compensate for emissions that would otherwise have been sequestered naturally.

\begin{figure}[H]
    \centering
    \includegraphics[width=\textwidth]{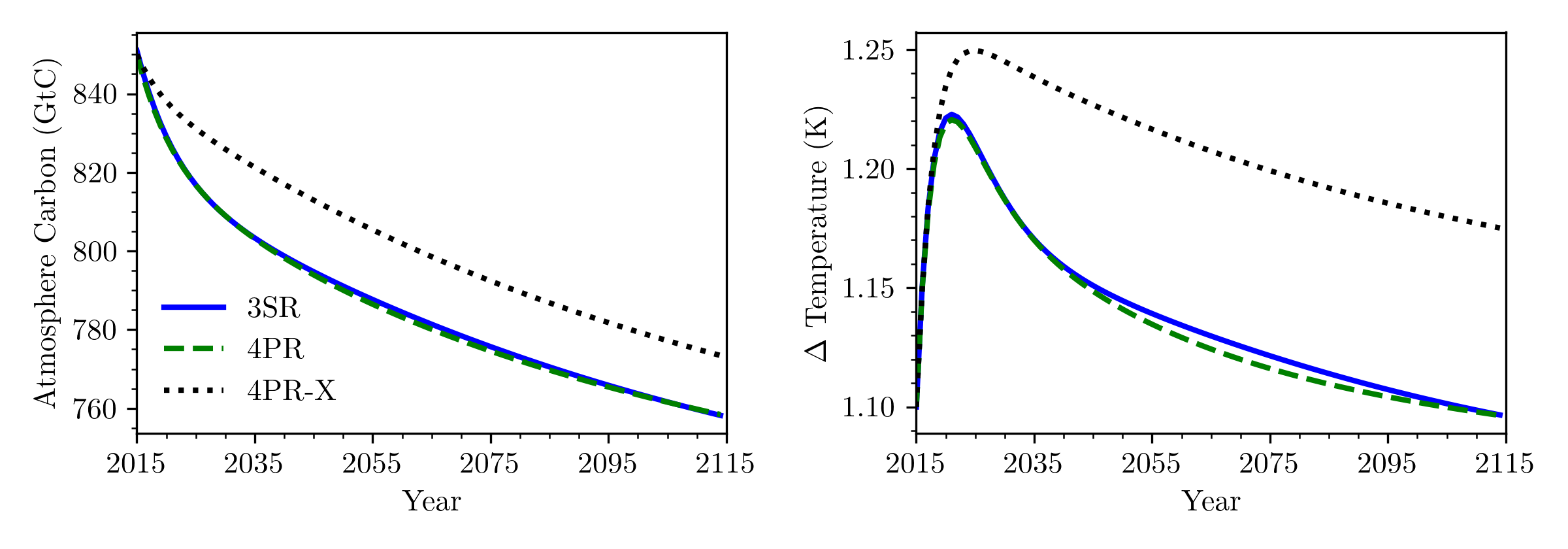}
    \caption{Atmospheric carbon concentrations (left panel) and atmospheric temperature (right panel) under the full abatement scenario ($\mu_t=1$).}
    \label{fig:ccs}
\end{figure}

We analyze this effect within the DICE-2016 framework by implementing a highly stylized policy that sets the mitigation rate to unity from the initial period, representing full deployment of the backstop technology and eliminating industrial CO$_2$ emissions. Figure~\ref{fig:ccs} plots the resulting atmospheric carbon and temperature trajectories for each emulator, and Table~\ref{tab:ccs_data} lists the corresponding temperature values.
%

\begin{table}[h!]
    \centering
    \begin{tabular}{c|c|c|c|c|c}
        \toprule
        Year & Variable & 3PR (1) & 4PR (2) & 4PR-X (3) & diff (1) \& (3) \\
        \midrule
        2020 & $\bb{T}^{\text{A}}_{2020}$ ($^\circ$C)  & 1.22 & 1.22 & 1.23  & 0.01 $^\circ$C (0.89\%) \\
        2050 & $\bb{T}^{\text{A}}_{2050}$ ($^\circ$C) & 1.14 & 1.14 & 1.22  & 0.08 $^\circ$C (6.72\%) \\
        2100 & $\bb{T}^{\text{A}}_{2100}$ ($^\circ$C) & 1.10 & 1.10 & 1.18  & 0.08 $^\circ$C (7.09\%) \\
        \bottomrule
    \end{tabular}
    \caption{Atmospheric temperature data ($\bb{T}^{\text{AT}}_{t}$) across different years. Absolute differences are $4$PR-X relative to $3$PR, with percentage differences in parentheses.}
    \label{tab:ccs_data}
\end{table}

%
Consistent with the BAU and optimal-policy experiments, adding a static land-biosphere reservoir leaves climate trajectories practically unchanged. By contrast, when the four-reservoir emulator incorporates a declining equilibrium land sink, it exhibits the same qualitative divergence observed earlier: even with full carbon capture in place, continued deforestation raises atmospheric temperatures by roughly 7\% relative to the static cases by 2050. 

Under the full mitigation, the $4$PR-X model projects higher atmospheric carbon and temperature than $3$SR, due to its reduced biospheric uptake.\footnote{In the $4$PR-X model, carbon emissions from land-use change are assumed to reduce the carbon storage capacity of the land biosphere concurrently and by an equal amount. The underlying reasoning is that, for example, deforestation not only results in carbon emissions, but it also means the forest will no longer be there as a carbon reservoir. The shrinking carbon-holding capacity of the land biosphere implies that a higher share of carbon emissions will be directed to the remaining reservoirs, notably the atmosphere. This effect of land-use change accumulates over time. Consequently, the difference between the $3$SR and $4$PR-X model, which are both calibrated against PI conditions, increases with time.} This underscores that even with ambitious CCS policies, failing to model biosphere dynamics may lead to overly optimistic projections of climate outcomes and underestimation of required mitigation levels.
Our findings indicate that CCS can succeed only if natural carbon sinks are preserved. Land degradation and deforestation erode the land‐biosphere’s sequestration capacity, undermining the net benefit of engineered carbon removal. Policies that protect or enhance terrestrial sinks are therefore an essential complement to large‐scale CCS deployment.

%

\section{Linking Global to Local Temperatures and Damages}
\label{sec:pattern_scaling}

Economic models that resolve climate impacts across space require equally resolved projections of key climatic drivers, in particular regional temperatures (in °C) and their future changes, to quantify local damages and related outcomes (see, e.g., \citealp{Krusell2022,Cruz2024,Desmet2024,Kotlikoff2024}). 
In this section, we present a computationally efficient procedure, grounded in state-of-the-art climate science, for deriving regional temperatures and temperature changes from global climate emulators (cf.\ Section~\ref{sec:3}).

A widely used and inexpensive way to obtain such projections is \emph{pattern scaling}, a form of statistical downscaling that expands a change in global mean temperature into a gridded warming pattern (see \citealp{Tebaldi2014,Kravitz2017,Lynch2017,Mathison2024,mathison-et-al:25}, and references therein).  
The resulting pattern can subsequently be aggregated to any geographical units the spatially-resolved models require.

Because both the global-mean temperature trajectory and the warming pattern pertain to an as-yet unobserved future, they must be taken from Earth-system model (ESM) simulations. The pattern is therefore not unique; it inherits the spread of the underlying ESM ensemble, just as an emulator for global mean temperature inherits the calibration choice of its driving ESM(s) (cf. Section~\ref{sec:4}). Selecting a particular pattern thus introduces an additional source of uncertainty driven by model choice.  When absolute temperatures (rather than anomalies) are needed, the pattern must be anchored to present-day observations, ensuring consistency with the empirical climate record.

In what follows, Section~\ref{sec:pattern_scaling_basics} formalizes the pattern-scaling method and states the governing equations. Section \ref{sec:pattern_scaling_temperatures} quantifies (i) the spread of temperature-change patterns across an ensemble of ESMs and (ii) the region-specific absolute temperatures projected for 2100. Building on these results, it also outlines a procedure for deriving region-specific temperatures that are consistent with both empirical observations and the ESMs.
Section~\ref{sec:pattern_warming_2100} links these local temperature projections to illustrative local impacts and damages in the year 2100.  

\subsection{Basics of Pattern Scaling}
\label{sec:pattern_scaling_basics}

Pattern scaling, introduced by \citet{Santer1990}, establishes a relationship between global and local temperature changes using large-scale ESM outputs. Specifically, it relates the change in global mean near-surface air temperature, $\Delta T^{\text{AT}}$, to the local mean near-surface air temperature change, $\Delta T^{\text{z}}$, at a specific location $z$ (representing areas potentially as fine as $1^{\circ}\times 1^{\circ}$). Temperature changes are typically computed after temporal smoothing, often by averaging over 10 or 20 years, to remove high-frequency interannual variability. A linear relationship is generally considered an adequate approximation for temperature, as well as for other variables such as precipitation (\citealp{Lynch2017,pfahl-et-al:17,mathison-et-al:25,munday-et-al:25}) and sea level change (\citealp{bilbao-et-al:15}).
Limitations of the linearity assumption can arise in the presence of highly localized radiative‐forcing agents, such as anthropogenic aerosols, or when multi-century time scales are considered (see, e.g., \citealp{rugenstein-et-al:16}); however, these issues are typically irrelevant for many economic applications.

Formally, the linear pattern-scaling approach used in this article expresses the local temperature change $\Delta T^{\text{z}}$, measured in °C, at location $z$ as:
\begin{equation}
\Delta T^{\text{z}} = \Delta T^{\text{AT}} \beta^\text{z},
\label{eq:pattern_scaling}
\end{equation}
where $\beta^\text{z}$ denotes the spatially resolved temperature change pattern. Because $\beta^\text{z}$ relates to future climate change, it cannot be derived from observations but must be obtained from model simulations, notably ESMs, thereby inheriting ESM-associated uncertainties.

For the subsequent analysis, this work relies on a publicly accessible library of temperature-change patterns $\beta^{\mathrm{z}}$ by \cite{Lynch2017}.\footnote{\url{https://github.com/JGCRI/CMIP5_patterns}.} They applied least squares regressions to CMIP5 data from future RCP8.5 scenarios across 41 different ESMs. For each model, a least squares regression relates the time series of annual global mean temperature change $\Delta T_t^{\text{AT}}$ to the time series of the location-specific ($z$), two-dimensional gridded latitude-longitude temperature change $\Delta T_t^\text{z}$ via:
\begin{equation}
\Delta T_t^\text{z} = \alpha^\text{z} + \beta^\text{z} \cdot \Delta T_t^{\text{AT}} + \epsilon_t^\text{z}.
\label{eq:how_to_get_beta}
\end{equation}
In this regression, the term $\beta^\text{z}$ represents the desired temperature change pattern. It is a two-dimensional field of regression slopes quantifying the temperature change at position $z$ relative to the global mean temperature change; thus, it has physical dimensions of degrees Celsius (local change) per °C (global change). The other terms in Equation~\eqref{eq:how_to_get_beta}, the $y$-intercept $\alpha^\text{z}$ (which \citealp{Lynch2017} assume to be zero) and the residual $\epsilon_t^\text{z}$, are not directly used in the pattern scaling application itself. The 41 scaling patterns derived from the 41 ESMs are qualitatively similar but quantitatively distinct, reflecting inter-model differences -- we provide concrete examples of this below. As there is often no clear basis for selecting one ESM pattern as superior, real-world applications should consider employing multiple patterns to assess the robustness of results against this pattern uncertainty.

Local damage functions, however, often require projections of future {\it absolute temperature} patterns, $T_\text{abs}^\text{z}$ (see, e.g., \citealp{Krusell2022,Desmet2024}, and references therein). Absolute temperatures are typically more challenging to quantify accurately than temperature changes, both in observations and models. ESMs, for instance, are primarily designed and evaluated based on their ability to simulate temperature changes relative to a baseline (e.g., pre-industrial) rather than absolute temperatures (e.g., \citealp{mauritsen-et-al:12}).

Obtaining projections of absolute temperatures using pattern scaling requires anchoring the {\it temperature change} pattern $\beta^\text{z}$ to an {\it absolute temperature pattern} $T_\text{abs,c}^\text{z}$ from a historical reference period (typically a 30-year climatological average). The future absolute temperature at time $t$ is then calculated as:
\begin{equation}
T_\text{abs,t}^\text{z} = T_\text{abs,c}^\text{z} + \Delta T_t^{\text{AT}} \cdot \beta^\text{z},
\label{eq:absolute_pattern_scaling}
\end{equation}
where $\Delta T_t^{\text{AT}}$ represents the global mean temperature change between the chosen historical reference period and the future time of interest, $t$. 

\begin{figure}
\centering
\centerline{
\includegraphics[width=0.7\linewidth]{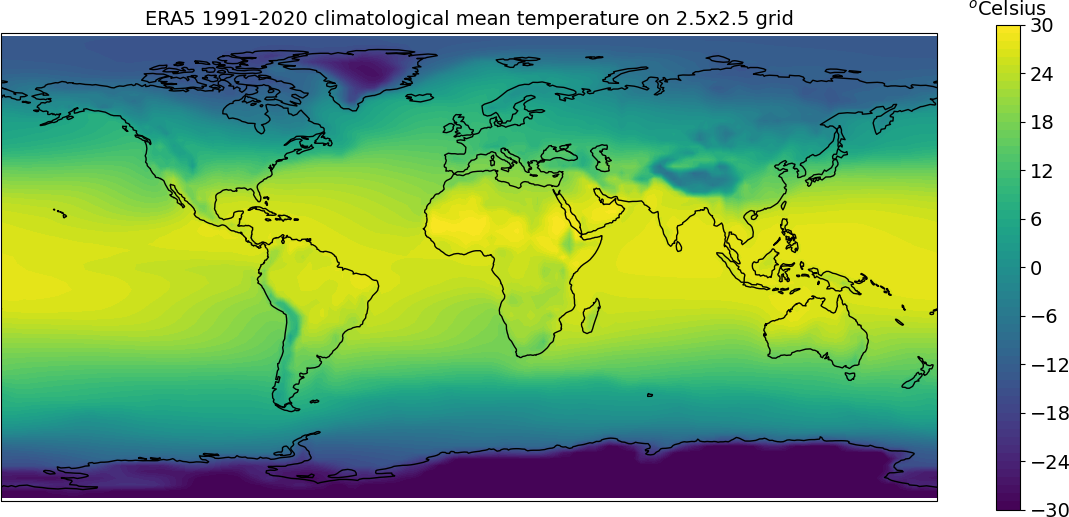}
}
\centerline{
\includegraphics[width=0.48\linewidth]{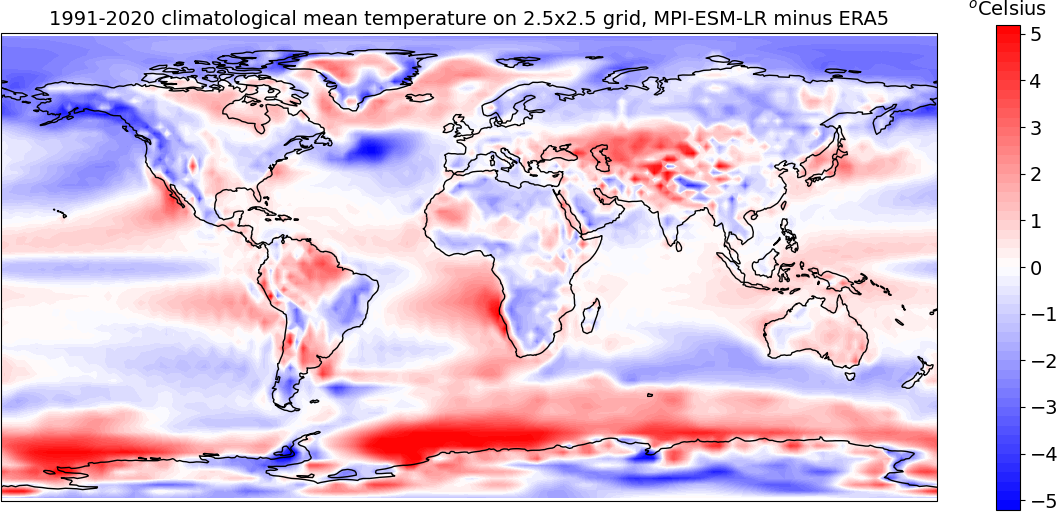}
\includegraphics[width=0.48\linewidth]{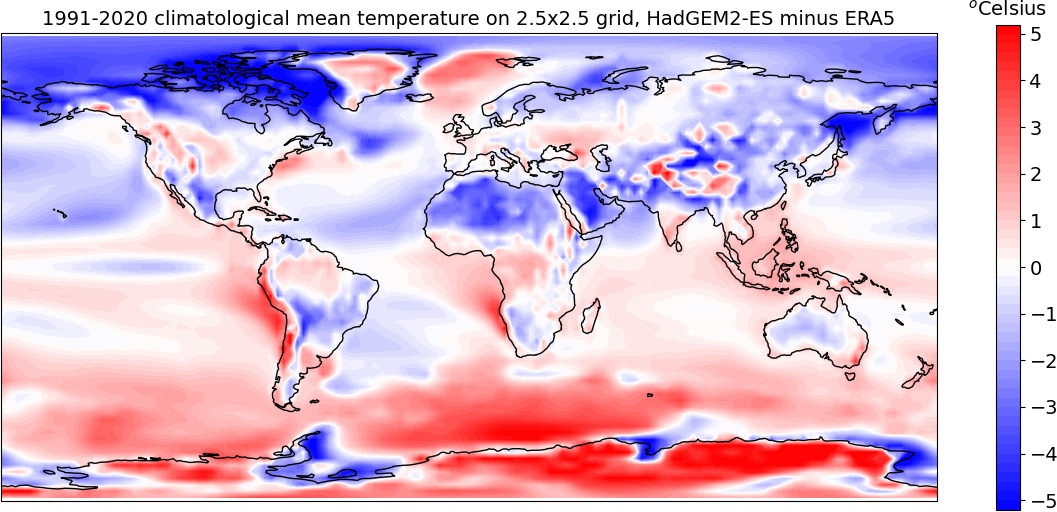}
}
\caption{Top panel: 30-year climatological mean near-surface air temperature (1991–2020) from the ERA5 reanalysis, taken as the observational reference.  
Bottom panels: temperature biases of two CMIP5 Earth-System Models with respect to that reference, shown as \emph{model climatology minus ERA5}.  
\textbf{Left:} MPI-ESM-LR (Max-Planck-Institute Earth System Model, low-resolution).  
\textbf{Right:} HadGEM2-ES (Hadley Centre Global Environmental Model, version 2 – Earth System).  
Positive values indicate regions where the ESMs are warmer than ERA5. 
}
\label{fig:era5_diff_clim}
\end{figure}

There is no single best choice for the historical anchor pattern $T_\text{abs,c}^\text{z}$. Since this pattern pertains to the historical period, options include observation-based datasets or model-based climatologies. For economic applications where proximity to real-world conditions is often crucial, using an observation-based historical absolute temperature dataset for $T_\text{abs,c}^\text{z}$ is a common and justifiable choice. Several publicly available gridded datasets of observed absolute surface air temperature exist.\footnote{Among them are re-analysis data from ERA5 (\url{https://www.ecmwf.int/en/forecasts/dataset/ecmwf-reanalysis-v5}), JRA-3Q (\url{https://jra.kishou.go.jp/JRA-3Q/index_en.html}), MERRA-2 (\url{https://gmao.gsfc.nasa.gov/reanalysis/merra-2}), and NCEP (\url{https://psl.noaa.gov/data/gridded/data.ncep.reanalysis2.html}) or for land only temperatures the CRU data (https://crudata.uea.ac.uk/cru/data/hrg).} Among the highest-regarded datasets currently available is the ERA5 reanalysis~\citep{hersbach-et-al:20}. This dataset is used below to derive the gridded 1991--2020 climatological mean absolute surface air temperature, $T_\text{abs,c}^\text{z}(ERA5)$, shown in the top panel of Figure~\ref{fig:era5_diff_clim}.
The global mean value of $T_\text{abs,c}^\text{z}(ERA5)$ over this period is 14.4 degrees Celsius, consistent with the global mean temperature estimate for 1991--2020 published by the European Union's Copernicus Climate Change Service.\footnote{\url{https://climate.copernicus.eu}.} 
Thus, in practical applications, local absolute temperature can be computed as
\begin{equation}
T_{\text{abs,t}}^{\text{z}} \;=\; T_\text{abs,c}^\text{z}(ERA5) \;+\; \Delta T^{\text{AT}}_t \beta^{\text{z}},
\label{eq:pattern_scaling_abs}
\end{equation}
where $\Delta T_t^{\text{AT}}$ denotes the global-mean temperature change provided by the climate emulator (cf.\ Section~\ref{sec:3} and Table~\ref{tab:bau}).

An alternative approach for choosing $T_\text{abs,c}^\text{z}$, described by~\cite{Lynch2017}, prioritizes \emph{internal consistency} within each ESM and anchors both, warming pattern and historical baseline, to output from the \textit{same} ESM model run. The ESM specific historical 30-year averaged (1961--1990) absolute-temperature climatologies, $T_\text{abs,c}^\text{z}(ESM)$, exhibit global-mean values ranging from 12.5 to 15.3 °C, compared with the observed global-mean estimate for 1961--1990 of $14.0 \pm 0.5$ °C~\citep{jones-et-al:99}. The offset between observations and a given ESM can be removed by rescaling: $T_\text{abs,c}^\text{z}(ESM) + \Delta T_{\text{obs,ESM}}\,\beta^\text{z}(ESM)$,  where $\Delta T_\text{obs,ESM} = 14.0 - T_\text{abs,c}^\text{z}(ESM)$. The resulting temperature fields are specific to the ESM's climate and generally differ from fields anchored directly to observations. Figure~\ref{fig:era5_diff_clim} illustrate the differences between the 1991--2020 climatologies based on ERA5 (top panel) and the appropriately rescaled climatologies based on two example ESMs (bottom panels), namely MPI-ESM-LR (Max-Planck-Institute Earth System Model, low-resolution) and HadGEM2-ES (Hadley Centre Global Environmental Model, version 2 – Earth System). Both models are well-established in climate science and display somewhat different warming patterns. In conclusion, various options exist for selecting $T_\text{abs,c}^\text{z}$, each potentially suitable for different purposes; the choice should be made carefully based on the specific requirements of the application.

Combining pattern scaling, as described, with an emulator for global mean temperature change, such as CDICE~\citep{folini2024climate}, provides a computationally efficient and transparent method for generating spatially resolved temperature fields suitable for input into damage functions. As detailed above, these temperature fields are subject to uncertainties stemming from the choice and calibration of the global temperature emulator, the selection of the scaling pattern $\beta^\text{z}$, and the choice of the anchoring pattern $T_\text{abs,c}^\text{z}$. A key advantage of this component-based approach is that the influence of each source of uncertainty can be explored separately and transparently, as demonstrated below. While other publicly accessible tools for pattern scaling exist (e.g., those presented by \citealp{Hernanz2023} and \citealp{Beusch2020}), they may be less readily suited for such component-wise sensitivity analysis due to differences in their structure or complexity.\footnote{See~\url{https://github.com/ahernanzl/pyClim-SDM} and \url{https://github.com/MESMER-group/mesmer-openscmrunner}, respectively.}

A technical summary of how to do pattern scaling reads as follows.
\begin{itemize}
  \item \textbf{Select a warming pattern.}  
        Choose $\beta^\text{z}$ from one of the available Earth-system models.

  \item \textbf{Select an absolute temperature pattern for anchoring.}  
        When absolute future temperature fields $T_\text{abs,t}$ are required, choose $T_\text{abs,c}^\text{z}$ from among observation-based data sets, for example ERA5 re-analysis.

  \item \textbf{Select a baseline period.}  
        Decide on the baseline period relative to which one global mean temperature change can be computed (employing your CE) and over which you average the absolute temperature fields needed for pattern scaling, for example, the 30-year period 1991–2020.

  \item \textbf{Compute future (absolute) temperatures.}  
        Combine the chosen $\beta^\text{z}$ and $T_\text{abs,c}^\text{z}$ with the projected global-mean warming $\Delta T_t^{\text{AT}}$ to obtain  
        \[
           T_\text{abs,t}^\text{z} \;=\; T_\text{abs,c}^\text{z} + \Delta T_t^{\text{AT}}\,\beta^\text{z} .
        \]
\end{itemize}

\subsection{Spatial Temperature Change Patterns}
\label{sec:pattern_scaling_temperatures}

This section illustrates the uncertainty associated with the parameter $\beta^\text{z}$ (see Equation~\ref{eq:pattern_scaling}) by demonstrating how a one-degree Celsius change in global mean temperature translates into region-specific warming, and by assessing the sensitivity of these estimates to the underlying Earth System Models (ESMs). Furthermore, the section addresses the implications of selecting $T_\text{abs,c}^\text{z}$, which is necessary to generate absolute temperature patterns. For illustration, we use two models from the 41 ESMs presented in \cite{Lynch2017}, namely MPI-ESM-LR and HadGEM2-ES (cf. Figure~\ref{fig:era5_diff_clim}). 

To ensure tractability, regional analyses often aggregate smaller areas (e.g., $1^\circ \times 1^\circ$ grid cells) into larger units. In this study, we employ the WGI v4 reference regions \citep{essd-12-2959-2020}. Defined on geographical and climatological criteria, these regions are widely used in recent IPCC reports and climate-modeling studies, yet they remain relatively uncommon in the economics literature.\footnote{See~\url{https://github.com/IPCC-WG1/Atlas/blob/main/reference-regions/IPCC-WGI-reference-regions-v4_coordinates.csv} for details.} 
Our analysis focuses exclusively on the land areas of these regions, as shown in Figure~\ref{fig:wgiv4_regions}. The regions are listed and explained in Table~\ref{tab:wgi_ref_regions_acronyms} of Appendix~\ref{APX:WGI}.
%
\begin{figure}[th!]
    \centering
    \includegraphics[width=0.9\linewidth]{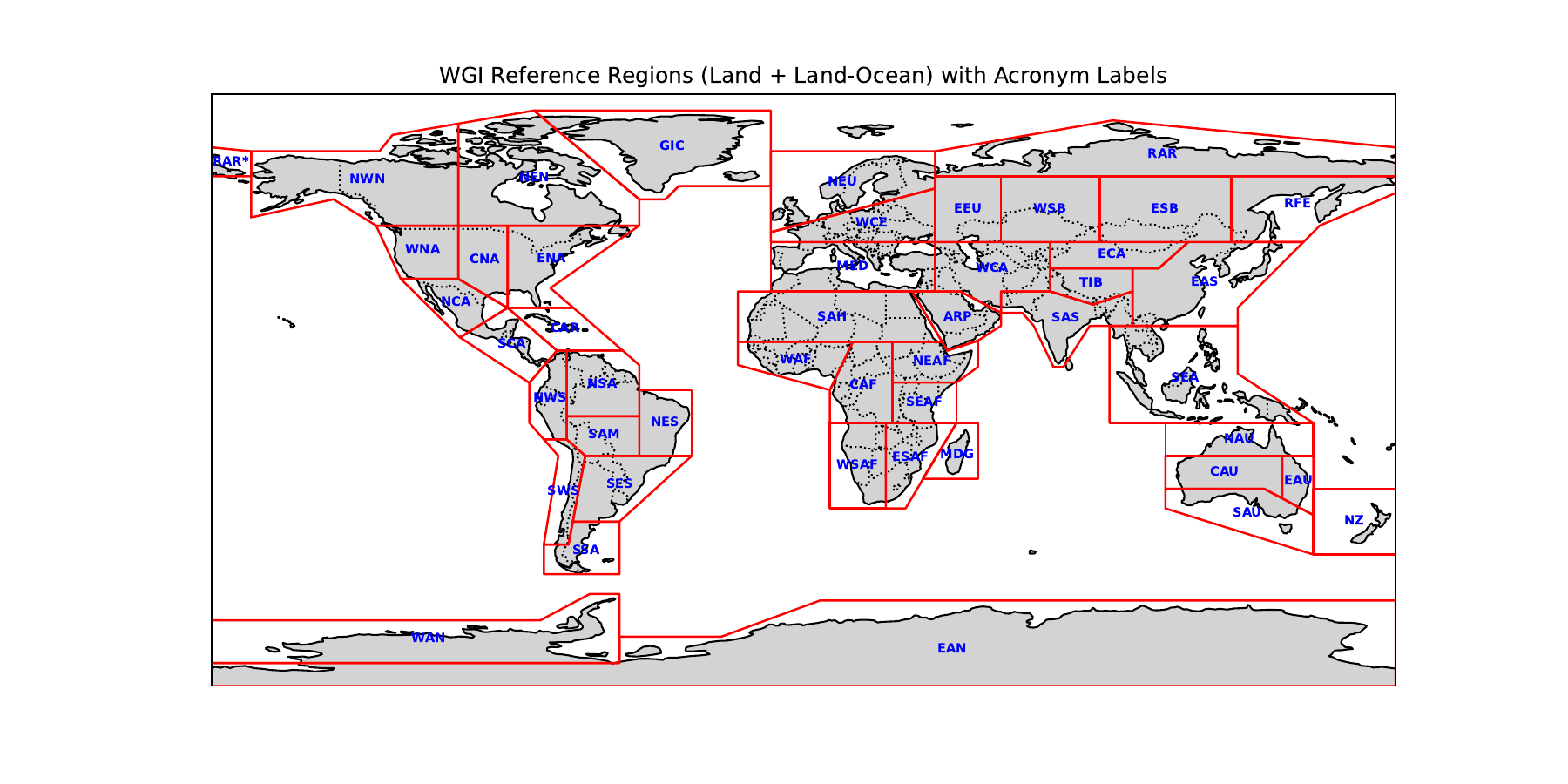}
    \caption{WGI (v4)  Reference Land and Land-Ocean Regions~\citep{essd-12-2959-2020}.}
    \label{fig:wgiv4_regions}
\end{figure}
%
The regional partition adopted here balances spatial resolution against computational tractability. Nevertheless, the optimal degree of regional aggregation ultimately depends on the specific objectives of the study at hand.
%
\begin{figure}[t!]
    \centering
    \centerline{
    \includegraphics[width=0.5\linewidth]{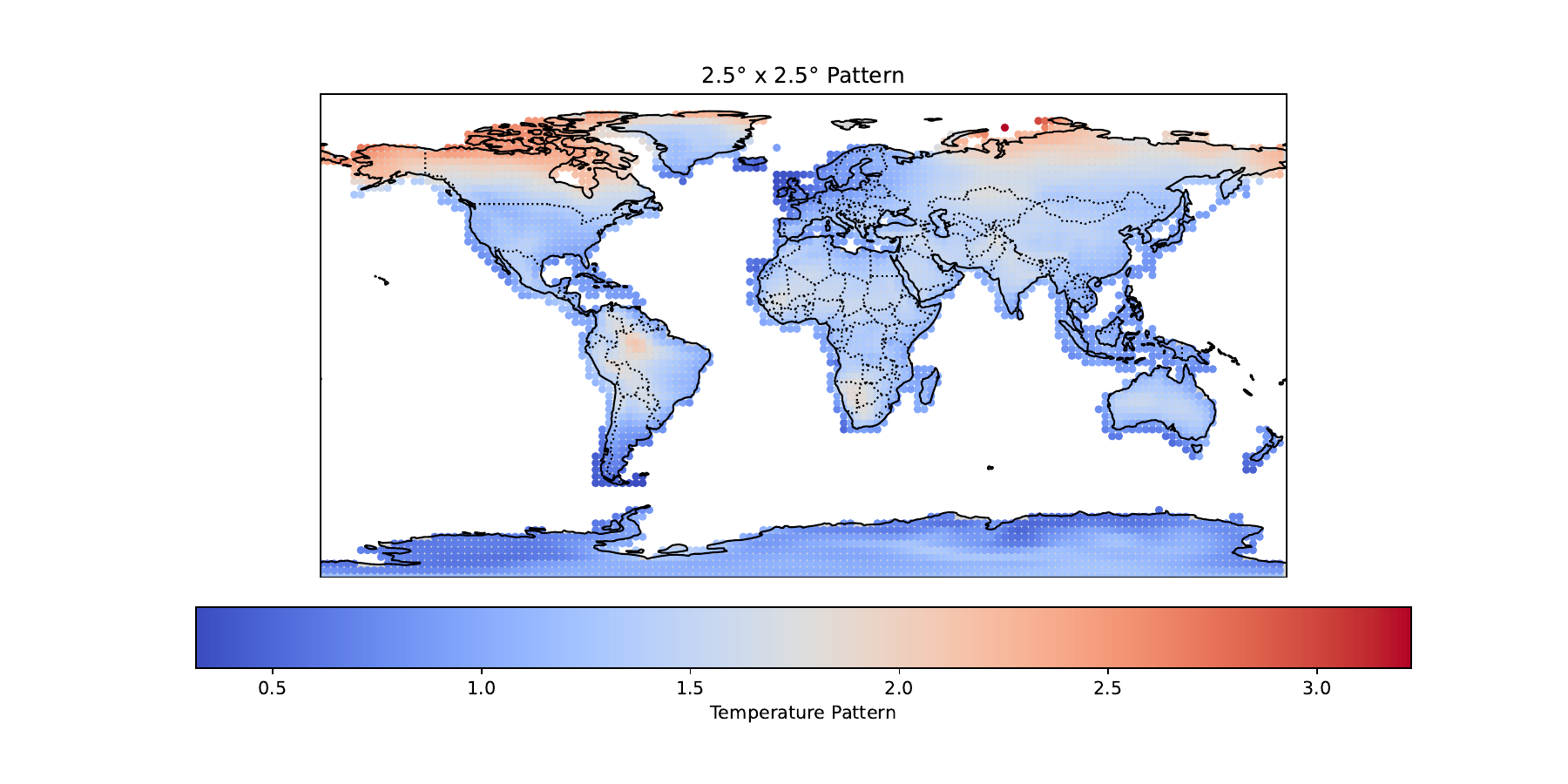}
    \includegraphics[width=0.5\linewidth]{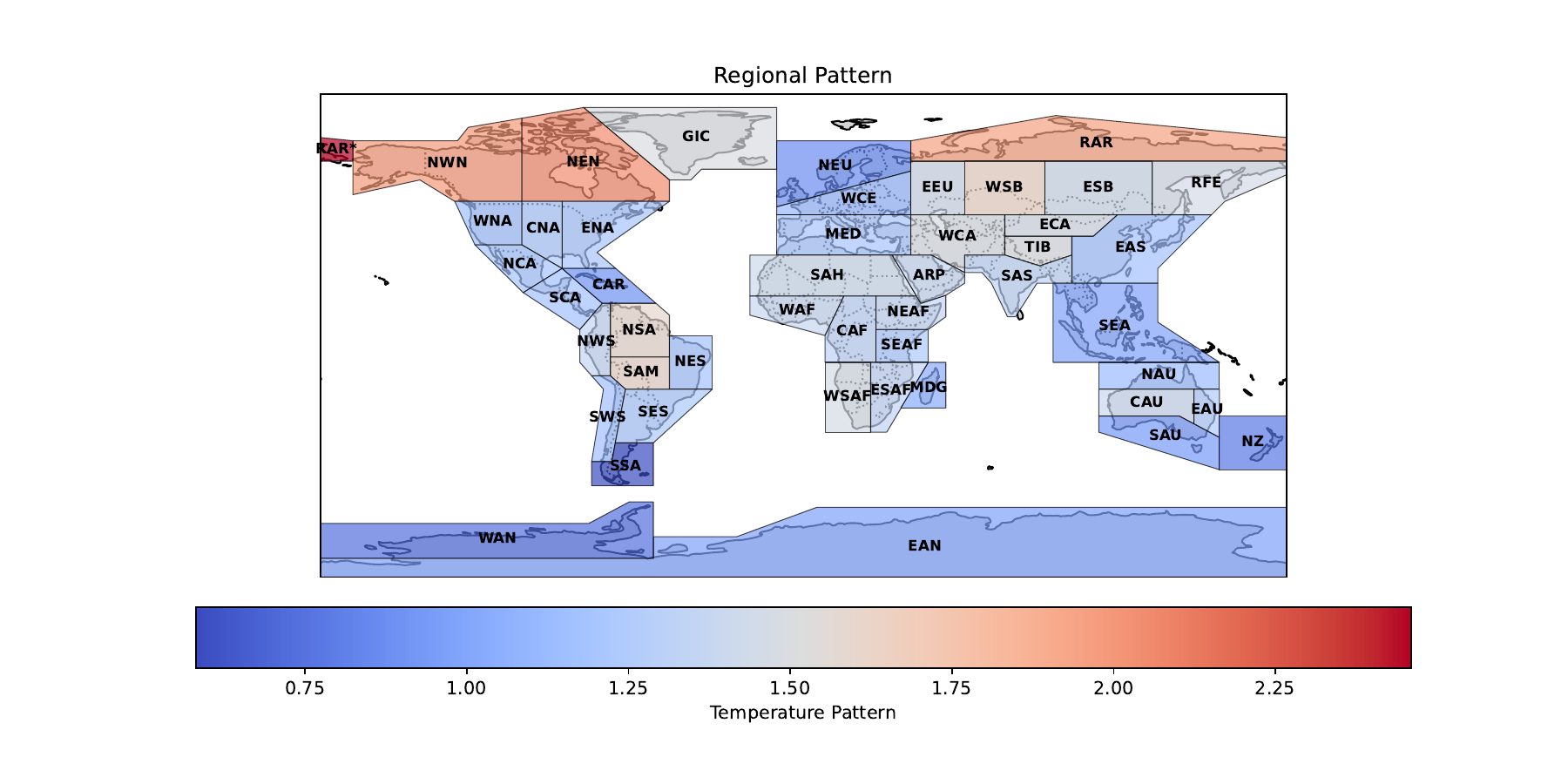}
    }
    \centerline{
    \includegraphics[width=0.5\linewidth]{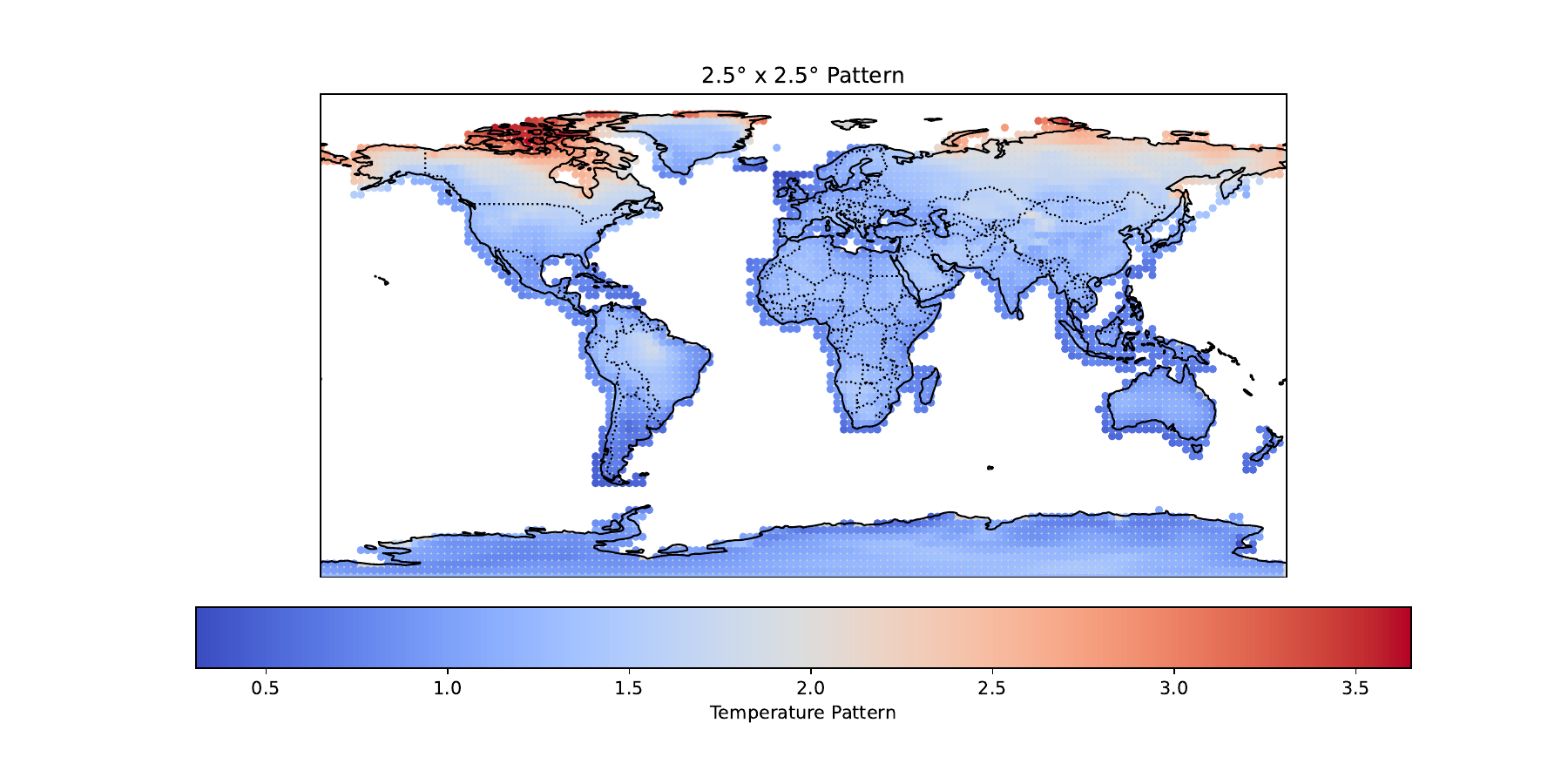}
    \includegraphics[width=0.5\linewidth]{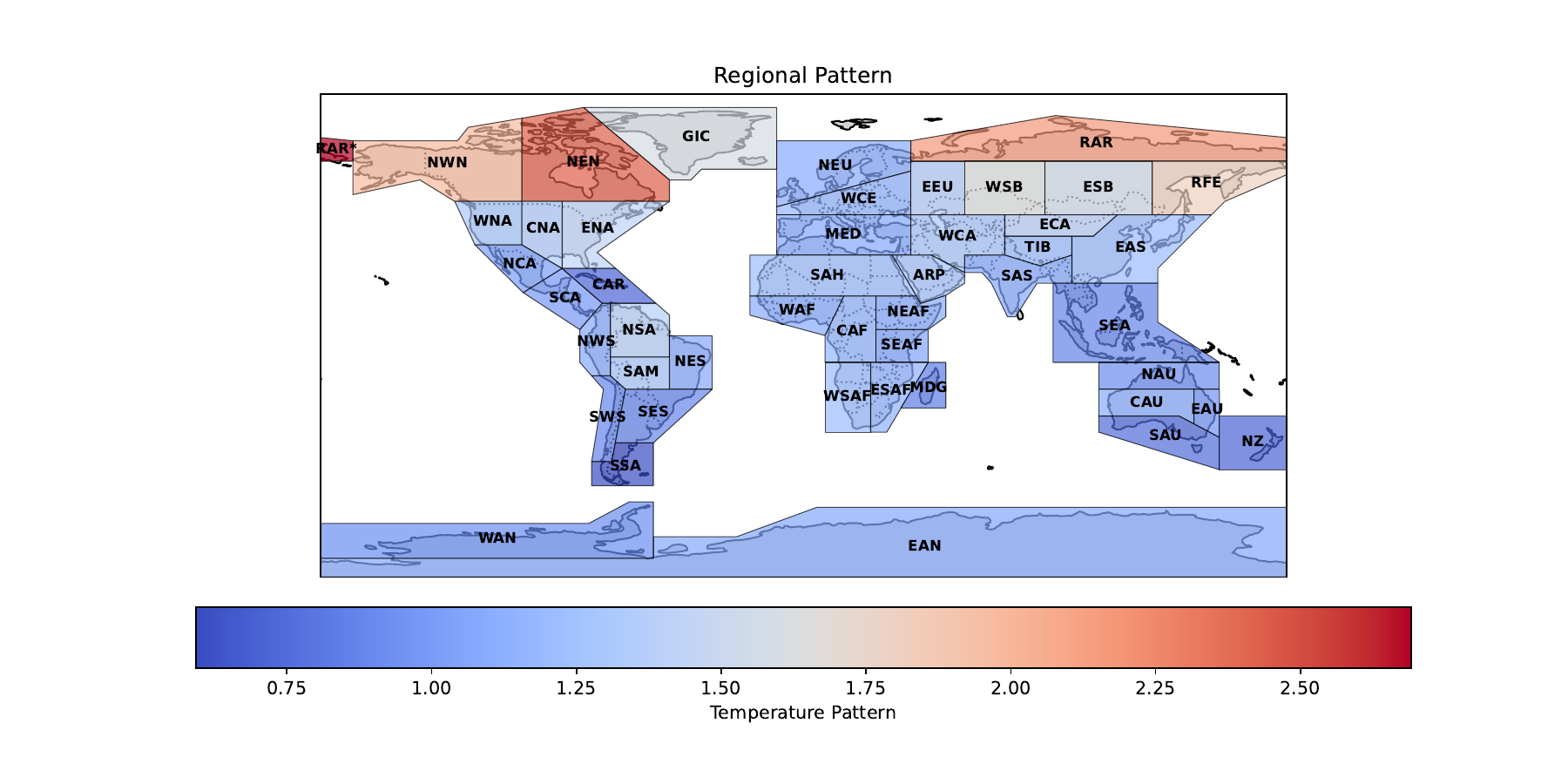}
    }
    \caption{Spatial warming patterns $\beta^\text{z}$ corresponding to a 1°C increase in global mean temperature, shown as gridded maps (left column) and regionally aggregated maps (right column). The patterns are generated using pattern scaling based on the ESMs MPI-ESM-LR (top row) and HadGEM2-ES (bottom row).}
    \label{fig:1deg_warming_pattern}
\end{figure}
%

Figure~\ref{fig:1deg_warming_pattern} shows the geographical warming patterns ($\beta^\text{z}$) per degree of global mean temperature increase for the MPI-ESM-LR and HadGEM2-ES climate models. Several key features are apparent: warming is particularly strong across Arctic regions and most land areas experience warming rates exceeding the global average ($\beta^\text{z}>1$). 
Significant spatial heterogeneity exists, leading to markedly different warming rates even in geographically neighboring regions, such as the contrast observed within South America between the South America Monsoon (SAM) and Northeast South America (NES) regions. Furthermore, the warming patterns differ notably between the two models. For instance, MPI-ESM-LR projects stronger warming than HadGEM2-ES in areas like western Brazil and parts of South Africa, while HadGEM2-ES shows greater warming in northern Canada. These inter-model discrepancies persist when aggregated to the reference region level. Compared to HadGEM2-ES, the MPI-ESM-LR model shows greater warming in Northern South America (NSA), the South America Monsoon (SAM) region, and Western Southern Africa (WSAF), whereas Northern Europe (NEN) warms more in HadGEM2-ES (by approximately $0.2$ to $0.4\,^{\circ}$C per degree of global warming). These differences may seem small, but they scale with global mean temperature change, and thus imply substantial cumulative impacts over time.

To illustrate this in more detail,
Table~\ref{tab:wgi_regions} presents a subset of numerical values for five economically relevant WGI regions.
The regions were 
selected for two reasons: (i) they display a strong warming amplification (\(\beta^{\text{z}}\!\ge\!1.30\) in either MPI-ESM-LR or HadGEM2-ES); and (ii) they coincide with economic blocs whose combined 2024 GDP exceeds USD 2 trillion.
\begin{table}[ht!]
\caption{Subset of IPCC WGI version-4 land-regions used in this study.  
Columns 1–2 list the official acronym and reference name.  
Columns 3–4 give the regional warming per degree of global mean warming, $\beta^{\text{z}}$, estimated by MPI-ESM-LR (“MPI’’) and HadGEM2-ES (“Had’’).  
Columns 5–6 show the minimum and maximum $\beta^{\text{z}}$ values across the full 41-member CMIP5 ensemble, thus indicating the structural spread among models.  
Columns 7–10 report the projected near-surface air temperature in the year 2100, $T_{\text{abs}}(2100)$, assuming a global mean warming since pre-industrial times of $\Delta T^{\text{AT}} = 2.65$ °C (scenario 4PR-X; cf. Figure~\ref{fig:opt}).  
Values labeled “MPI+M’’ and “Had+H’’ use each model’s own 1961–1990 climatology, $T_{\text{abs},c}^{\text{z}}(\mathrm{ESM})$, as the baseline; values labeled “MPI+E’’ and “Had+E’’ employ the ERA5 1991–2020 climatology, $T_{\text{abs},c}^{\text{z}}(\mathrm{ERA5})$. All temperatures are expressed in degrees Celsius.}
\label{tab:wgi_regions}
\begin{footnotesize}
\centering
\begin{tabular}{llcccccccc}
\toprule
  &   & \multicolumn{4}{|c|}{$\beta$} & \multicolumn{4}{|c|}{$T_\text{abs}(2100)$} \\
Acronym & RefName & MPI & Had & Min & Max & MPI+M & Had+H & MPI+E & Had+E\\
\midrule
CNA & C.\ North America  & 1.22 & \textbf{1.35} & 0.89 & 1.84 & 14.50 & 14.18 & 14.51 & 14.71 \\
ECA & E.\ Central Asia   & \textbf{1.46} & 1.38 & 1.09 & 1.61 & 10.93 &  7.11 &  8.49 &  8.37 \\
ARP & Arabian Peninsula  & \textbf{1.38} & 1.27 & 1.00 & 1.50 & 29.24 & 26.52 & 28.80 & 28.63 \\
EEU & E.\ Europe         & \textbf{1.43} & 1.36 & 1.06 & 1.98 &  8.84 &  7.69 &  7.03 &  6.93 \\
SAS & South Asia         & \textbf{1.33} & 1.09 & 0.86 & 1.33 & 28.02 & 25.68 & 27.63 & 27.25 \\
\bottomrule
\end{tabular}
\end{footnotesize}
\end{table}

 Even within this limited sample, the regional-warming factors vary appreciably: for the two illustrative ESMs the range is \(1.22\!-\!1.46\) (MPI-ESM-LR) and \(1.09\!-\!1.38\) (HadGEM2-ES), implying that these economies are projected to warm roughly 10–45 \% faster than the globe as a whole. The full 41-model ensemble widens the spread to \(0.86\!-\!1.98\), underscoring that the model choice for the warming pattern $\beta^\text{z}$ adds a further uncertainty of up to \(\sim\!1\)\,°C per degree of global warming. 

Turning to absolute-temperature projections for 2100, given in the right-hand columns of Table~\ref{tab:wgi_regions}, we find that the choice of the anchoring baseline (model-internal climatology versus ERA5) can shift regional means by up to about 2.5–3 °C. For example, East-Central Asia warms to 10.9 °C when anchored to the MPI-ESM-LR climatology but to only 8.5 °C when the same pattern is anchored to ERA5, a difference of 2.4 °C. Regional temperatures in the two columns using the same \( T_\text{abs,c}^\text{z}(\text{ERA5}) \) agree much better than those in the columns based on \( T_\text{abs,c}^\text{z}(\text{ESM}) \). Although the table is not exhaustive, the five regions serve as representative testbeds for sensitivity analyses that combine strong climate signals with large economic stakes. More details on $\beta^\text{z}$ and its dependence on different regions and ESMs are provided in Appendix~\ref{APX:G}, Figure~\ref{fig:warming_all_regs_mods}.

In summary, the above data show that typical values for regionally averaged warming patterns $\beta^\text{z}$ are in a range of \(1.5 \pm 0.5\) °C per °C of global mean warming. Extreme values outside a range of 0.5–2.5 °C per°C do exist (cf.~Figure~\ref{fig:warming_all_regs_mods}). When realistic absolute future temperature patterns are required, the choice of \( T_\text{abs,c}^\text{z} \) is crucial, and we advocate the use of observation-based climatologies for this purpose. For large global-mean temperature changes, differences in \( \beta^\text{z} \) will eventually dominate, as they scale with the magnitude of the change.

\subsection{Regional Damages From Global Warming by 2100}
\label{sec:pattern_warming_2100}
Having discussed how global mean temperature trajectories translate into local temperatures, we now
examine how these varying local climate outcomes lead to heterogeneous economic impacts across regions. The specification of spatially resolved damage functions is central to various policy questions.
A common approach is to model damages as a function of local warming  (cf.\ Section~\ref{sec:pattern_scaling_temperatures}).
Following \citet{Krusell2022}, many models assume an inverted U-shaped relationship between a region’s absolute temperature and productivity, calibrated so that the spatially resolved damage model replicates aggregate global damage estimates.\footnote{This specification, also employed by \citet{Hassler2023} and \citet{Kotlikoff2024}, can be parameterized to match aggregate damage patterns identified in the literature~\citep{Desmet2024}.}
This approach aligns with evidence from climate science that relates current population density, crop production, and GDP to absolute temperature (see, e.g.,~\citealp{xu-et-al:20}).

In what follows, we adopt the formalism of \citet{Krusell2022} to map the outputs of our global climate emulators and pattern-scaling procedure into region-specific damage estimates.\footnote{As discussed in Section~\ref{sec:pattern_scaling_temperatures}, absolute future temperatures are harder to quantify than temperature anomalies. We therefore apply the “ERA5-correction’’ introduced in expression Equation~\eqref{eq:pattern_scaling_abs}.}  
Within this framework, the climate component of regional total factor productivity (TFP), $\tilde{D}(T_t^{\text{z}})$, is assumed to follow  
\begin{equation}
\tilde{D}(T_t^{\text{z}}) =
\begin{cases} 
(1 - d) \exp\left(-\kappa^+(T_t^{\text{z}} - T^*)^2\right) + d & \text{if } T_t^{\text{z}} \geq T^*, \\[4pt]
(1 - d) \exp\left(-\kappa^-(T_t^{\text{z}} - T^*)^2\right) + d & \text{if } T_t^{\text{z}} < T^*,
\end{cases}
\label{eq:damages_local}
\end{equation}
with parameter values reported in Table~\ref{tab:parameters_KrSM}.  
\begin{table}[th!]
\centering
\begin{tabular}{|c|l|c|}
\hline
\textbf{Parameter} & \textbf{Description} & \textbf{Value} \\ \hline
$d$ & Lower bound of $\tilde{D}(T_t^{\text{z}})$ as $T_t^{\text{z}}$ diverges from $T^*$ & 0.02 \\ \hline
$T^*$ & Optimal regional temperature where $\tilde{D}(T_t^{\text{z}})$ is maximized & 11.58 \\ \hline
$\kappa^+$ & Rate of decline of $\tilde{D}(T_t^{\text{z}})$ for $T_t^{\text{z}} \geq T^*$ & 0.00311 \\ \hline
$\kappa^-$ & Rate of decline of $\tilde{D}(T_t^{\text{z}})$ for $T_t^{\text{z}} < T^*$ & 0.00456 \\ \hline
\end{tabular}
\caption{Parameter values for the function $\tilde{D}(T_t^{\text{z}})$.}
\label{tab:parameters_KrSM}
\end{table}
%
Regional damages attributable to global warming are then calculated as the relative change of the climate component of TFP,   
\begin{equation}
\label{eq:damages_ks}
    D(T_t^{\text{z}})=\frac{\tilde{D}(T_t^{\text{z}})}{\tilde{D}(T^{\text{z}}_{\text{Baseline}})} - 1,
\end{equation}
where $T^{\text{z}}_{\text{Baseline}}$ denotes the local absolute temperature in the baseline period.\footnote{Many studies (see, e.g.,~\citealp{Krusell2022};~\citealp{Kotlikoff2024}, and references therein) use Nordhaus’ GEcon database or similar sources to construct GDP-weighted regional temperature patterns (\url{https://gecon.yale.edu}, GEcon 4.0 for 2005). GDP weighting ensures that grid cells with greater economic activity exert a larger influence on the regional mean. For example, Canada’s average is drawn toward its populous southern corridor, whereas an unweighted mean would be dominated by sparsely populated Arctic cells. A detailed comparison of weighting schemes is beyond the scope of this article.} Values $D(T_t^{\text{z}}) > 0$ indicate an increase in TFP as the local temperature changes from $T^{\text{z}}_{\text{Baseline}}$ to $T_t^{\text{z}}$, values $D(T_t^{\text{z}}) < 0$ indicate a decrease in TFP.

\begin{figure}[ht!]
    \centering
    \centerline{
    \includegraphics[width=0.48\linewidth]{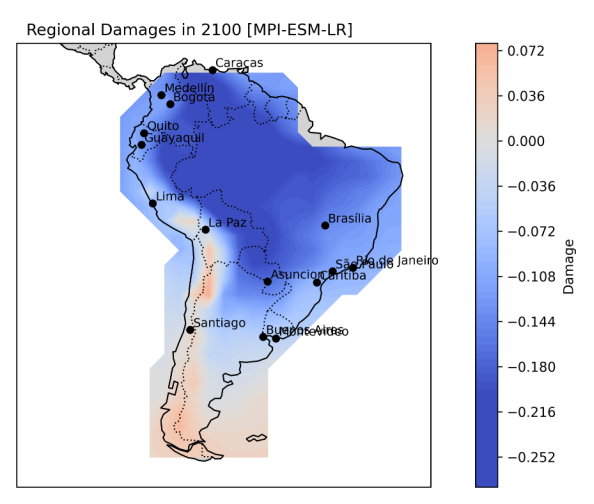}
    \includegraphics[width=0.48\linewidth]{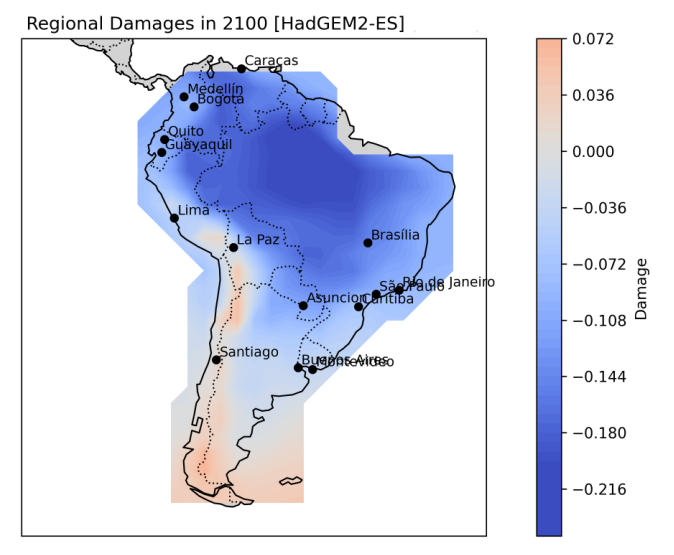}
    }
    \caption{Damages, expressed as \( \tilde{D}(T_{2100}^{\text{z}})/\tilde{D}(T_{\text{Baseline}}^{\text{z}}) - 1 \), assuming a global-mean warming of 2.65 °C since pre-industrial times ($\approx$ 1.55 °C since 2015) under the 4PR-X scenario. Absolute temperatures stem from ERA5 (\(T_{\text{abs,c}}^{\text{z}}\)); pattern factors \( \beta^{\text{z}} \) are taken from MPI-ESM-LR (left) or HadGEM2-ES (right). Red (blue) colors indicate an increase (decrease) in TFP due to global warming by 2100. }
    \label{fig:SouthAmericaDamageMaps}
\end{figure}

As this section is illustrative rather than exhaustive, we restrict our attention to South America. Specifically, we analyze two alternative regional temperature fields for the year 2100. Both fields are anchored to the ERA5 absolute‐temperature climatology for 1991–2020, denoted \(T_{\text{Baseline}}^{\text{z}}\), and use warming patterns \( \beta^{\text{z}} \) from either HadGEM2-ES or MPI-ESM-LR. Table~\ref{tab:temperature_projection} lists present-day ERA5 temperatures together with 2100 projections for 15 of South America's largest cities. Corresponding damages (climate induced relative changes in TFP) are illustrated in Figure~\ref{fig:SouthAmericaDamageMaps}.

\begin{table}[ht!]
    \centering
    \renewcommand{\arraystretch}{1.2}
    \begin{footnotesize}
    \begin{tabular}{l l c c c}
        \toprule
        \textbf{City} & \textbf{Country} & \textbf{ERA5 [°C]} & \textbf{HadGEM2-ES 2100 [°C]} & \textbf{MPI-ESM-LR 2100 [°C]} \\
        \midrule
        São Paulo     & Brazil     & 21.21 & 23.04 & 23.19 \\
        Buenos Aires  & Argentina  & 16.61 & 17.73 & 17.97 \\
        Rio de Janeiro & Brazil    & 22.24 & 23.60 & 23.71 \\
        Lima          & Peru       & 17.34 & 18.74 & 19.43 \\
        Bogotá        & Colombia   & 20.45 & 22.00 & 22.44 \\
        Santiago      & Chile      & 9.67  & 11.13 & 11.54 \\
        Caracas       & Venezuela  & 27.03 & 28.80 & 29.17 \\
        Quito         & Ecuador    & 21.94 & 23.60 & 24.03 \\
        La Paz        & Bolivia    & 11.63 & 13.43 & 13.97 \\
        Brasília      & Brazil     & 24.44 & 26.38 & 26.34 \\
        Medellín      & Colombia   & 20.45 & 22.00 & 22.44 \\
        Guayaquil     & Ecuador    & 20.46 & 21.87 & 22.17 \\
        Asunción      & Paraguay   & 22.84 & 24.51 & 25.36 \\
        Montevideo    & Uruguay    & 16.45 & 17.34 & 17.75 \\
        Curitiba      & Brazil     & 18.94 & 20.52 & 20.85 \\
        \bottomrule
    \end{tabular}
    \end{footnotesize}
    \caption{Temperature (in °C) “today’’, ERA5 1991–2020 mean, and as projected for 2100 in selected South-American cities.}
    \label{tab:temperature_projection}
\end{table}

The city data reveal a wide present-day temperature span, from 9.67 °C in Santiago to 27.03 °C in Caracas. Because Santiago’s current climate lies below the optimal temperature of 11.58 °C (see Equation~\ref{eq:damages_local}), moderate warming could raise local productivity. Such statements hinge on accurate present-day temperatures, underscoring the need to anchor pattern scaling in observation-based data such as ERA5. Between now and 2100, the projected warming ranges from roughly 1 °C to 2 °C. Model choice matters: MPI-ESM-LR yields larger increases for several cities, for example, Asunción warms by 0.85 °C, about 50\% more than under HadGEM2-ES, whereas others, such as Brasília, show similar warming between models.

\begin{figure}[ht!]
    \centering
    \includegraphics[width=\linewidth]{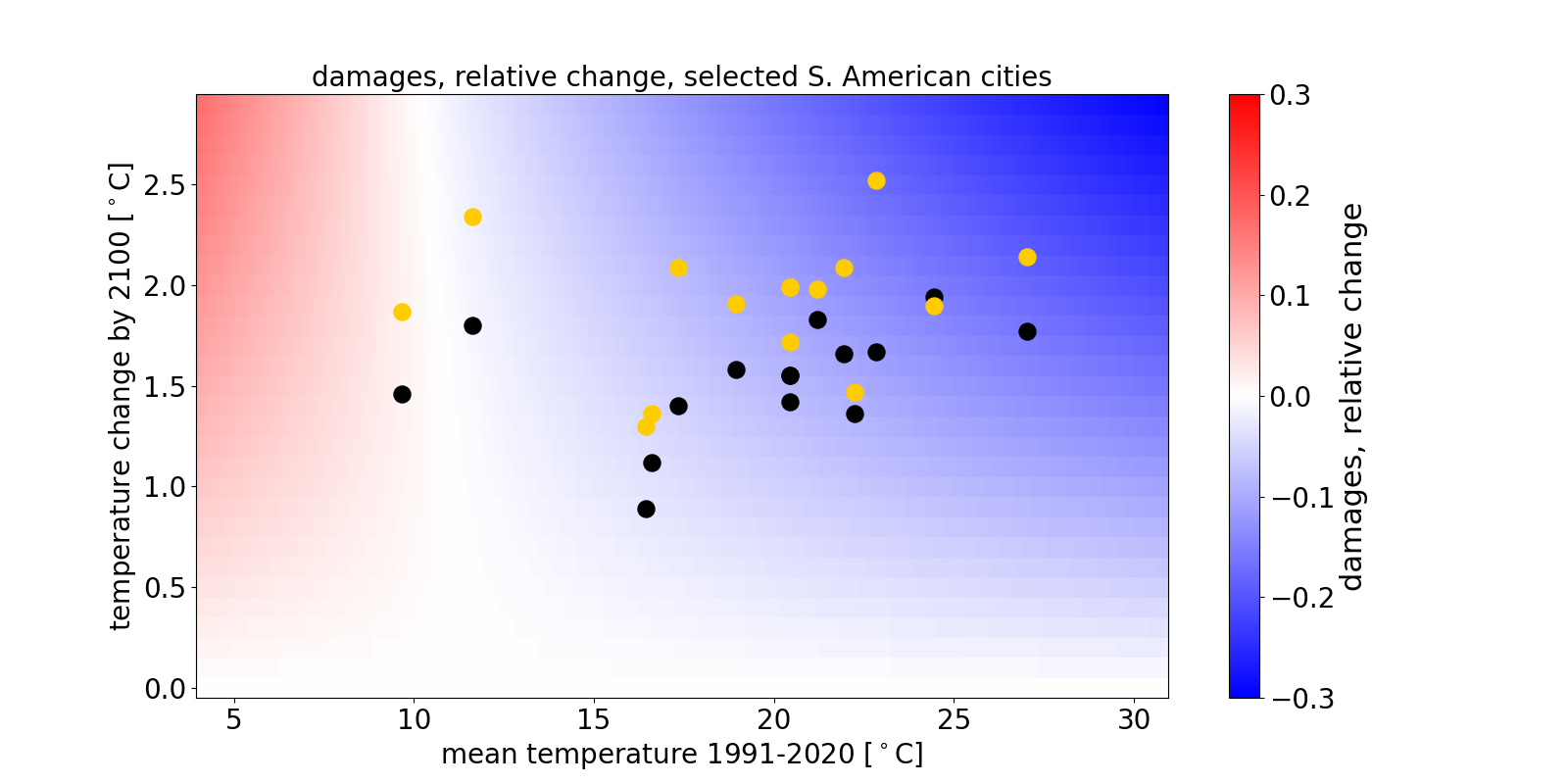}
    \caption{Relative change in damages \( \tilde{D}(T_{2100}^{\text{z}})/\tilde{D}(T_{\text{Baseline}}^{\text{z}}) - 1 \) as a function of present-day (x-axis) and future (y-axis) temperatures, based on expression Equation~\eqref{eq:damages_ks}. Coloured points correspond to the cities listed in Table~\ref{tab:temperature_projection}, with HadGEM2-ES (black) and MPI-ESM-LR (yellow) projections for 2100. Red (blue) colors indicate an increase (decrease) in TFP.}
    \label{fig:damages_illustration}
\end{figure}

Figure~\ref{fig:SouthAmericaDamageMaps} shows projected 2100 regional damages under our to ESMs.
The figure reveals that most of South America experiences net welfare losses, whereas segments of the Andes and the far south enjoy modest gains. Importantly, the projected damages are sensitive to model undertainty: the MPI-ESM-LR map shows deeper blue tones (indicating larger losses) than its HadGEM2-ES counterpart. The spatial distribution differs: For instance, dark-blue areas are centred on the Amazon under HadGEM2-ES but extend into Paraguay and Colombia under MPI-ESM-LR (cf.\ Table~\ref{tab:temperature_projection}). Such discrepancies highlight the need for caution when interpreting spatially resolved yet regionally averaged results.
 
When the damage function Equation~\eqref{eq:damages_ks} is used to classify relative winners and losers under global warming, the two factors on the right-hand side of the pattern-scaling Equation~\eqref{eq:pattern_scaling} interact in a non-trivial manner. Figure~\ref{fig:damages_illustration} illustrates this interaction. The absolute present-day temperature (shown on the x-axis), introduced via \( T_\text{abs,c}^\text{z} \), determines whether a location can benefit from an increase in the global mean temperature: gains are possible only if \( T_\text{abs,c}^\text{z} < 11.58\,^\circ\mathrm{C} \). It also determines the impact of any given future temperature change. For example, a warming of 2.5\,°C results in different relative changes of damages depending on present-day conditions, having a smaller effect when today’s temperature is around 15\,°C than when the present day temperature is 25\,°C. Put differently: how uncertainties in future warming ($\beta^\text{z}$) translate into relative changes in damages also depends on present-day temperatures ($T_\text{abs,c}^\text{z}$).  While \( T_\text{abs,c}^\text{z} \) can be measured reliably from observation-based products such as ERA5, uncertainty in the warming pattern, derived from the emulator-based global-mean change multiplied by \( \beta^\text{z} \), is unavoidable because it depends on the chosen ESM.

Therefore, these two components of pattern scaling play different roles. Observation-based absolute temperatures \( T_\text{abs,c}^\text{z}(\text{ERA5}) \) anchor the analysis in the present-day climate and establish the potential for benefits, whereas the model-dependent warming patterns \( \beta^\text{z} \) drive the spread of outcomes. The transparency of the pattern-scaling framework makes this distinction explicit.


\section{Conclusion}
\label{sec:conclusion}

This paper presents an adaptable computational framework that enables IAM practitioners to construct transparent, computationally efficient climate box emulators, with up to four reservoirs, tailored to their research questions. By focusing on linear box-model structures, the framework supports rapid calibration while preserving interpretability, physical consistency, and seamless integration into economic and policy analyses.

To illustrate how calibration targets, emulator configuration, and hyperparameter selection affect economic outcomes, we compare three carbon-cycle emulators calibrated to pre-industrial and present-day conditions. Our results show that incorporating a dynamically evolving land-biosphere reservoir materially alters climate-economic projections.
The PI-calibrated \(4\)PR model, which treats the land reservoir as static, behaves similarly to the simpler \(3\)SR configuration, whereas the \(4\)PR-X variant, which endogenizes land-use change, yields markedly higher atmospheric \(\mathrm{CO}_{2}\) concentrations and temperatures under both RCP and business-as-usual trajectories. These climate differences raise the optimal social cost of carbon and underscore the need for stronger mitigation, highlighting the policy relevance of deforestation and urbanization.

Neglecting land-use dynamics therefore risks systematic underestimation of future atmospheric carbon burdens, temperature rise, and the carbon price required to meet climate targets. Carbon-capture‐and‐storage experiments corroborate that higher carbon taxation (or equivalent measures) becomes necessary whenever land management is imperfect or trends toward net deforestation.

Extending the analysis with linear pattern scaling links global mean warming to regional outcomes, revealing that local temperature responses typically range from 50 \% to 250 \% of the global mean. Although this spatial resolution enriches impact assessments, it also inherits the structural uncertainty of the underlying Earth system model patterns.

A distinguishing strength of the framework is its transparency. In contrast to many machine-learning emulators that replicate Earth system model output, our constrained least-squares calibration enforces mass balance, non-negativity, and dynamical stability, and is accompanied by automated validation tests that expose how calibration choices propagate to economic indicators. 
All model variants are openly available, via a simple Python API, at \url{https://github.com/ClimateChangeEcon/Building_Interpretable_Climate_Emulators_forEconomics}, enabling effortless exploration of structural and parametric uncertainty.

\clearpage
\appendix


\section{Additional Numerical Experiments}
\label{sec:appendix:additional_experiments}

\begin{figure}[ht!]
\begin{minipage}{0.5\textwidth}
	\noindent
	\centering
	\hspace{2em}\small{\textsc{Fossil Fuels \& Industry}}
\end{minipage}%
\begin{minipage}{0.5\textwidth}
	\noindent
	\centering
	\small{\textsc{Land-Use Change}}
\end{minipage}
\vspace{-1em}
    \includegraphics[width=\textwidth]{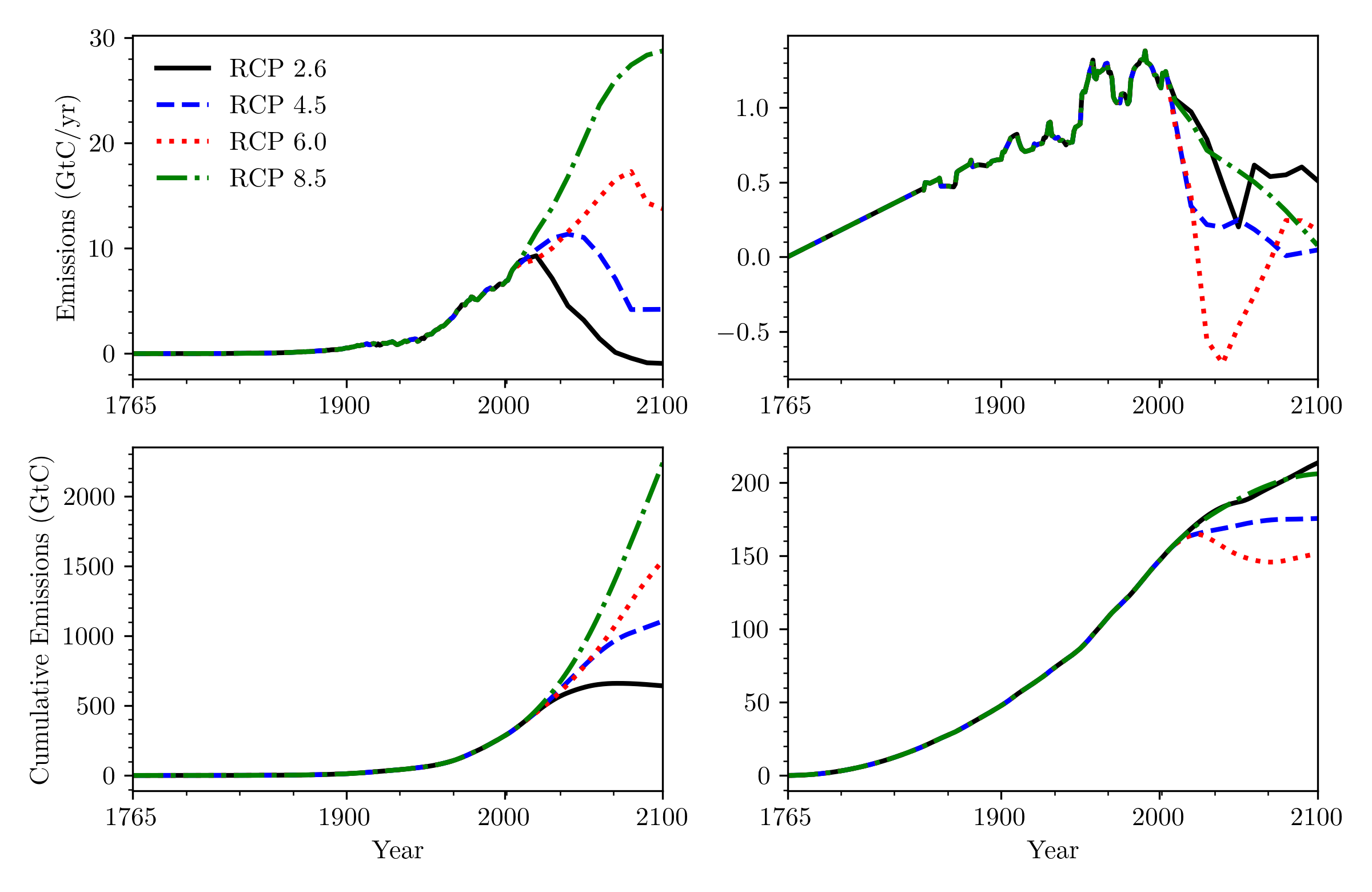}
    \caption{
    Annual and cumulative CO$_2$ emissions: total fossil-fuel and industrial sources (left panels) versus land-use-change sources (right panels) across RCP scenarios $2.6$, $4.5$, $6.0$, and $8.5$.
    }
\label{fig:4}
\end{figure}

This Appendix provides additional supporting numerical experiments for the PI-calibrated emulators from the climate science literature, designed to systematically test and validate the full climate emulator framework. This framework integrates the carbon-cycle and temperature models described in Sections~\ref{sec:3.1} and~\ref{sec:3.2}, respectively. In Appendix~\ref{sec:4.1}, we describe a set of standardized atmospheric-perturbation tests that enable consistent comparisons across models.
Next, in Appendix~\ref{sec:4.2} we present simulations of global CO$_2$ concentrations and temperature change under Representative Concentration Pathway (RCP) emission scenarios~$2.6$, $4.5$, $6.0$, and $8.5$.
In Figure~\ref{fig:4}, we present test RCP emissions scenarios, which are categorized into fossil-fuel, industrial, and land-use-change categories.
We note that the land-use-change component is particularly relevant for the $4$PR-X model.

 \subsection{Atmospheric Perturbation Tests} \label{sec:4.1}
%
Our first numerical experiment follows a standard pulse-decay test from the climate-science literature (cf.~\citealp{folini2024climate}, and references therein).
The test evaluates the atmospheric CO$_2$ decay trajectory after an atmospheric pulse emission, as described in~\cite{joos2013carbon,zickfeld-et-al:21}. 
We next reproduce the zero-emission-commitment experiment of~\cite{macdougall-et-al:20} to probe the emulator's behavior once emissions cease.
Together, these experiments highlight key properties of the linear carbon cycle, especially scale invariance and symmetry, while revealing both the strengths and limitations of our climate emulator.
Because land-use emissions are excluded from the protocol, the $4$PR-X configuration is not considered here.

 \subsubsection{Pulse Decay}
 %
As a calibration benchmark, we simulate the decay of a $100$ GtC pulse that is injected instantaneously into the atmosphere and then allow the system to evolve freely.
The resulting decay trajectories from the $3$SR and $4$PR configurations are compared with the pulse‐decay benchmarks outlined in Section~\ref{sec:3.1}.
To span the plausible range of carbon-cycle responses, we also apply the weighted operator defined by $-1 \leqslant \alpha \leqslant 1$ (see Section~\ref{sec:3.1.2}). 
Note that $\alpha = 0$ yields the multi-model-mean dynamics, while the extrema correspond to $\alpha = -1$, which accelerates decay (larger transfer rates), and $\alpha = 1$, which slows it (smaller transfer rates).
For additional context, we include two externally calibrated extreme models, CLIMBER2–LPJ and MESMO.\footnote{
    These trajectories demonstrate that an integrated assessment modeler wishing to probe extreme carbon-cycle outcomes can do so at negligible computational cost by simply setting $\alpha = -1$ or $\alpha = 1$ instead of recalibrating the entire emulator, for example when aiming to reproduce a MESMO-like response.
}
We note that the simulated values for $t \leqslant 250$ are in-sample because they were used during calibration, whereas values for $t > 250$ are out-of-sample and therefore provide a predictive validation of the model.
%
\begin{figure}[ht!]
\noindent 
\begin{minipage}{0.5\textwidth}
	\noindent
	\centering
	\hspace{3em}\small{\textsc{3SR}}
\end{minipage}%
\begin{minipage}{0.5\textwidth}
	\noindent
	\centering
	\small{\textsc{4PR}}
\end{minipage}
\vspace{-1em}
    \includegraphics[width=\textwidth]{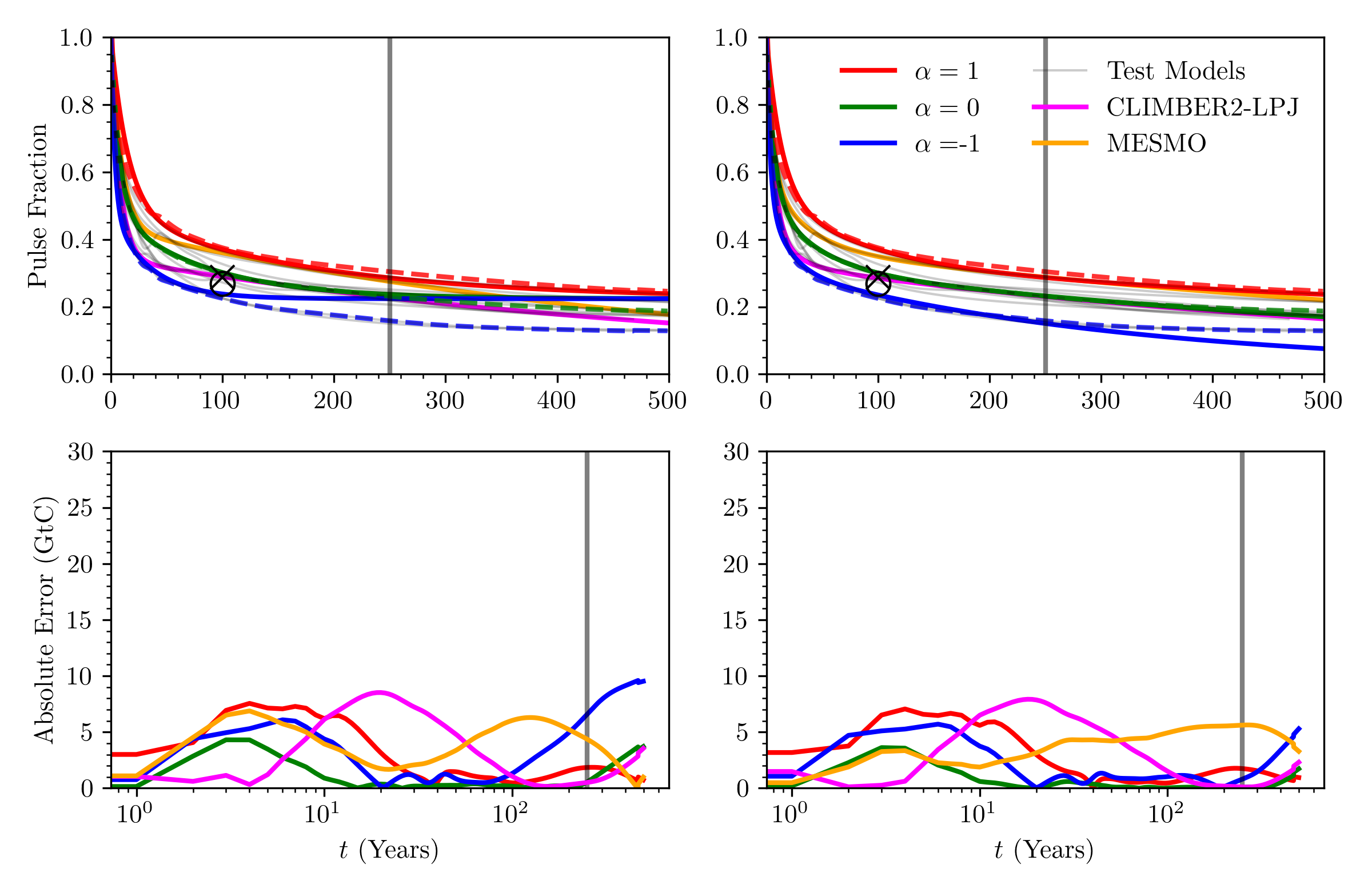}
    \caption{
    Emulated pulse fraction (top panels) and absolute error (bottom panels) for a $100$ GtC atmospheric pulse under PI equilibrium conditions, simulated with the $3$SR and $4$PR configurations.
    Grey lines reproduce the multi-model test ensemble of \cite{joos2013carbon}; the dashed curve is the ensemble mean and two standard deviations (see $\mu$-,$\mu^+$-, and $\mu^-$-benchmark in Figure~\ref{fig:2}).
    Solid lines show emulator results for the indicated $\alpha$ values, along with the two models CLIMBER2–LPJ and MESMO.
    Points left of the vertical gray line at $t = 250$ are in-sample (calibration), while those to the right are out-of-sample (prediction).
    The ``$\times$'' and ``$\bigcirc$'' symbols mark the $+100$ GtC and $-100$ GtC pulse fractions at $t = 100$ year reported by \cite{zickfeld-et-al:21}.
    }
    \label{fig:5}
\end{figure}

\begin{figure}[!htbp]
    \centering
\textsc{Pulse Decay: Fitted Model Reservoir Absorption}\vspace{2em}
\noindent 
\begin{minipage}{0.5\textwidth}
	\noindent
	\centering
	\hspace{3em}\small{\textsc{3SR}}
\end{minipage}%
\begin{minipage}{0.5\textwidth}
	\noindent
	\centering
	\small{\textsc{4PR}}
\end{minipage}
    \includegraphics[width=\textwidth]{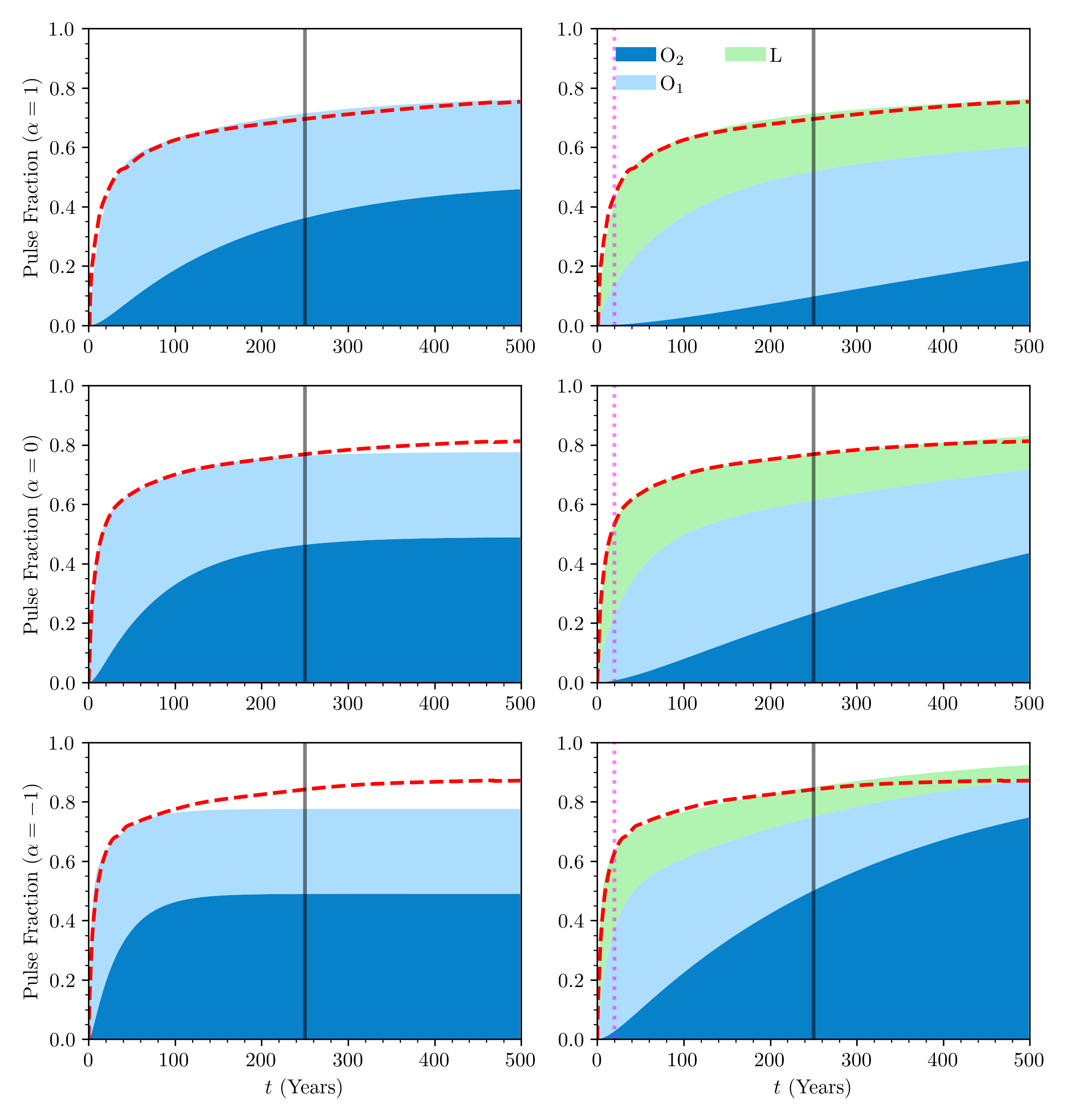}
    \caption{
    Fraction of a $100$ GtC pulse absorbed by various reservoirs over a $500$-year period using the weighted operator $\bb{A}^\alpha$ for the $3$SR (left column) and $4$PR (right column) model configurations; see Section~\ref{sec:3.1.2} for further details.
    The benchmark simulation with $\alpha = 0$ represents the multi-model mean benchmark ($\mu$-benchmark), while $\alpha = 1$ ($\mu^+$-benchmark) and $\alpha = -1$ ($\mu^-$-benchmark) correspond to one standard deviation above and below the multi-model mean, respectively.
    Calibration methodology and fitted parameters are provided in Table~\ref{tab:1} and follow the procedure described in Section~\ref{sec:3.1}.
    }
    \label{fig:B.2}
\end{figure}

Figure~\ref{fig:5} shows that, for $\alpha = 0$, the $3$SR and $4$PR configurations yield comparable error metrics for $t < 250$ when compared to the benchmark results of \cite{joos2013carbon}.
In Figure~\ref{fig:B.2} we show the cumulative uptake of the pulse by the individual reservoirs.
For $t \geqslant 250$, a significant difference emerges between the two model configurations, as the $3$SR model cannot capture the fast decay trajectories associated with the extreme cases.
The primary distinction is that the $4$PR model can represent short-, medium-, and long-term dynamic timescales while the $3$SR model can only represent two timescales, namely short- and medium-term dynamics (see, e.g., Table~\ref{tab:1}).
This limitation of the $3$SR configuration becomes more pronounced when $\alpha \neq 0$. 
Overall, the $4$PR emulator exhibits higher accuracy than $3$SR, particularly in out-of-sample predictions.

Due to the  linearity of the carbon-cycle emulator, it is both homogeneous and symmetric: the atmospheric fraction of a pulse, in absolute value, is independent of the pulse's magnitude and sign.
Consequently, pulse emissions of $1000$, $100$, or $-100$ GtC all follow the same fractional trajectory shown in Figure~\ref{fig:5}.
However, this does not accurately reflect real-world dynamics.
A reservoir cannot be depleted below zero, yet its mass can grow arbitrarily large, breaking the symmetry that linear theory implies.
The work of~\citet{zickfeld-et-al:21} illustrated this asymmetry with the UVic ESCM 2.9 model; their results for $\pm 100$ GtC pulse fractions remaining in the atmosphere are annotated in Figure~\ref{fig:5}.
Although the asymmetry is small for a $\pm 100$ GtC perturbation, it becomes more pronounced as the magnitude of the introduced pulse increases.


\begin{figure}[!htbp]
    \centering
    \noindent
    \textsc{Pulse Decay: Effects Penalty Functions}\vspace{2em}
\begin{minipage}{0.5\textwidth}
	\noindent
	\centering
	\hspace{3em}\small{\textsc{3SR}}
\end{minipage}%
\begin{minipage}{0.5\textwidth}
	\noindent
	\centering
	\small{\textsc{4PR}}
\end{minipage}
    \includegraphics[width=\textwidth]{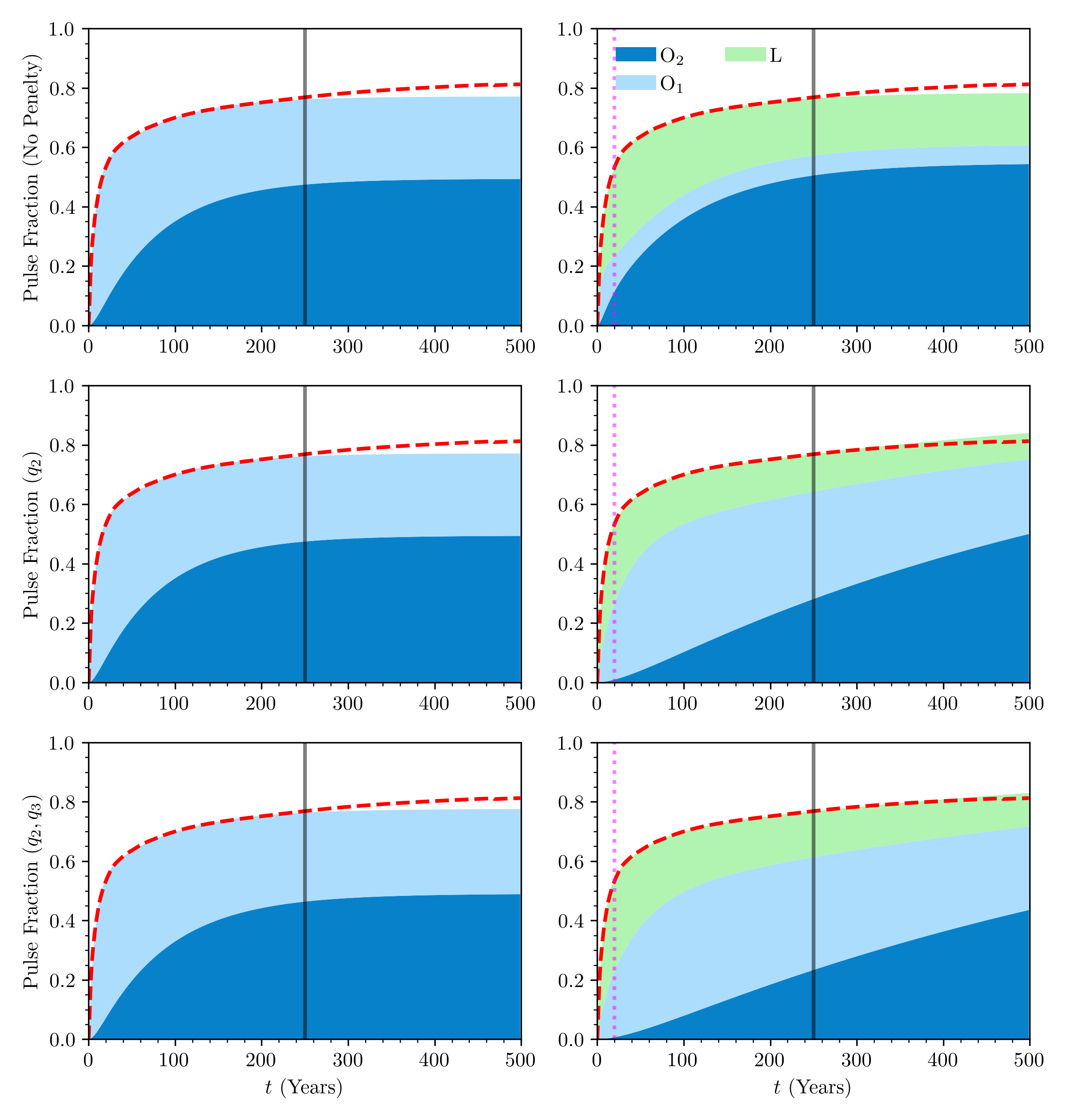}
    \caption{
    Relative atmospheric fraction of a $100$ GtC pulse over $500$ years, illustrating the influence of the penalty terms for Equilibrium-Mass Variability ($q_2$) and Reservoir-Absorption Ratios ($q_3$), simulated with the $3$SR (left panels) and $4$PR (right panels) emulators.
    The experiment follows the setup outlined in Section~\ref{sec:3}; implementation details are provided in Section~\ref{sec:3.1.1}.}
    \label{fig:B.1}
\end{figure}

Figure~\ref{fig:B.1} and~\ref{fig:B.2} are simulations based on a $100$ GtC pulse decay. The model used for this simulation is discussed in detail in Sections~\ref{sec:3.1}. 
The results show the fraction of the pulse absorbed by different reservoirs in the $3$SR and $4$PR model configurations.
Here, $\text{O}_2$, $\text{O}_1$, and $\text{L}$ represent the upper ocean, deep ocean, and land biosphere reservoirs, respectively, with the dashed red line indicating the $\mu$-benchmark for the remaining carbon mass.
The experiment is based on PI conditions starting from $1765$, with a dashed red line serving as the benchmark simulation derived from various test models of differing complexities (for further details, see \citealp{joos2013carbon}).
The vertical gray line marks the time span of $250$ years used for fitting the model parameters. Note that the simulation results to the left and right of this line correspond to in-sample and out-of-sample results, respectively.
The pink vertical line signifies the $20$-year mark at which the $q_1$ penalty function enforces an approximately equal mass of carbon absorption between the oceans and the land biosphere.
Note that the $q_1$ penalty function is only relevant in the $4$PR model configuration.  
%

\subsubsection{Zero Emissions}
\label{sec:5.1.2}
%
This experiment evaluates the emulator's performance using the Zero-Emissions Commitment Model Intercomparison Project (ZECMIP) protocol, which quantifies the committed temperature response after anthropogenic CO$_2$ emissions cease \citep{macdougall-et-al:20}.
Emissions are ramped so that atmospheric CO$_2$ increases by $1\%$ per year until cumulative emissions reach $1000$ GtC.
At that point, emissions are set to zero, and the coupled model is allowed to evolve freely.
%
\begin{figure}[ht!]
\begin{minipage}{0.5\textwidth}
	\noindent
	\centering
	\hspace{3em}\small{\textsc{3SR}}
\end{minipage}%
\begin{minipage}{0.5\textwidth}
	\noindent
	\centering
	\small{\textsc{4PR}}
\end{minipage}
\vspace{-1em}
\includegraphics[width=\textwidth]{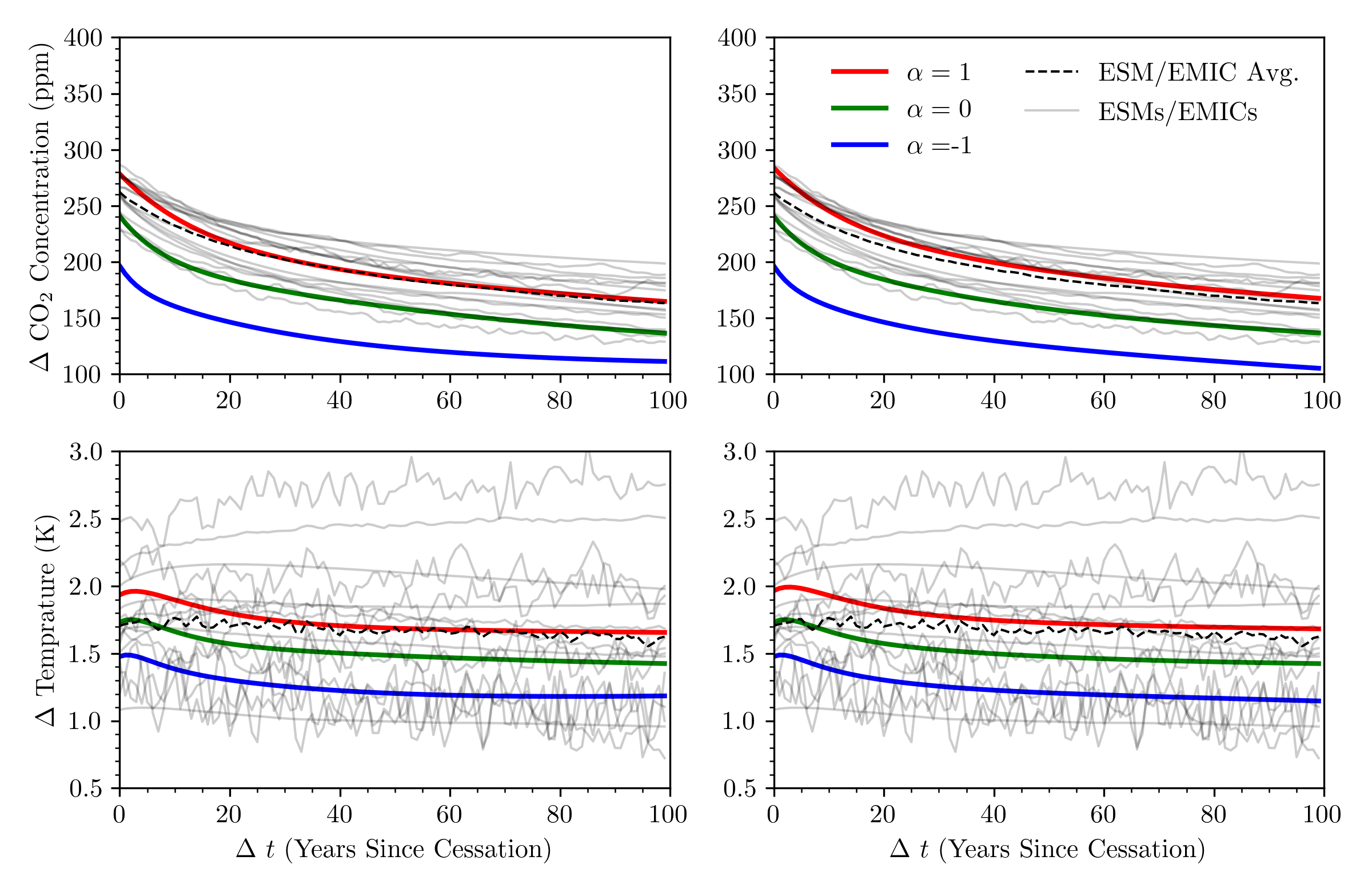}
\caption{
    Trajectories of atmospheric CO$_2$ concentration (top panel) and global-mean temperature (bottom panel) after emissions cease, shown for the $3$SR and $4$PR emulators.
    Thick solid lines correspond to $\alpha = -1$, $0$, and $1$.
    Thin solid lines show simulations from \cite{macdougall-et-al:20} using ESMs and EMICs, with the dashed thin line indicating the average across these simulations.
}
\label{fig:6}
\end{figure}

\begin{figure}[!htbp]
    \centering
    \textsc{Zero Emissions: Commitment Model Intercomparison Project}\vspace{2em}
\noindent 
\begin{minipage}{0.5\textwidth}
	\noindent
	\centering
	\hspace{3em}\small{\textsc{3SR}}
\end{minipage}%
\begin{minipage}{0.5\textwidth}
	\noindent
	\centering
	\small{\textsc{4PR}}
\end{minipage}
    \includegraphics[width=\textwidth]{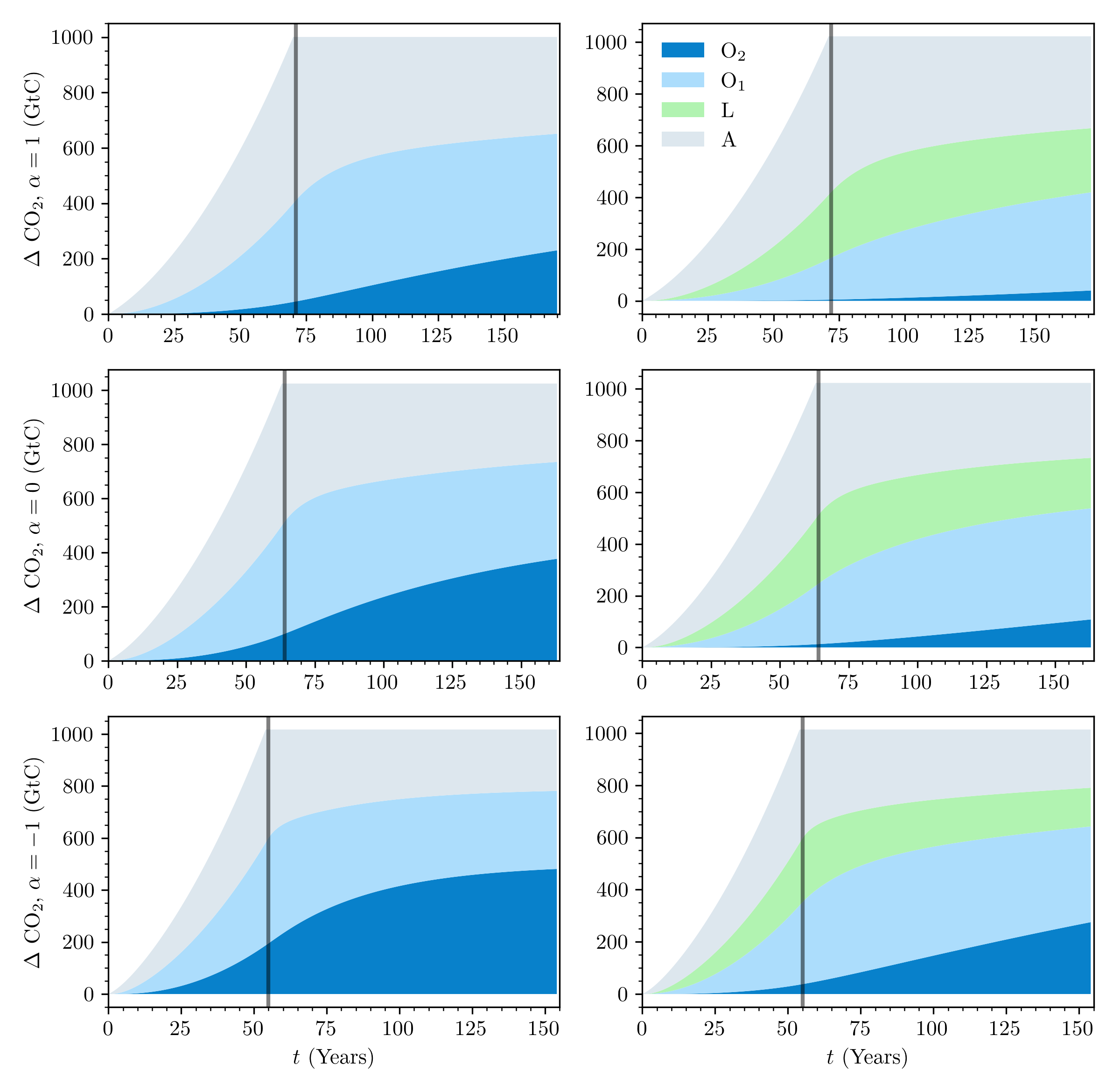}
    \caption{
    Change in $\text{CO}_2$ mass within each reservoir relative to the PI equilibrium condition, using the weighted operator $\bb{A}^\alpha$ in the $3$SR (left panels) and $4$PR (right panels) model configurations.
    The simulations are based on the experimental setup described in Appendix~\ref{sec:5.1.2}.
    Model calibration and fitted parameters are detailed in Table~\ref{tab:1}, following the procedure outlined in Section~\ref{sec:3.1}.
     }
    \label{fig:B.3}
\end{figure}

Figure~\ref{fig:6} shows the simulated changes in atmospheric CO$_2$ concentration and global mean temperature after cessation, along with simulations from ESMs and EMICs for comparison.
Additional detail is provided in Figure~\ref{fig:B.3}, which breaks down the cumulative carbon uptake by each reservoir for both emulator configurations.
Both emulators reproduce the overall decline in atmospheric carbon but show a faster decay in atmospheric CO$_2$ concentrations than other ESMs and EMICs.
With this said, the $4$PR configuration provides a slightly better fit. 
%
The figure also illustrates a fundamental limitation of any linear carbon-cycle emulator: it cannot capture non-linear feedbacks such as the carbon–climate and carbon–concentration couplings \citep{joos2013carbon,arora-et-al:20}.
%
This constraint leads to systematic biases when cumulative carbon emissions become massive and feedbacks become relevant. This is the case for the cumulative emissions from pre-industrial times to the present day.
When the emulator is calibrated under PI conditions, it removes carbon from the atmosphere too rapidly and efficiently, producing the initial underestimate visible in Figure~\ref{fig:6} (green curve below the dashed mean).
Recalibrating the emulator under PD conditions reverses the bias: the trajectory shown in Figure~\ref{fig:6_pd} in the Appendix now overshoots the Earth systems model of intermediate complexity (ESM–EMIC) ensemble mean at the moment emissions cease.


Figure~\ref{fig:B.3} presents results from the experiments outlined in Appendix~\ref{sec:4.1}, following the ZECMIP protocol~\citep{macdougall-et-al:20}. These experiments examine the impact of carbon emissions and subsequent zero-emission states on climate variables over time.
The results correspond to a \emph{Type A} test, where simulations begin under PI conditions and emissions are introduced to achieve a 1\% increase in atmospheric carbon concentration per year until total emissions reach $1000$ GtC.
At this point, symbolized by the gray vertical line, emissions cease, and the system evolves without further carbon inputs. Note that the cessation point varies across simulations.
We also denote the atmosphere as $\text{A}$ in the ``Pulse Decay'' description.
Results are shown for the weighted operator $\bb{A}^\alpha$ with $\alpha = 1$, $0$, and $-1$, as described in Section~\ref{sec:3.1.2}.

\subsection{Representative Concentration Pathway} \label{sec:4.2}

\begin{figure}[h!]
    \includegraphics[width=\textwidth]{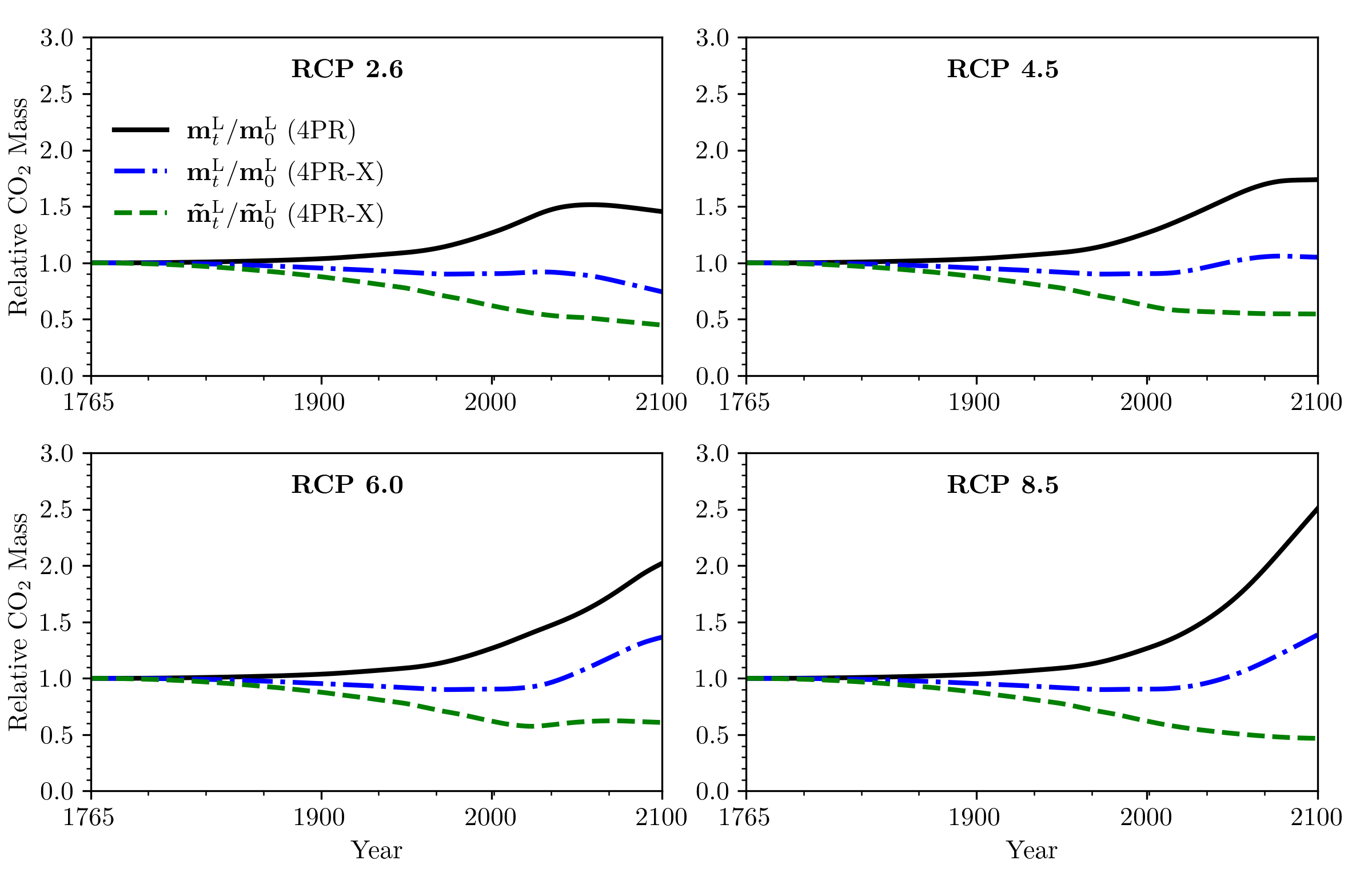}
    \vspace{-1em}
    \caption{
    The time dependent carbon mass as well as the equilibrium mass of the land biosphere, denoted as $\bb{m}_t^{\text{L}}$ and $\bbt{m}_{t}^{\text{L}}$, respectively, are shown as fractions of their initial values ($t=0$) for the $4$PR and $4$PR-X case.
    Results are provided for both constant and time-dependent simulations across each RCP scenario.
    Initial equilibrium conditions are based on the PI calibration values listed in Table~\ref{tab:1}.
    These results pertain specifically to the $4$PR land biosphere model, where the ratio $\bb{m}_t^{\text{L}}/\bbt{m}_t^{\text{L}}$ is relevant only in time-dependent operator simulations, equating to $1$ in constant operator simulations.
}
    \label{fig:7}
\end{figure}

This section assesses the three calibrated climate emulators against the RCP forcing scenarios.
Figure~\ref{fig:7} plots the change in land-biosphere carbon mass relative to the PI baseline $t=0$ for the $4$PR and $4$PR-X configurations.
%
%
The $4$PR-X emulator shows a declining equilibrium land-biosphere mass, $\bbt{m}^{\text{L}}_{t}$ (green dashed line), due to land-use-related emissions, as defined by Equation~\eqref{eq:lb_m_stp} and detailed in Section~\ref{sec:2.1} on Time-dependent Land Capacity.
Consequently, the normalized land-biosphere content, $\bbt{m}^{\text{L}}_{t}/\bbt{m}^{\text{L}}_{0}$, is much lower in $4$PR-X (blue dash-dotted line) than in $4$PR (black line), indicating reduced carbon uptake and higher atmospheric concentrations, consistent with the theoretical expectations in Section~\ref{sec:2.1}.
In addition, the relatively flat land-biosphere carbon uptake in the $4$PR-X model up to present times (e.g., $2015$) aligns with estimates from~\cite{IPCC_carbon_cycle}, who report an approximate $10\%$ increase since the PI period.
The associated annual and cumulative emission trajectories are shown in Figure~\ref{fig:4}. The carbon absorption in the different reservoirs for the various RCP scenarios is provided in Figure~\ref{fig:B.4} in the Appendix.
We emphasize that the $4$PR-X is an emulator rather than a fully physical model; its performance depends strongly on the factor $r$ introduced in Section~\ref{sec:2.1}.

Because a box emulator redistributes emitted carbon in proportion to reservoir equilibrium masses, the shrinking \(\boldsymbol{\tilde m}^{\mathrm L}_{\,t}\) equilibrium mass in the \(4\)PR-X model gradually reduces the fraction absorbed by the land biosphere. The carbon fraction going to the other reservoirs must consequently increase in concert, notably also the carbon fraction that ends up in the atmosphere.  
Thus the time dependence of \(\boldsymbol{\tilde m}^{\mathrm L}_{\,t}\) gradually takes the emulator from its PI calibration, where only a small fraction of emitted carbon remains in the atmosphere, toward a PD calibration, where a larger fraction of emitted carbon remains in the atmosphere, because the uptake by other reservoirs, notably the land biosphere, has become less efficient. While the PI-calibrated \(4\)PR model systematically draws too much carbon out of the atmosphere under PD and future conditions, the \(4\)PR-X emulator largely corrects this bias and is therefore applicable across PI, PD, and future states.

%

%
%
\begin{figure}[ht]
    \includegraphics[width=\textwidth]{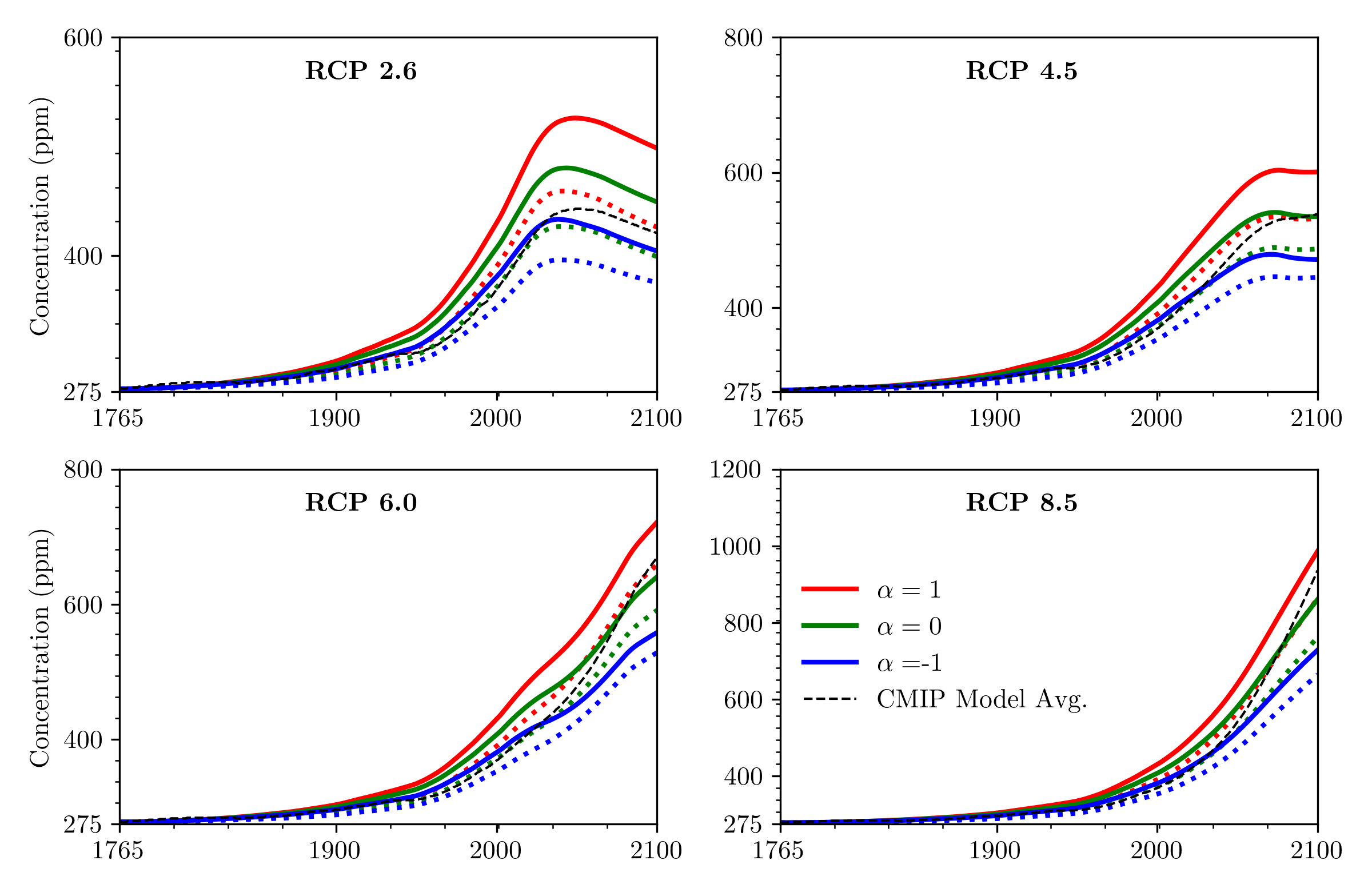}
\includegraphics[width=\textwidth]{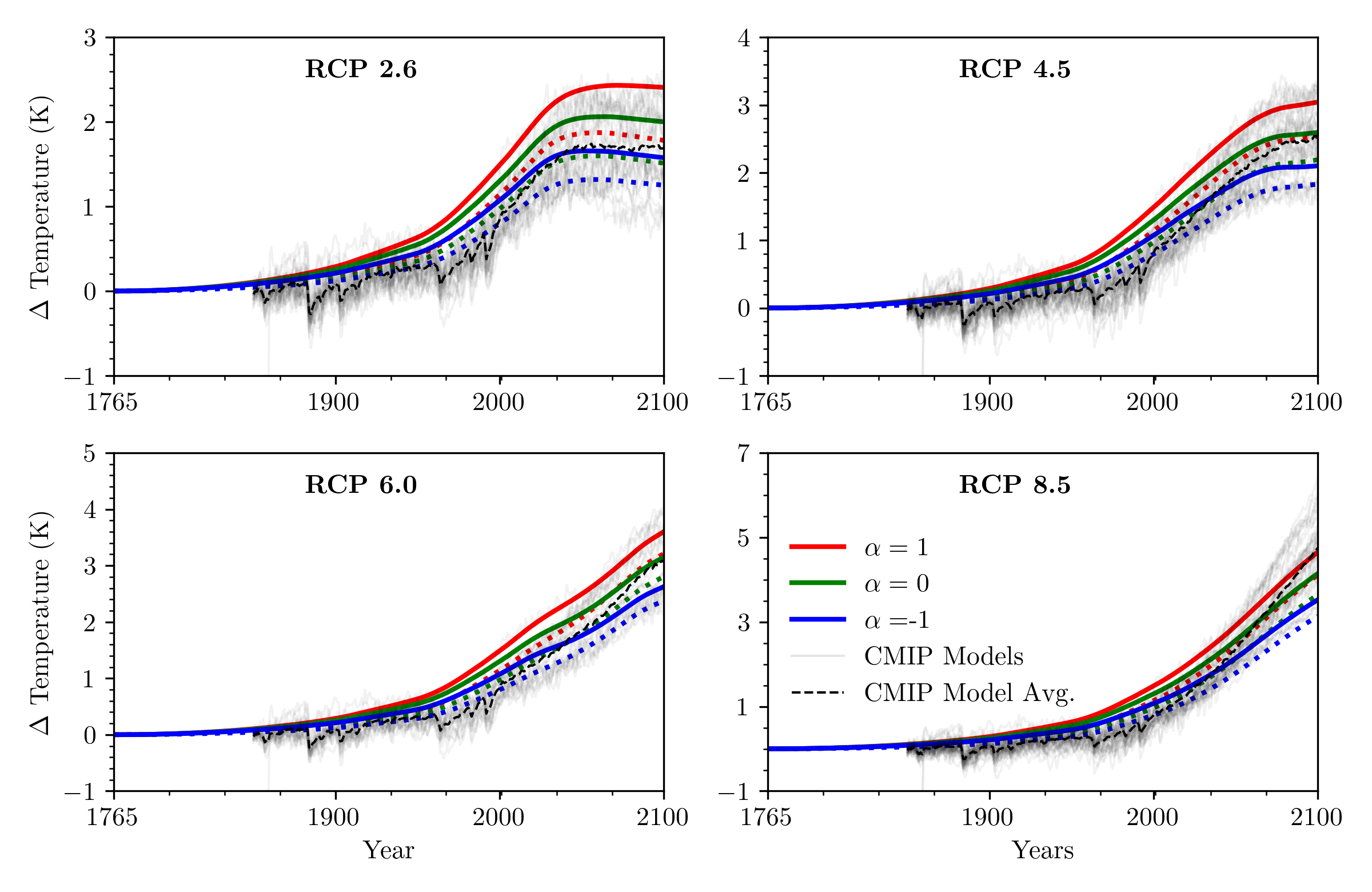}
\caption{
    %
    %
    %
    The top four panels illustrate the simulated atmospheric CO$_2$ concentration, followed by the next four panels, which show the global-mean temperature change. 
    A CCE should be able to reproduce atmospheric CO$_2$ concentration as used in the RCP scenarios of CMIP5. This goal is met by the $4$PR-X emulator (solid green line; red and blue lines indicate extreme calibrations, see text), which is calibrated against PI conditions and whose equilibrium land-biosphere mass is updated dynamically in proportion to land-use emissions, as illustrated in Figure~\ref{fig:7}. A time-constant emulator calibrated to PD conditions, as in CDICE~\citep{folini2024climate}, performs similarly well under future emissions. By contrast, the time constant $3$SR and $4$PR CCEs calibrated to PI conditions remove too much  CO$_2$ from the atmosphere (dotted lines). The added value of the $4$PR-X CCE lies in the fact that despite its PI calibration, it smoothly transitions from PI conditions, against which it is calibrated, onto PD and then future conditions.
    }
    \label{fig:8}
\end{figure}
%

Figure~\ref{fig:8} contrasts the atmospheric CO$_2$ concentration and global-mean temperature trajectories produced by the PI-calibrated $3$SR/$4$PR (which overlap in the visualization), and $4$PR-X emulators across the RCP $2.6$, $4.5$, $6.0$, and $8.5$ scenarios.
Earlier experiments showed that the $4$PR emulator captures long-term pulse-decay dynamics more accurately than the $3$SR model.
However, this advantage has limited impact in the RCP experiments: both configurations follow nearly identical trajectories (both are visualized as the dotted lines in Figure~\ref{fig:8}) and systematically underestimate atmospheric CO$_2$, as they are calibrated to PI conditions.
In contrast, the $4$PR-X runs generate higher concentrations and larger temperature responses; the time-dependent operator reduces the equilibrium land-biosphere sink and therefore leaves more carbon in the atmosphere.

Focusing on the global-mean temperature, for the baseline choice $\alpha = 0$ and comparing with CMIP ensemble means, the $4$PR-X emulator reproduces RCP $4.5$ and $6.0$ remarkably well.
This highlights the importance of an evolving land-biosphere capacity for present-day and moderate-future forcing levels.
At the extremes, CMIP outcomes for $2100$ are best matched by $\alpha = -1$ under RCP 2.6 and $\alpha = 1$ under RCP 8.5.
Varying $\alpha$ thus offers a computationally cheap way to explore stronger negative or positive feedbacks: lower (higher) values yield lower (higher) atmospheric burdens.


Figure~\ref{fig:B.4} depicts the changes in reservoir masses relative to PI conditions for the RCP simulations. The lines correspond to the multi-model mean calibration (\(\alpha = 0\)) and contrast the \(4\)PR emulator with a constant operator against the \(4\)PR-X emulator with a time-dependent operator. The \(3\)SR configuration behaves similarly to the \(4\)PR constant-operator run, except that the carbon that would reside in the land biosphere in \(4\)PR is instead partitioned into the ocean reservoirs.

\newpage
\begin{figure}[!htbp]
    \centering
    \textsc{RCP: Reservoir Absorption (constant operator)}
    \noindent 
    \includegraphics[width=\textwidth]{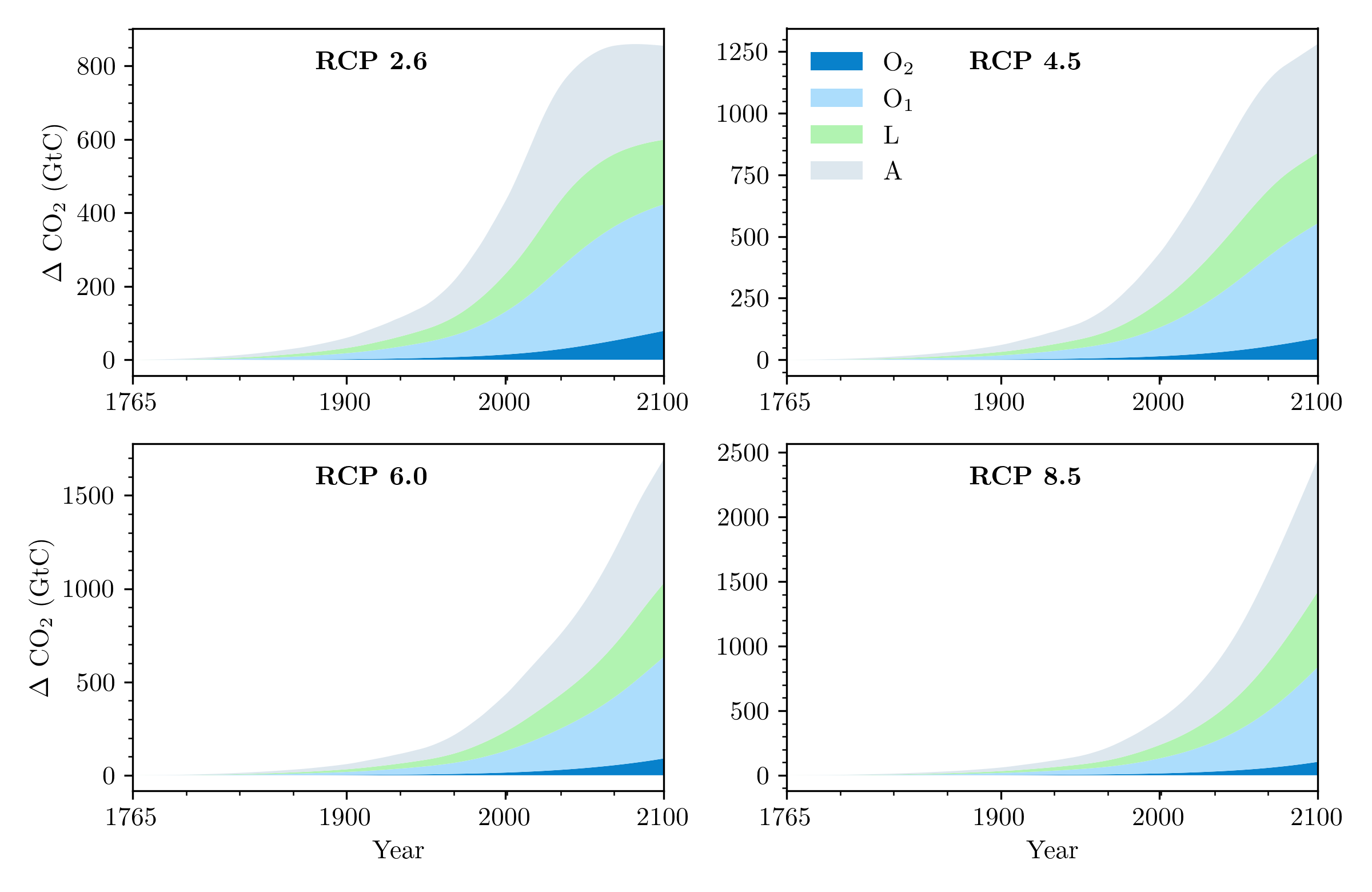}
    \newline
    \newline
    \textsc{RCP: Reservoir Absorption (Time-dependent operator)}
    \includegraphics[width=\textwidth]{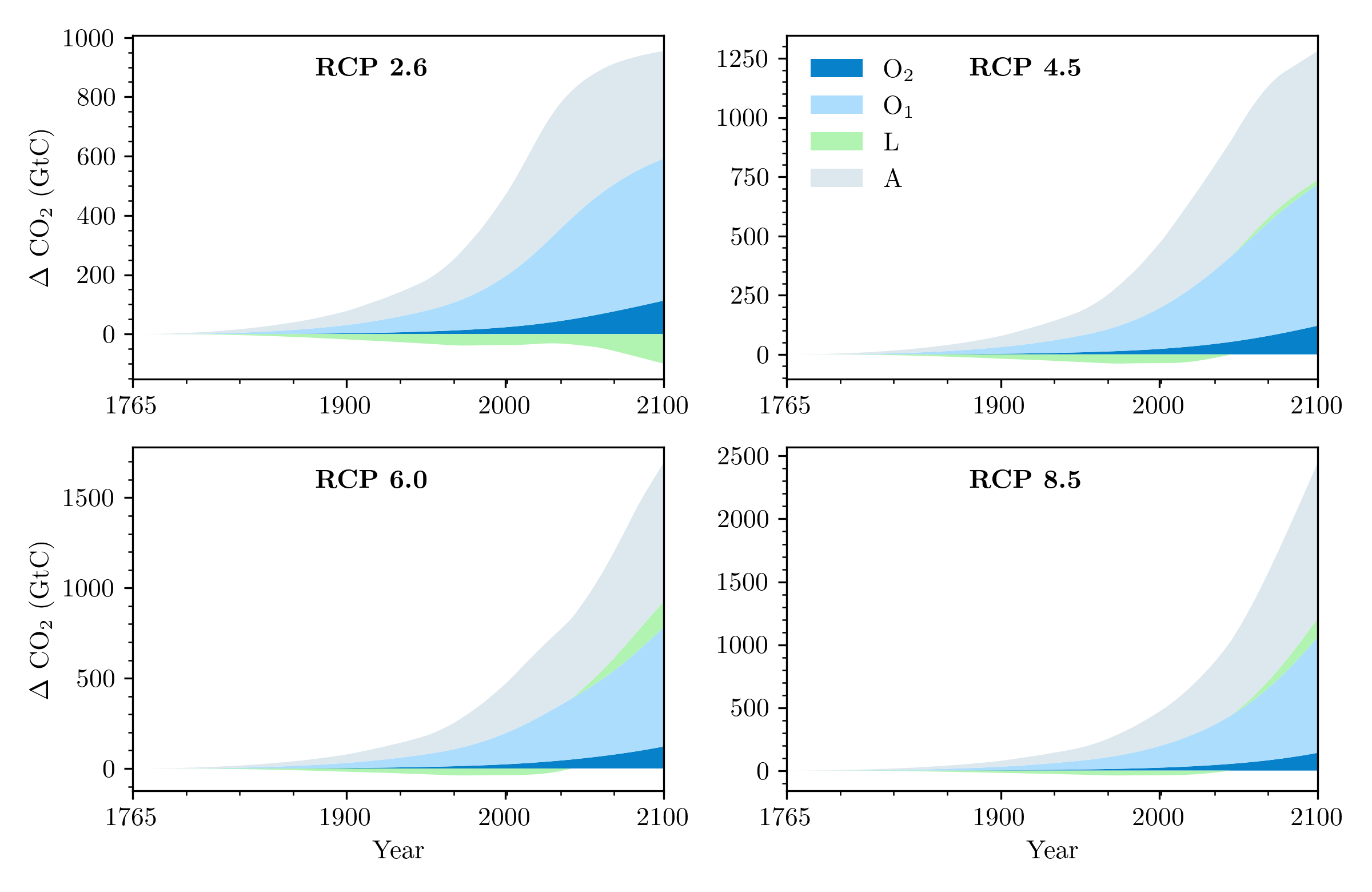}
    \vspace{-1em}
    \caption{Reservoir \(\mathrm{CO}_2\) anomalies relative to the pre-industrial equilibrium, simulated with the \(4\)PR emulator: constant operator (top panels) versus time-dependent operator (bottom panels).  All runs follow the experimental protocol of Section~\ref{sec:4} and use \(\alpha = 0\).}
    \label{fig:B.4}
\end{figure}
\section{Present-Day Carbon Cycle Parameters and Simulations }\label{APX:C}

The main text and Appendix~\ref{sec:appendix:additional_experiments} report results calibrated to pre-industrial conditions.  
In this appendix, we present the corresponding analysis 
re-calibrated with present-day data from \cite{joos2013carbon}.

\subsection{Climate Emulator Calibration and Simulations}
%
The calibration procedure mirrors that described in Section~\ref{sec:3.1}, ensuring consistency across both PI and PD baselines.  

\begin{figure}[H]
\noindent 
 \vspace{-1em}
     \centering
    \includegraphics[width=\textwidth]{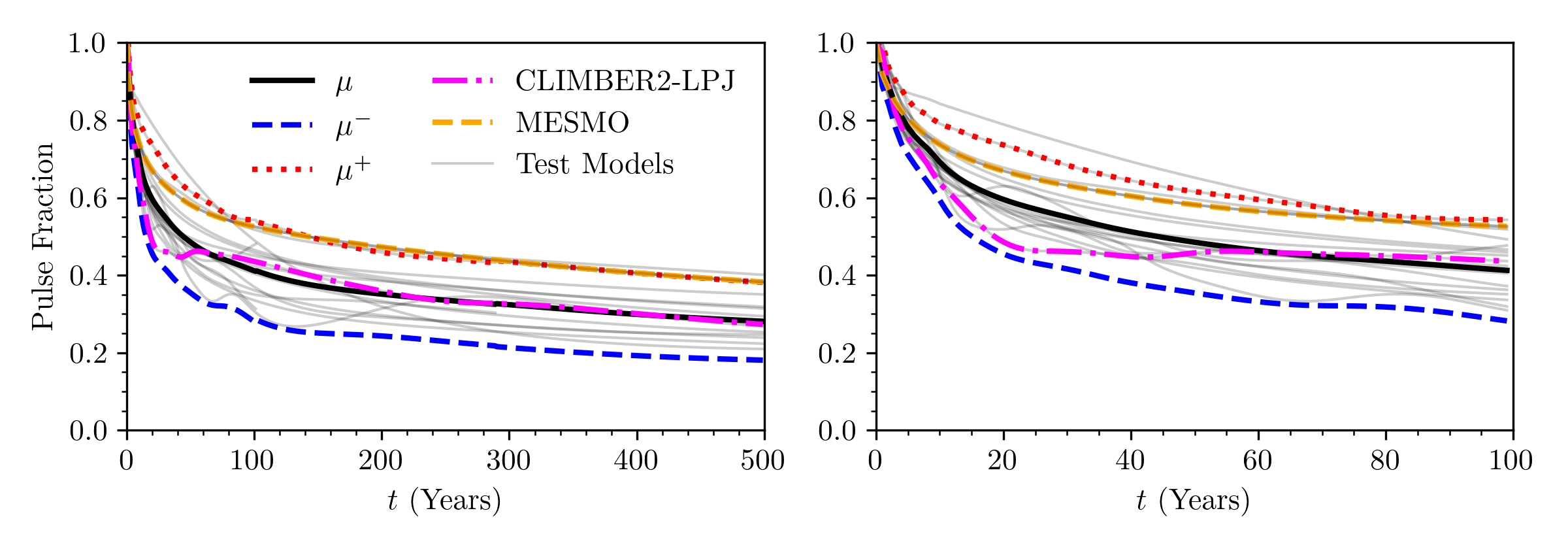}
    \caption{
    Analogous to Figure~\ref{fig:2} (see its caption for further details), the figure presents simulated atmospheric CO$_2$ decay after a $100$ GtC pulse under PD equilibrium conditions with experimental data from \cite{joos2013carbon}.
    }
    \label{fig:2_pd}
\end{figure}

\begin{figure}[H]
\noindent 
 \vspace{-1em}
     \centering
    \includegraphics[width=0.9\textwidth]{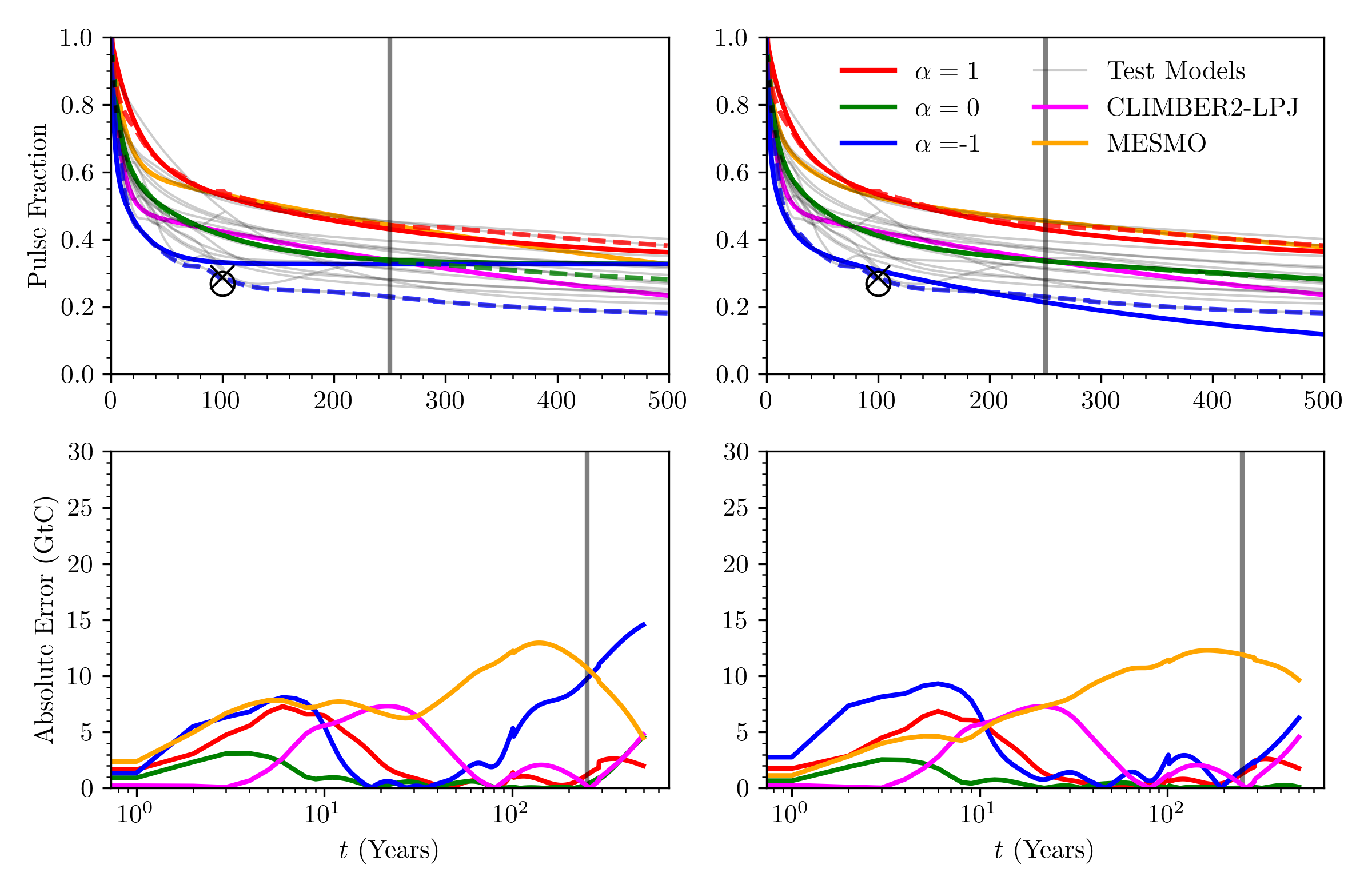}
    \caption{ 
    Analogous to Figure \ref{fig:5} (see its caption for further details), this figure shows the emulated pulse fraction (top panels)  and absolute error (bottom panels) for a $100$ GtC atmospheric pulse under PD equilibrium conditions, simulated with the $3$SR and $4$PR configurations.
    }
    \label{fig:6_pd}
\end{figure}

\begin{figure}[H]
\noindent 
 \vspace{-1em}
     \centering
    \includegraphics[width=0.9\textwidth]{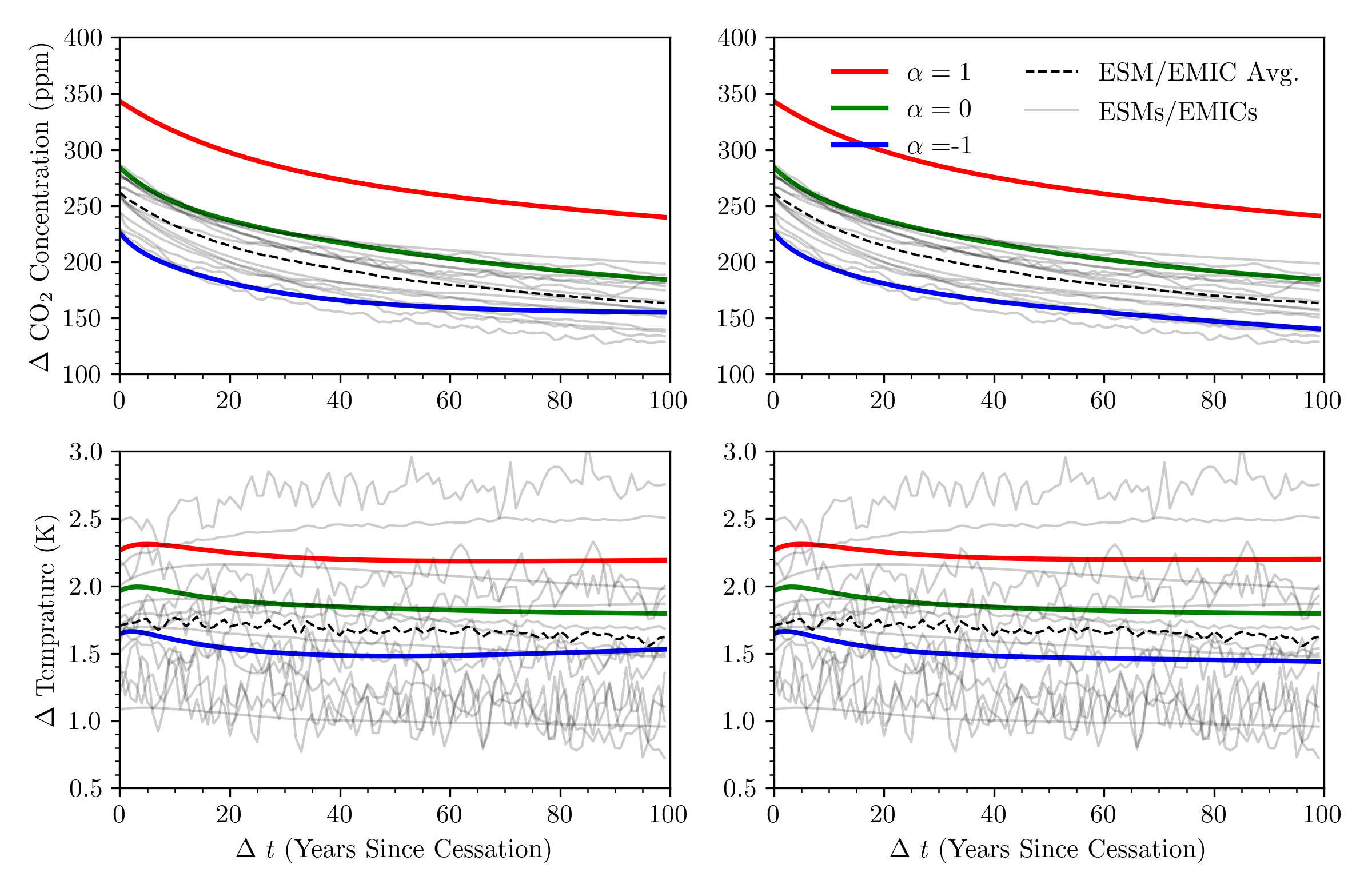}
    \caption{
     Analogous to Figure \ref{fig:6} (see its caption for further details), this figure shows the trajectories of atmospheric CO$_2$ concentration (top panel) and global-mean temperature (bottom panel) after emissions cease, shown for the $3$SR and $4$PR emulators.
     This experiment follows the procedure of~\citealp{macdougall-et-al:20} and uses model calibration based on PD conditions; see Appendix~\ref{sec:5.1.2} for details.
     }
    \label{fig:6_pd2}
\end{figure}

Table~\ref{tab:1_pd} (analogous to Table~\ref{tab:1}) lists the estimated parameters.  
Figure~\ref{fig:2_pd}, corresponding to Figure~\ref{fig:2}, compares the multi-model mean ($\mu$) and its two-standard-deviation envelope ($\mu^{+},\mu^{-}$) from \citet{joos2013carbon} with two illustrative Earth-system models, CLIMBER2–LPJ and MESMO, that bracket very fast and very slow pulse-decay responses.  
Figure~\ref{fig:6_pd} (analogous to Figure~\ref{fig:6}) presents the emulated pulse fraction (top) and the associated absolute error (bottom) for a 100 GtC atmospheric pulse under present-day equilibrium conditions, simulated with the 3SR and 4PR configurations.  
Finally, Figure~\ref{fig:6_pd2} (also analogous to Figure~\ref{fig:6}) shows the simulated post-cessation trajectories of atmospheric CO$_2$ concentration and global-mean temperature, alongside results from ESMs and EMICs for comparison.

%
\begin{table}[H]
\setlength{\tabcolsep}{5pt}
\centering
\small
\begin{tabular}{cccccccccc||cc}
& 
\multicolumn{9}{c}{\textsc{Fitted parameter values - Present Day}} &  \\
&
	$\bb{a}^{\mu}_{\text{A}   \to \text{O}_1}$ & 
	$\bb{a}^{\mu}_{\text{O}_1 \to \text{O}_2}$ & 
	$\bb{a}^{\mu}_{\text{A}   \to \text{L}  }$ &
	$\bbt{m}^{\mu}_{\text{A}}$    &
	$\bbt{m}^{\mu}_{\text{O}_1}$  &
	$\bbt{m}^{\mu}_{\text{O}_2}$  &
	$\bbt{m}^{\mu}_{\text{L}}$   &
	$\bb{c}^{\mu^+}$ &
	$\bb{c}^{\mu^-}$ &
$\bb{m}_{\text{O}}^{t_e}/\bb{m}_{\text{L}}^{t_e} $ & 
$\bb{\tau}$
\\[4pt]
\toprule
\toprule
\textit{Lower}& ${\expnum{1}{-6}}$ & ${\expnum{1}{-6}}$  & ${\expnum{1}{-6}}$ & ${589}$ &  ${\expnum{1}{-6}}$    & ${\expnum{1}{-6}}$  & ${\expnum{1}{-6}}$ & ${\expnum{1}{-6}}$  & ${1}$ &  &\\
\textbf{3SR} & $\num{0.0530045}$ & $\num{0.0140693}$ & - & $589$ & $433 $    & $781 $  & - & $\num{0.3389668728501305}$ & $\num{3.321293553434418}$ & - & $7/78$ \\
\textbf{4PR} & $\num{0.0126925}$ & $\num{0.0014693}$ & $\num{0.0441067}$ & $589$ & $769$ & $37,\!185$ & $242 $ & $\num{0.3364479930454331}$ & $\num{3.7008616665615586}$ & $\num{0.66093}$&$6/53/1416$\\
\textit{Upper}& ${0.3}$ & ${0.3}$ & ${0.3}$  & ${589}$ & ${1,800}$ & ${74,\!200}$ & ${1,\!100}$ & ${1}$ & ${5}$ &  &\\
\end{tabular}
\caption{
Analogous to Table~\ref{tab:1} (see its caption for further details), this table presents the model fitting parameters calibrated using the PD datasets from~\cite{joos2013carbon}, with parameter definitions and calibration procedures as described therein.
}  
\label{tab:1_pd}
\end{table}

\subsection{Economic Results}
\begin{figure}[H]
 \begin{subfigure}[b]{0.5\linewidth}
     \centering
     \includegraphics[width=0.95\linewidth]{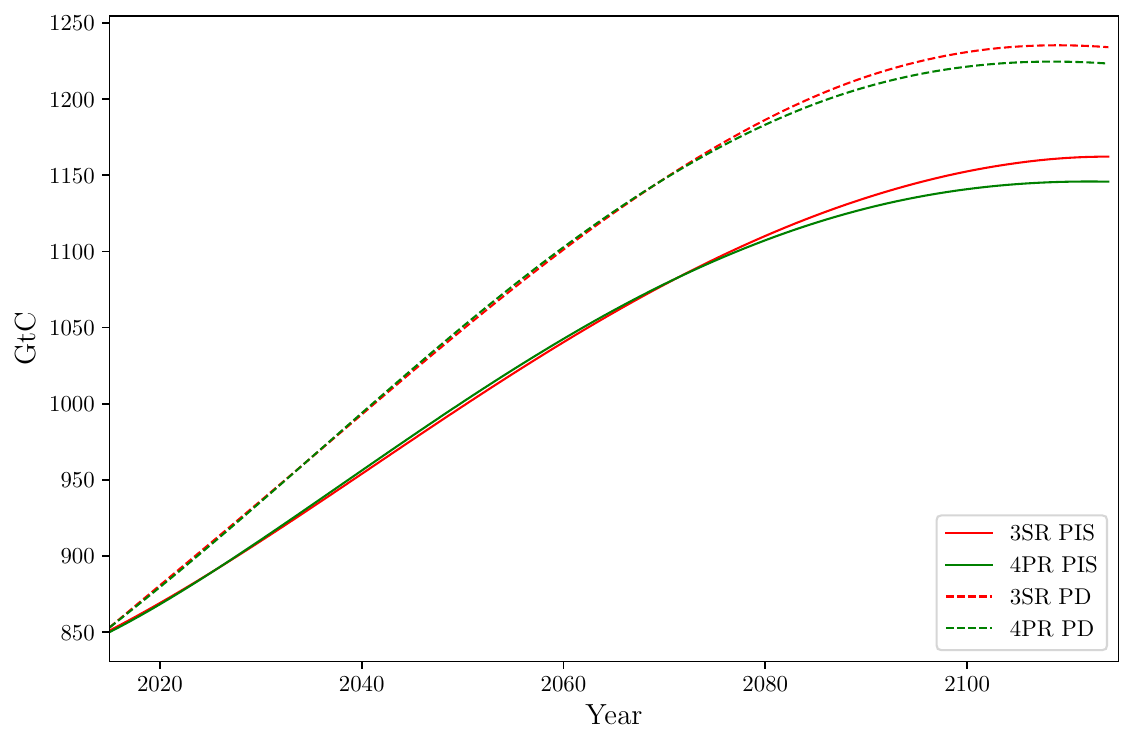}
   \caption{Mass of carbon in the atmosphere} 
     \label{fig:pd_massAT}
  \end{subfigure}
  \begin{subfigure}[b]{0.5\linewidth}
     \centering
     \includegraphics[width=0.95\linewidth]{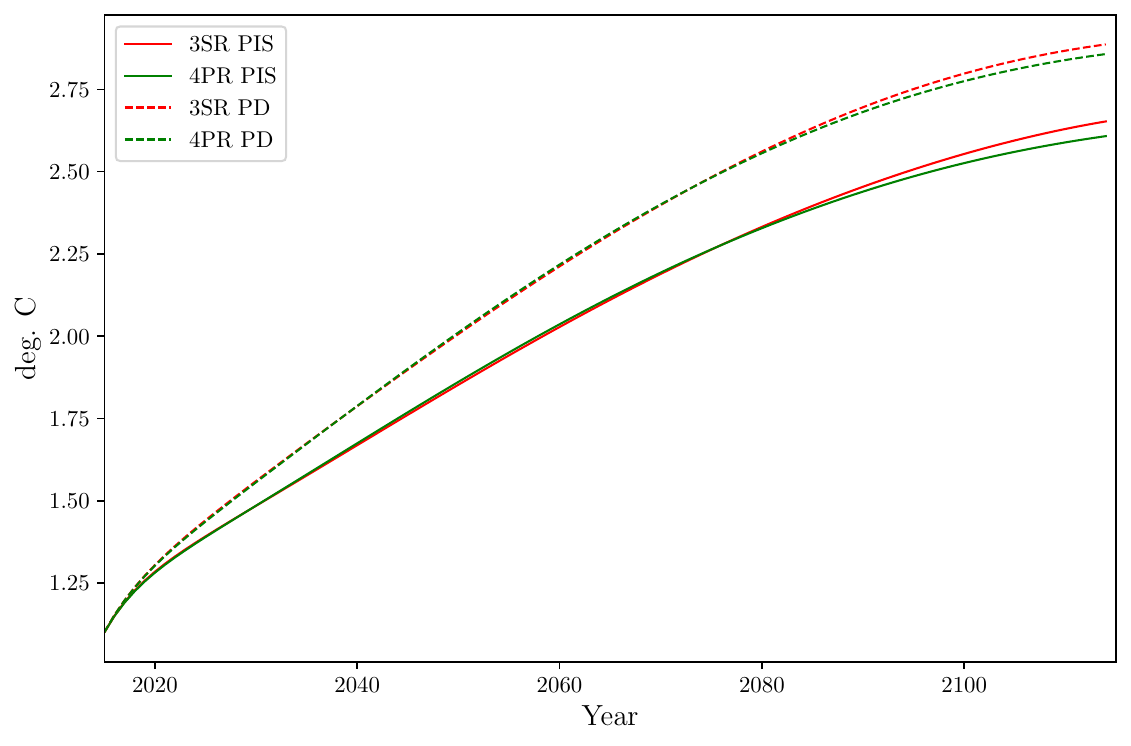}
   \caption{Temperature in the atmosphere} 
     \label{fig:pd_TAT}
  \end{subfigure}
     \caption{Optimal concentrations in the atmosphere (left panel) as well as the atmospheric temperature (right panel) for the present-day simulations.}
    \label{fig:pd_climate}
 \end{figure}

\begin{figure}[H]
 \begin{subfigure}[b]{0.5\linewidth}
     \centering
     \includegraphics[width=0.95\linewidth]{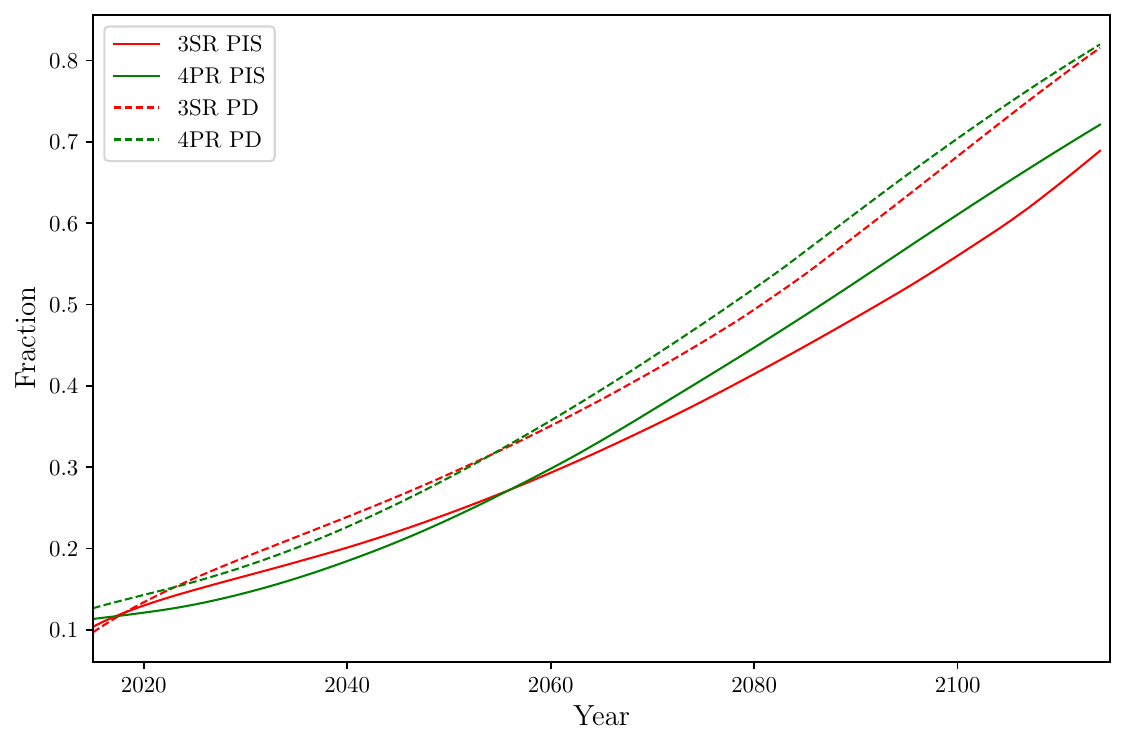}
   \caption{Abatement} 
     \label{fig:pd_muy}
  \end{subfigure}
  \begin{subfigure}[b]{0.5\linewidth}
     \centering
     \includegraphics[width=0.95\linewidth]{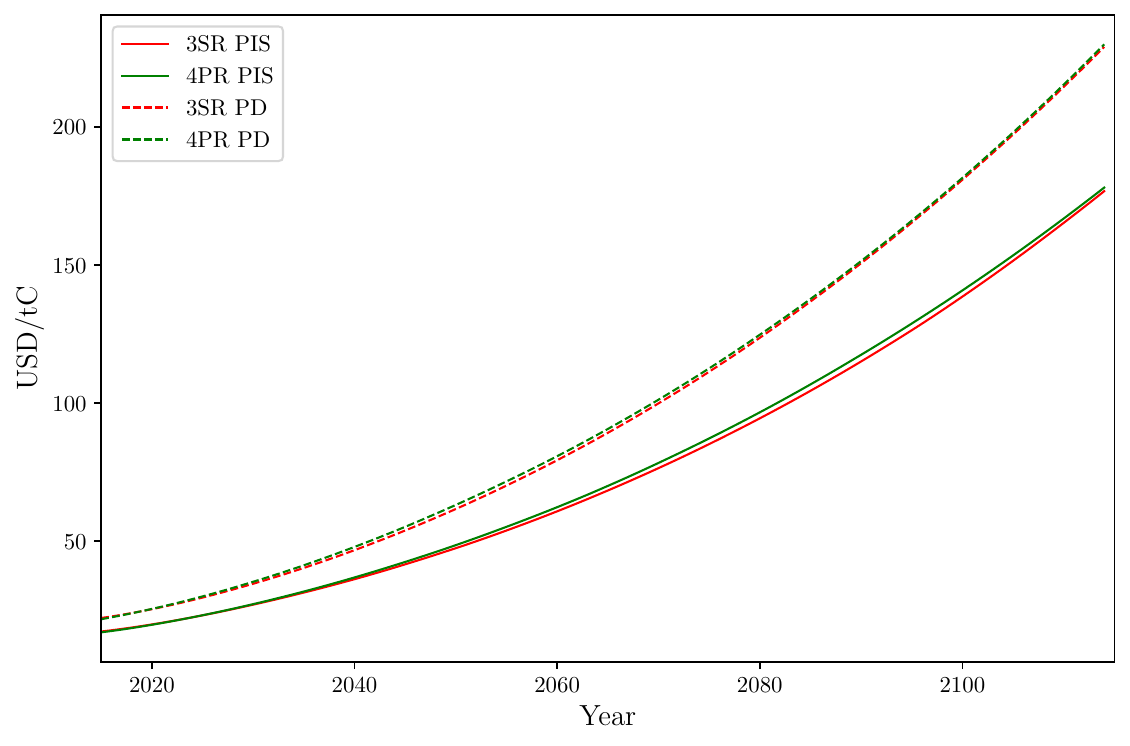}
    \caption{SCC}
     \label{fig:pd_scc}
     \end{subfigure}
     \caption{Optimal abatement (left panel) and the SCC (right panel) for present-day simulations.}
    \label{fig:pd_econ}
 \end{figure}

This appendix compares the economic implications of the PD simulations with those of the PI baseline.
Figure~\ref{fig:pd_climate} shows the trajectories of atmospheric carbon mass (\cref{fig:pd_massAT}) and global-mean temperature (\cref{fig:pd_TAT}); Figure~\cref{fig:pd_econ} presents the optimal mitigation rate (\cref{fig:pd_muy}) and the social cost of carbon (SCC; \cref{fig:pd_scc}).  Dashed lines correspond to PD initial conditions, solid lines to PI conditions, and each pair contrasts the \(3\)SR and \(4\)PR emulators.

With PD initialization, the two emulators behave almost identically; however, the higher initial atmospheric burden leads to greater carbon accumulation and, correspondingly, higher temperatures.

This more pronounced climate response to the emissions results in the need for higher mitigation efforts and the increased SCC.

\section{A Practitioner’s Guide to Using Climate Emulators} \label{APX:D}


This appendix supplies additional information on the IAM used in the numerical experiments of Section~\ref{sec:4} as well as summarizes practically relevant information on the application of climate emulators.
First, for the temperature equations, as well as a comprehensive description of the parametrization of exogenous drivers, namely, labor supply, total factor productivity, emission intensity, and the remaining economic parameters, we refer the reader to \cite{folini2024climate}. 


Second, for the current numerical experiment, the equilibrium values $(\bbt{m}^{\mu})$ and initial values $(\b{m}^{\mu}_{\text{ini}})$ for each carbon cycle model are summarized in the table \cref{table:model_calibration} below:
\begin{table}[H]
\centering
{\begin{tabular}{ c c c   } 
\toprule
 Model & $\bbt{m}^{\mu}=\left(\bbt{m}^{\mu}_{\text{A}},\bbt{m}^{\mu}_{\text{O}_1}, \bbt{m}^{\mu}_{\text{O}_2}, \bbt{m}^{\mu}_{\text{L}} \right)$ & $\b{m}^{\mu}_{\text{ini}}=\left(\b{m}^{\mu}_{\text{A}},\b{m}^{\mu}_{\text{O}_1}, \b{m}^{\mu}_{\text{O}_2}, \b{m}^{\mu}_{\text{L}} \right)$  \\ 
  \midrule
 {3SR}  & (589, 752, 1289, -) & (850, 983, 1377, -)  \\ 
 {3SR-MESMO}  & (589, 820, 36790, -) & (850, 1062, 36823, -)  \\ 
 {3SR-CLIMBER2-LPJ}  & (589, 1179, 37142, -) & (854, 1531, 37181, -)  \\ 
 {4PR}  & (589, 1078, 37220, 387) & (850, 1237, 37236, 531)  \\ 
 {4PR-MESMO}  & (589, 763, 37227, 323) & (850, 901, 37241, 446)  \\ 
 {4PR-CLIMBER2-LPJ}  & (589, 845, 37375, 470) & (854, 1041, 37403, 637)  \\ 
 {4PR-X}  & (589, 1078, 37220, 258) & (850, 1236, 37235, 371) \\ 
 \bottomrule
\end{tabular}}
\caption{Labeling of the models and their respective parameterization and initial values.
Notice that the land-biosphere equilibrium masses are different for $4$PR-X compared to the $4$PR model, as land-use-related emissions have been taken into account (see, e.g., Section~\ref{sec:2.1} for further details).
}
\label{table:model_calibration}
\end{table} 
We clarify here how the initial values were obtained.
Starting from the equilibrium carbon masses representative of the 1750 carbon cycle, we integrated each emulator forward under the historical fossil-fuel and land-use emissions trajectory.
The integration stopped when the simulated atmospheric concentration reached the observed present-day level, 380 ppm \(\mathrm{CO}_2\) in 2005 or 401 ppm in 2015.

In the \(4\)PR-X emulator the equilibrium carbon mass of the land-biosphere reservoir is updated continuously.
Each unit of land-use emission (e.g., deforestation) reduces the reservoir's equilibrium capacity by the same amount: a \(1\;\mathrm{GtC}\) land-use release in a given year (see Figure~\ref{fig:3}) lowers \(\bbt{m}^{\mu}_{\text{L}}\) by \(1\;\mathrm{GtC}\).
As this sink contracts, more carbon accumulates in the other reservoirs, so the \(4\)PR-X model reaches the present-day atmospheric concentration faster than the static configurations.
For consistency, all emulators are ultimately initialized at the same present-day atmospheric \(\mathrm{CO}_2\) level.

Third, to facilitate direct implementation, we explicitly formulate the model below. Incorporating the calibrated parameter values for the for the $\mu$-benchmark reported in Table~\ref{tab:1}, we derive the corresponding operator representation of the climate emulator as follows. We also provide the matrices of transfer coefficients for the $3$SR, $4$PR and $4$PR-X models below:

\begin{align}
\underset{3\text{SR}}{\bb{A}}=
\begin{bmatrix}
-0.0769 & 0.0602 & 0 \\
\textbf{0.0769} & -0.0712 & 0.00638 \\
0 & \textbf{0.0109} & -0.00638
\end{bmatrix}
\quad
\underset{4\text{PR}}{\bb{A}}=
\begin{bmatrix}
-0.0821 & 0.0114 & 0 & 0.0934 \\
\textbf{0.0208} & -0.0139 & 0.0001 & 0 \\
0 & \textbf{0.0025} & -0.0001 & 0 \\
\textbf{0.0613} & 0 & 0 & -0.0934
\end{bmatrix}
\end{align}
Note that the values in bold correspond to the fitted parameters reported in Table~\ref{tab:1}.
The equilibrium masses $\bbt{m}$ appear only in the definition of the upper-triangular entries, which are computed according to Equation~\eqref{eq:4}.
For example, $\bb{A}_{2,1} = \bb{A}_{1,2} \cdot \bbt{m}_{\text{O}_1}/\bbt{m}_{\text{A}}$, consistent across both the $3$PR and $4$PR model configurations.
The $4$PR-X model follows the same structure, except that here the equilibrium masses vary over time, resulting in a time-dependent operator:
\begin{align}
\underset{\text{4PR-X}}{\bb{A}_{t}} =
\begin{bmatrix}
-0.0821 & 0.0114 & 0 & 0.0613 \cdot \frac{\bbt{m}_{\text{A}}}{\bbt{m}_t^{\text{L}}} \\
\textbf{0.0208} & -0.0139 & 0.0001 & 0 \\
0 & \textbf{0.0025} & -0.0001 & 0 \\
\textbf{0.0613} & 0 & 0 & -0.0613 \cdot \frac{\bbt{m}_{\text{A}}}{\bbt{m}_t^{\text{L}}}
\end{bmatrix}
\end{align}
In some studies, the operator is defined such that it absorbs the identity matrix:  
$\mathbf{A}\,\to\,\mathbf{A}+\mathbf{I}$.  
With this convention, the updated Equation~\eqref{eq:1} reads as $\mathbf{m}_{t+1}= (\mathbf{A}+\mathbf{I})\,\mathbf{m}_{t}$, such that a unity term appears on the diagonal and Equation~\eqref{eq:NEW_ALL} states the identity matrix explicitly.
\section{Sensitivity Analysis} \label{APX:E}

This appendix presents two sensitivity analyses that are standard in integrated assessment modeling, both conducted under the pre-industrial baseline used in the main text.  
First, we examine a steeper climate-damage function to investigate whether higher potential damages elicit different responses from the carbon-cycle emulators.  
Second, we vary the discount rate by jointly adjusting the intertemporal elasticity of substitution and the pure rate of time preference while keeping the average real interest rate unchanged; see \cite{folini2024climate} for methodological details.  
Each exercise is performed with the three carbon-cycle representations analyzed earlier: the three-reservoir model (\(3\)SR), the four-reservoir model (\(4\)PR), and the dynamic four-reservoir model (\(4\)PR-X).

\subsection{Sensitivity to Damages}

\begin{figure}[H]
 \begin{subfigure}[b]{0.5\linewidth}
     \centering
     \includegraphics[width=1.0\linewidth]{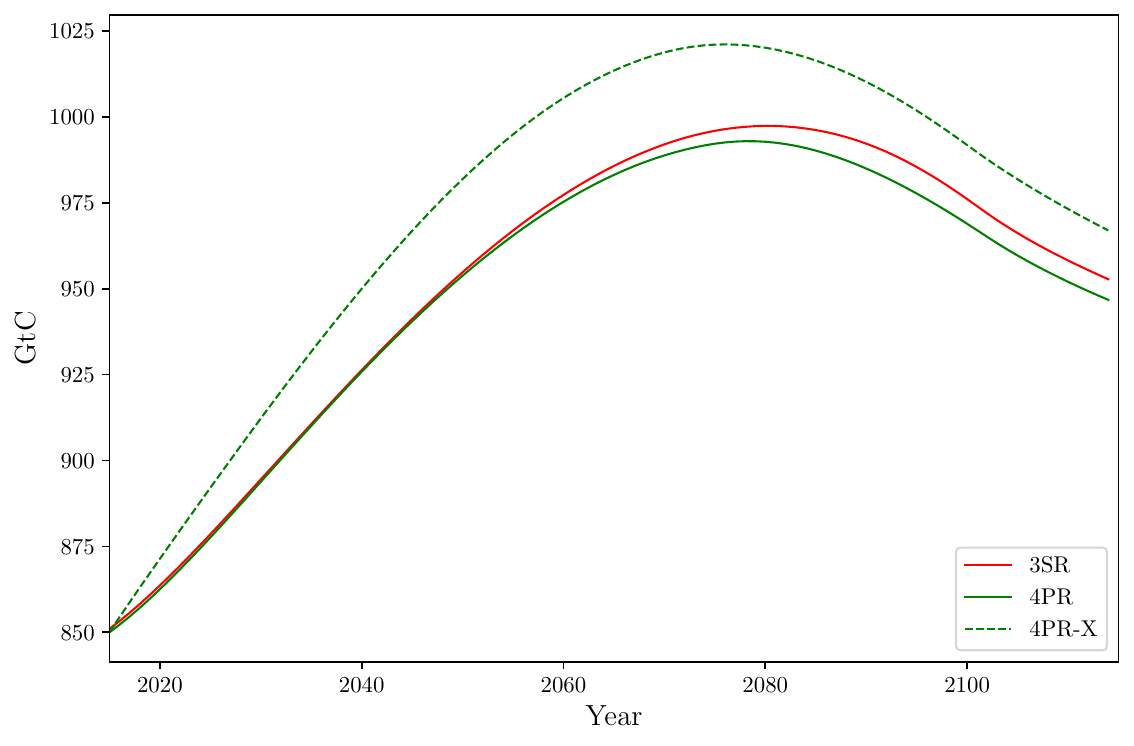}
   \caption{Mass of carbon in the atmosphere} 
     \label{fig:massAT}
  \end{subfigure}
  \begin{subfigure}[b]{0.5\linewidth}
     \centering
     \includegraphics[width=1.0\linewidth]{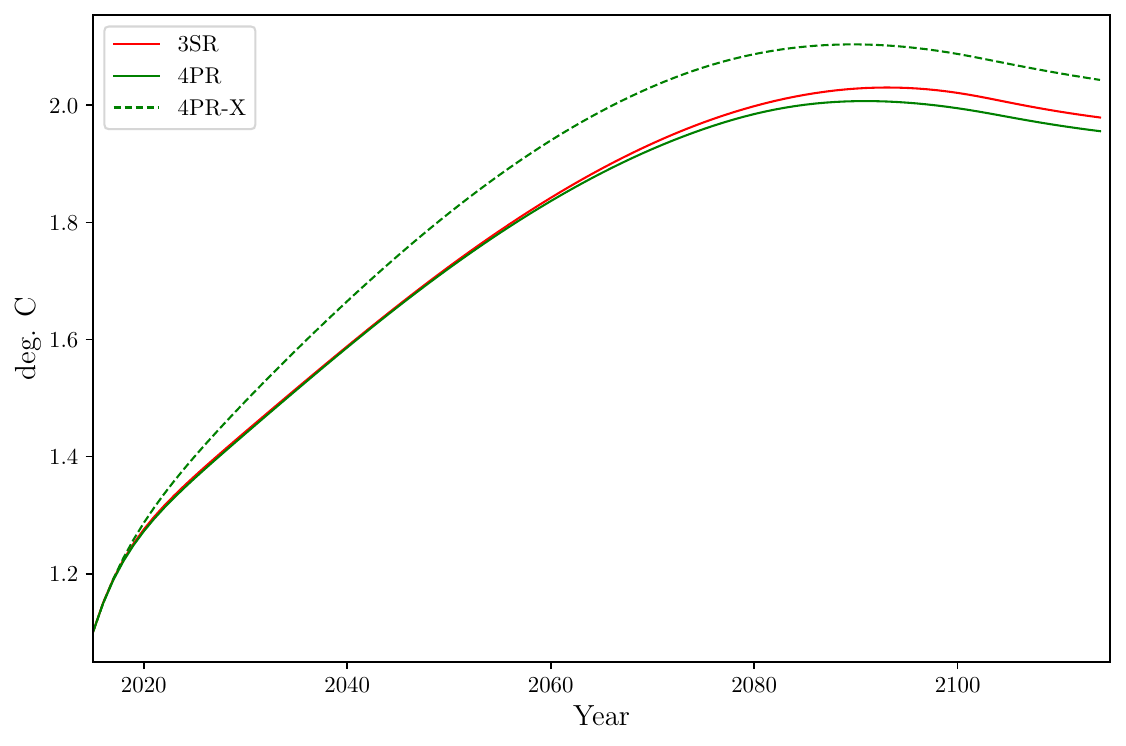}
   \caption{Temperature in the atmosphere} 
     \label{fig:TAT}
  \end{subfigure}
     \caption{Optimal atmospheric CO$_2$ concentrations (left panel) and atmospheric temperature (right panel) in the higher damages scenario.}
    \label{fig:higherdam_climate}
 \end{figure}

\begin{figure}[H]
 \begin{subfigure}[b]{0.5\linewidth}
     \centering
     \includegraphics[width=1.0\linewidth]{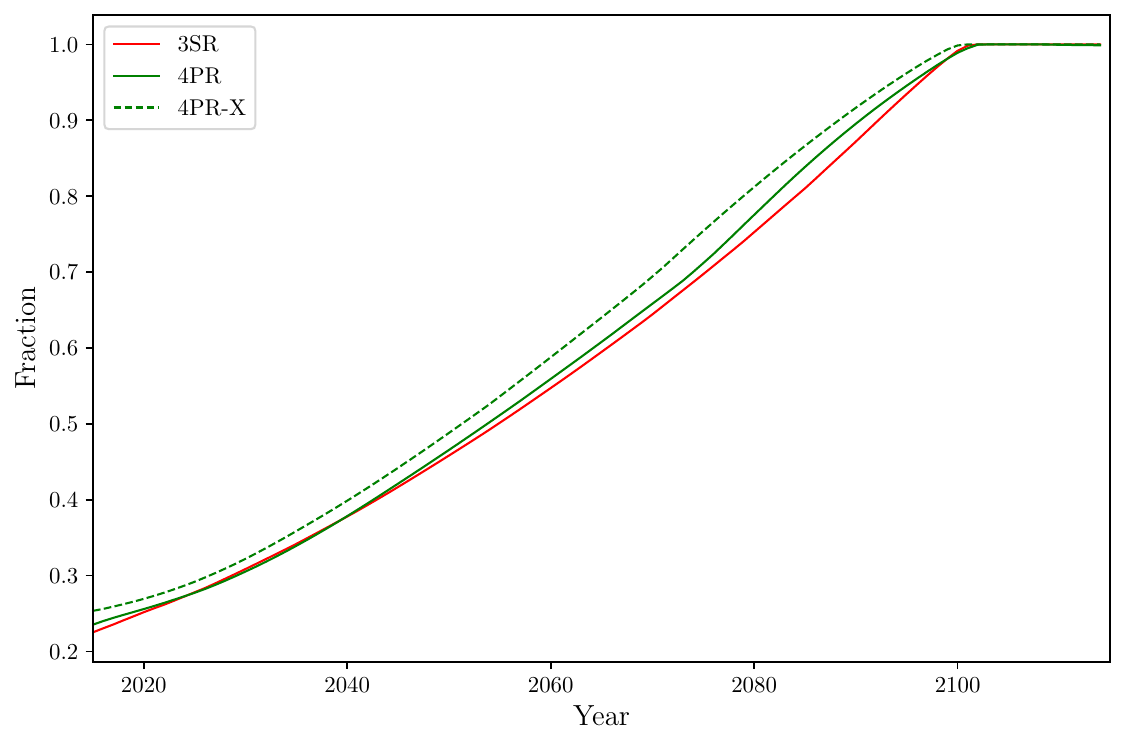}
   \caption{Abatement} 
     \label{fig:muy}
  \end{subfigure}
  \begin{subfigure}[b]{0.5\linewidth}
     \centering
     \includegraphics[width=1.0\linewidth]{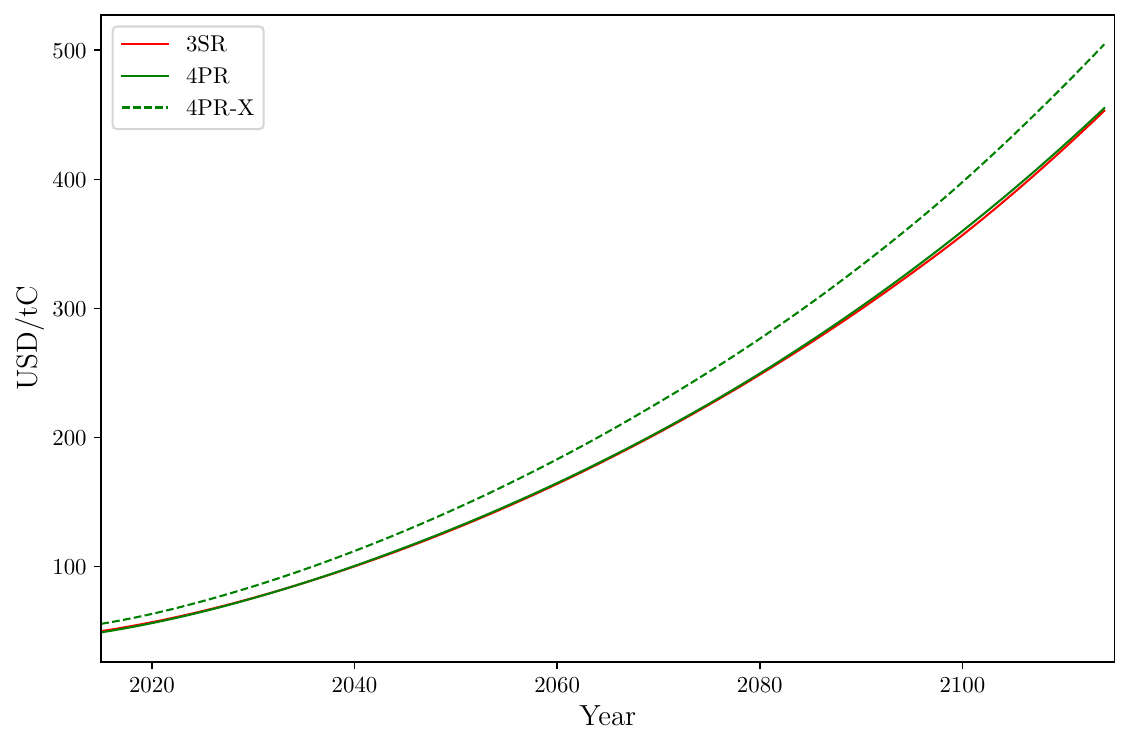}
    \caption{SCC}
     \label{fig:scc}
     \end{subfigure}
     \caption{Optimal abatement levels (left panel) and Social Cost of Carbon (SCC) (right panel) in the higher damages scenario.}
    \label{fig:higherdam_econ}
 \end{figure}

For the increased damages scenario, we assume damages to be three times higher than in the benchmark cases. The simulation results for this scenario are shown in \cref{fig:higherdam_climate} and \cref{fig:higherdam_econ}.

In comparison with the benchmark optimal case (Figure~\ref{fig:opt}), the prospect of higher damages induces markedly stronger mitigation in every scenario, reaching full abatement within 80 years, whereas the standard-damage case attains full mitigation only after 100 years.  The intensified mitigation trajectory yields lower atmospheric carbon concentrations and temperatures; however, the relative performance of the three carbon-cycle emulators remains unchanged from the benchmark case.

\subsection{Sensitivity to the Discount Rate}

\begin{figure}[H]
 \begin{subfigure}[b]{0.5\linewidth}
     \centering
     \includegraphics[width=1.0\linewidth]{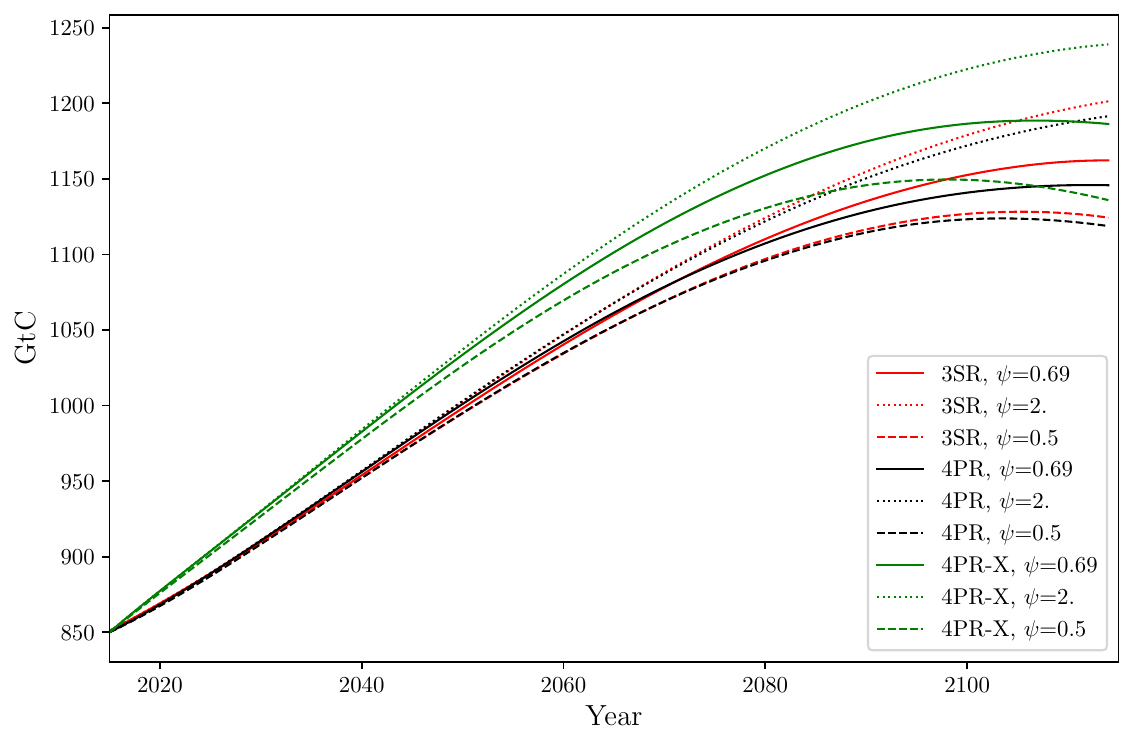}
   \caption{Mass of carbon in the atmosphere} 
     \label{fig:massAT}
  \end{subfigure}
  \begin{subfigure}[b]{0.5\linewidth}
     \centering
     \includegraphics[width=1.0\linewidth]{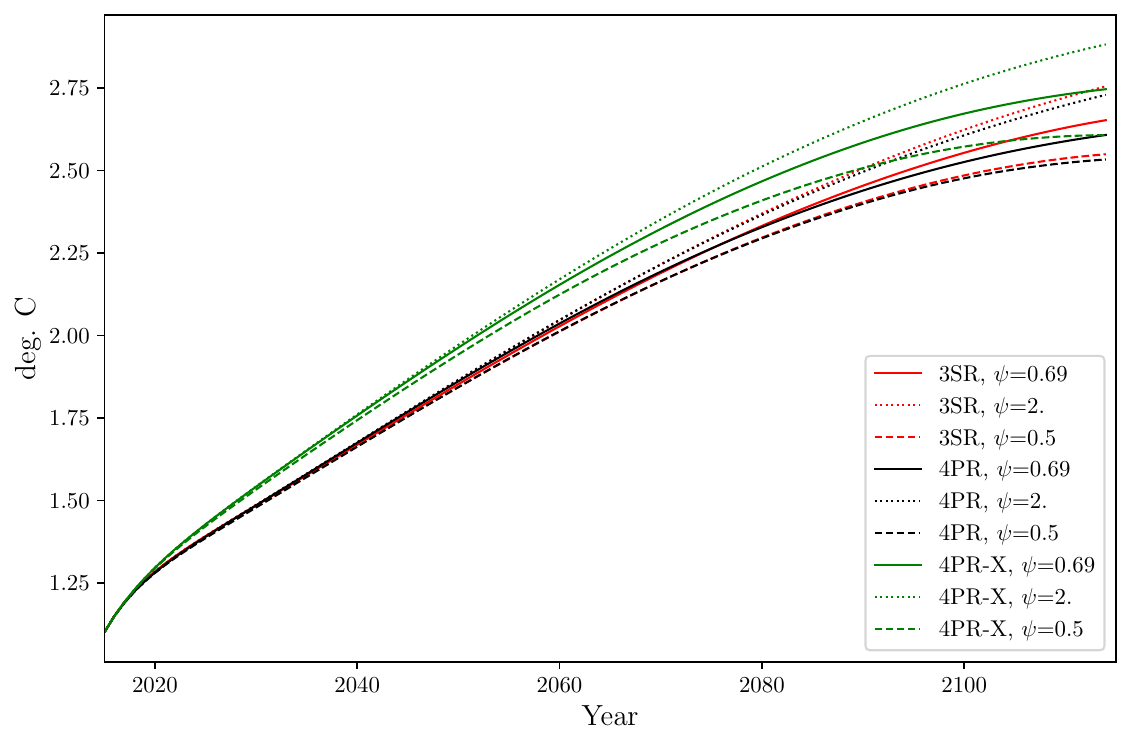}
   \caption{Temperature in the atmosphere} 
     \label{fig:TAT}
  \end{subfigure}
     \caption{Optimal atmospheric carbon concentrations (left panel) and atmospheric temperature (right panel) for varying discount rate values.}
    \label{fig:beta_climate}
 \end{figure}

\begin{figure}[H]
 \begin{subfigure}[b]{0.5\linewidth}
     \centering
     \includegraphics[width=1.0\linewidth]{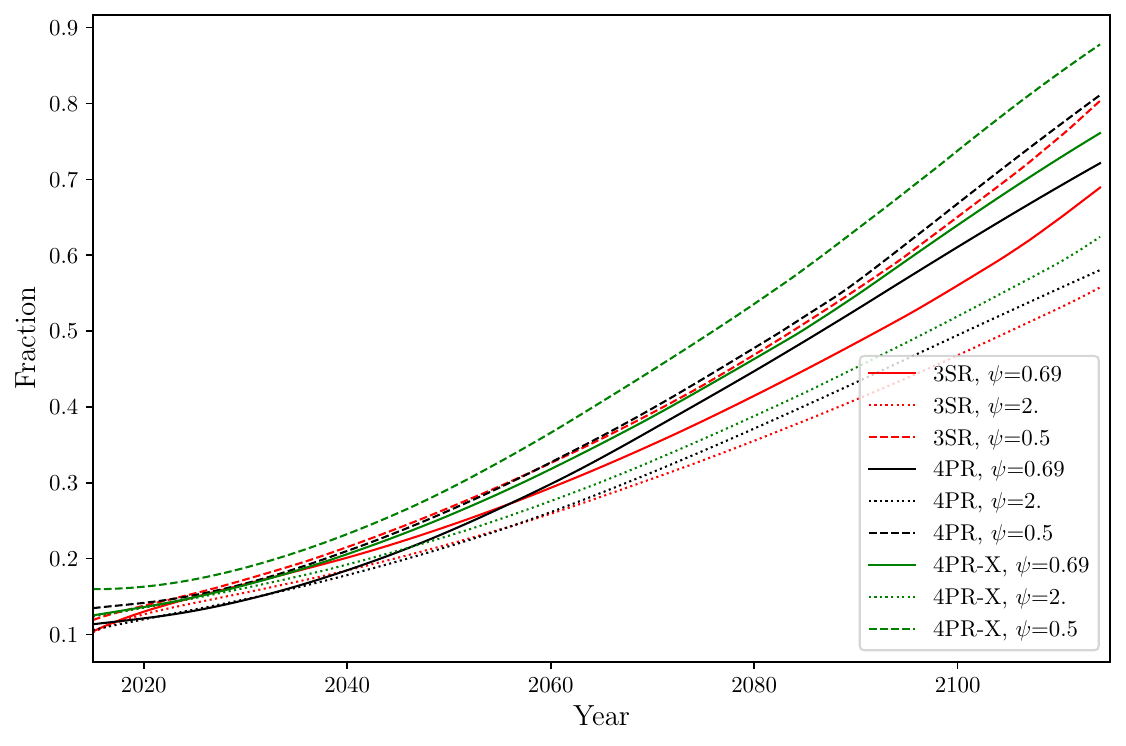}
   \caption{Abatement} 
     \label{fig:muy}
  \end{subfigure}
  \begin{subfigure}[b]{0.5\linewidth}
     \centering
     \includegraphics[width=1.0\linewidth]{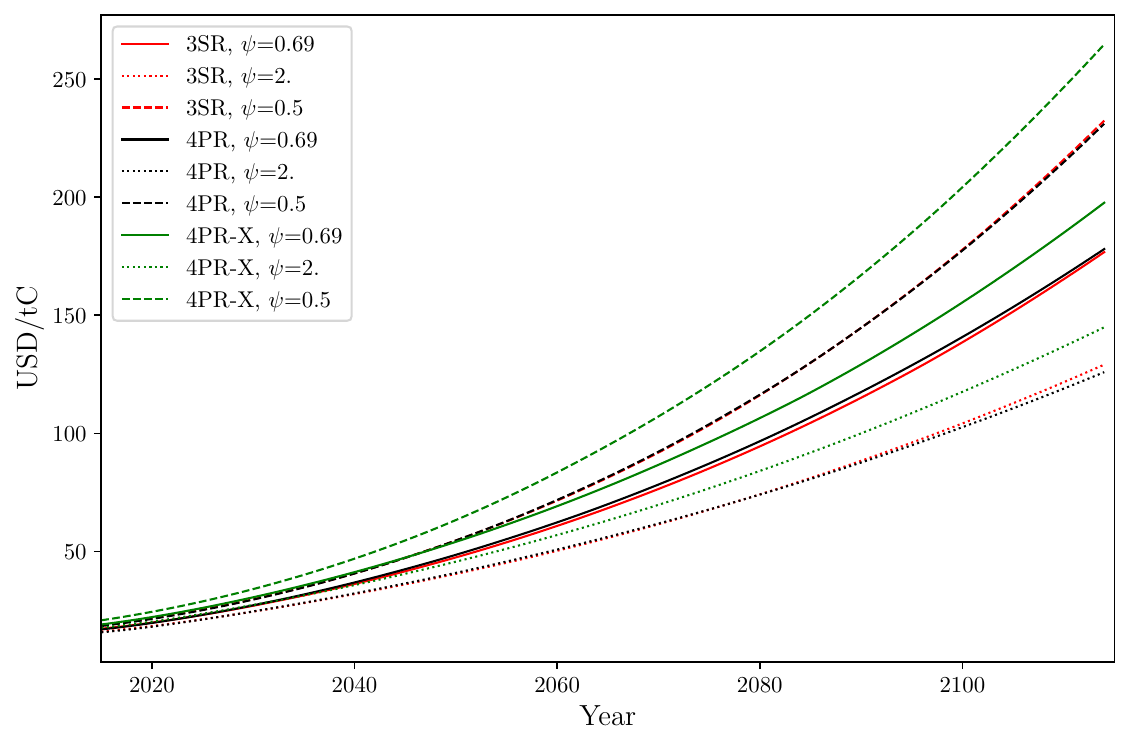}
    \caption{SCC}
     \label{fig:scc}
     \end{subfigure}
     \caption{Optimal abatement levels (left panel) and Social Cost of Carbon (SCC) values (right panel) across different discount rates.}
    \label{fig:beta_econ}
 \end{figure}

\cref{fig:beta_climate,fig:beta_econ} shows the effect of alternative discount rates on the three carbon-cycle configurations. Varying the discount rate produces only modest changes in atmospheric carbon and temperature trajectories, and the resulting SCC curves largely overlap. Differences between the \(3\)SR and \(4\)PR emulators remain small, and the \(4\)PR-X configuration behaves much as in Section~\ref{sec:optimal_mitigatoin}.


%

\newpage
\section{IPCC WGI v4 Reference Land Regions} \label{APX:WGI}
%

Table~\ref{tab:wgi_ref_regions_acronyms} lists the IPCC WGI v4 Reference Land-Regions—a standardized set of geographic polygons defined by IPCC Working Group I (version 4) that partition the global land surface into climatically coherent units and serve as a common spatial framework for regional assessments in the WGI Atlas.

\begin{table}[ht]
\caption{IPCC WGI AR6 reference \emph{land} regions~\citep{essd-12-2959-2020}. Acronyms follow the IPCC Atlas convention. The raw data are available at \url{https://github.com/IPCC-WG1/Atlas/blob/main/reference-regions/IPCC-WGI-reference-regions-v4\_coordinates.csv}.}%
\label{tab:ipcc_land_regions}
\label{tab:wgi_ref_regions_acronyms}
\centering
\begin{footnotesize}
\begin{tabular}{ll  ll  ll}
\toprule
\multicolumn{2}{c}{North America} & \multicolumn{2}{c}{Central \& South America} & \multicolumn{2}{c}{Europe \& Mediterranean}\\
\cmidrule(r){1-2}\cmidrule(r){3-4}\cmidrule{5-6}
GIC & Greenland / Iceland            & NCA & N.\ Central America      & NEU & N.\ Europe\\
NWN & N.W.\ North America            & SCA & S.\ Central America      & WCE & West \& Central Europe\\
NEN & N.E.\ North America            & CAR & Caribbean                & EEU & Eastern Europe\\
WNA & W.\ North America              & NWS & N.W.\ South America      & MED & Mediterranean\\
CNA & C.\ North America              & NSA & N.\ South America        &  & \\
ENA & E.\ North America              & NES & N.E.\ South America      &  & \\
     &                               & SAM & South American Monsoon   &  & \\
\midrule
\multicolumn{2}{c}{South America (cont.)} & \multicolumn{2}{c}{Africa} & \multicolumn{2}{c}{Middle East \& Asia}\\
\cmidrule(r){1-2}\cmidrule(r){3-4}\cmidrule{5-6}
SWS & S.W.\ South America            & WAF  & Western Africa           & SAH & Sahara\\
SES & S.E.\ South America            & CAF  & Central Africa           & ARP & Arabian Peninsula\\
SSA & S.\ South America              & NEAF & N.\ Eastern Africa       & WCA & W.\ Central Asia\\
     &                               & SEAF & S.\ Eastern Africa       & ECA & E.\ Central Asia\\
     &                               & WSAF & W.\ Southern Africa      & TIB & Tibetan Plateau\\
     &                               & ESAF & E.\ Southern Africa      & EAS & East Asia\\
     &                               & MDG  & Madagascar               & SAS & South Asia\\
     &                               &      &                          & SEA & S.E.\ Asia\\
\midrule
\multicolumn{2}{c}{Australasia} & \multicolumn{2}{c}{High-latitude Asia} & \multicolumn{2}{c}{Polar regions}\\
\cmidrule(r){1-2}\cmidrule(r){3-4}\cmidrule{5-6}
NAU & N.\ Australia                  & ESB & East Siberia             & RAR & Russian Arctic\\
CAU & C.\ Australia                  & RFE & Russian Far East         & EAN & East Antarctica\\
EAU & E.\ Australia                  &     &                          & WAN & West Antarctica\\
SAU & S.\ Australia                  &     &                          &     & \\
NZ  & New Zealand                    &     &                          &     & \\
\bottomrule
\end{tabular}
\end{footnotesize}
\end{table}
%

\section{Regional warming} \label{APX:G}

%
\begin{figure}[th!]
    \centering
    \centerline{
    \includegraphics[width=\linewidth]{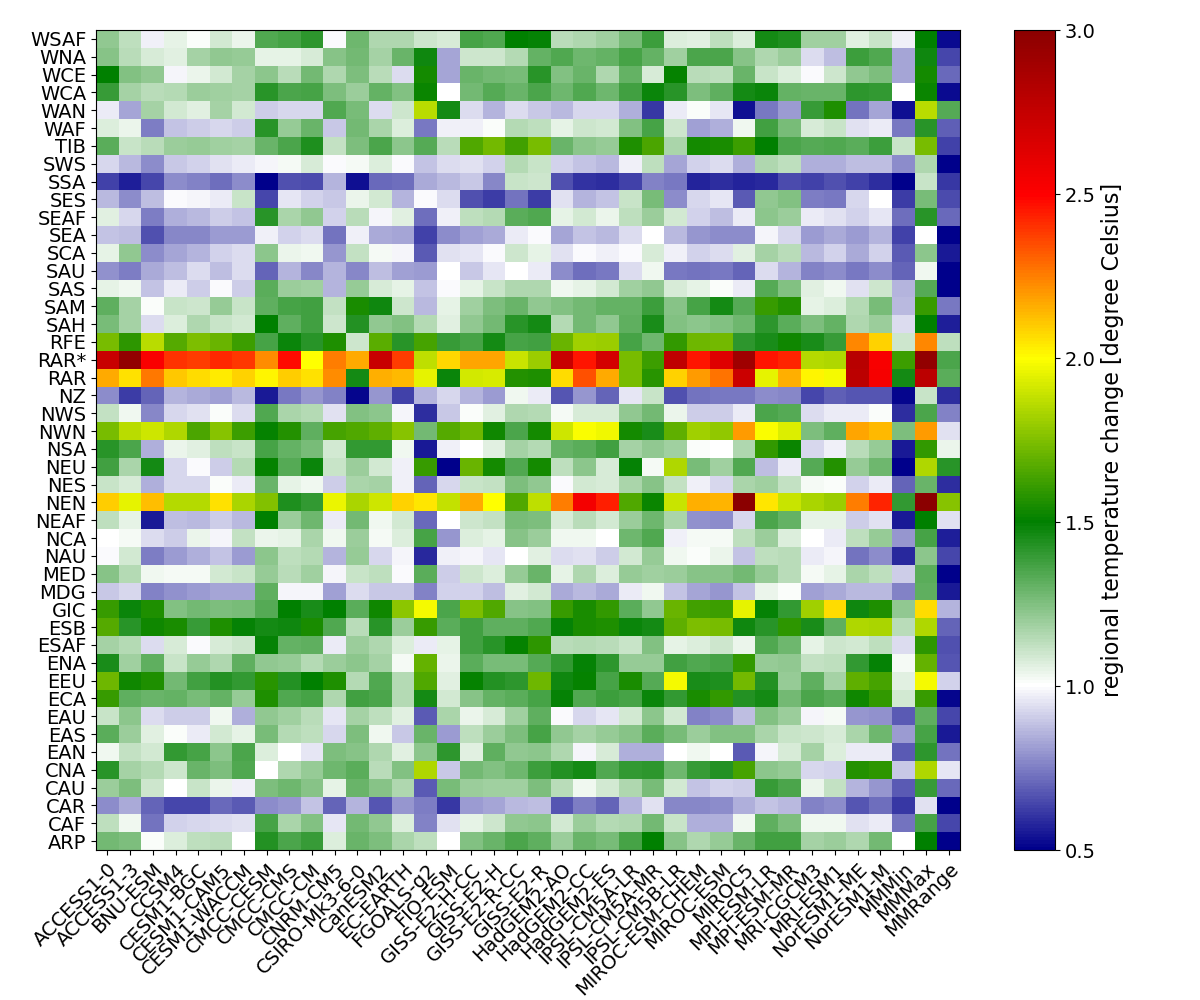}
    }
    \caption{Regional warming $\beta^\text{z}$ (color-coded in degrees Celsius) for 46 regions (y-axis) across 41 models (x-axis), including the multi-model minimum, maximum, and range, under a global mean temperature increase of 1 °C.}
    \label{fig:warming_all_regs_mods}
\end{figure}
A graphical representation of the regional average $\beta^\text{z}$, color-coded by ESM (x-axis) and region (y-axis), is shown in Figure~\ref{fig:warming_all_regs_mods}. Green hues dominate the figure, indicating that a regional temperature change of $1.5 \pm 0.5$ °C per 1 degree of global mean temperature change is prevalent across most regions and ESMs. Horizontally extended patches in blue and white correspond to regions, notably in the southern parts of the globe (SAS to SWS in the figure), where temperature changes are smaller than the global mean temperature change of 1.0 °C, irrespective of the specific ESM. 
In the vertical direction, where all regions for an individual ESM align, fewer prominent structures are observed, indicating that most ESMs span a broad range of regional temperature changes, from blue colors ($\beta^\text{z} = 0.5$) to red colors ($\beta^\text{z} = 2.5$). One exception is the IPSL model family, whose ESMs predominantly feature greenish colors, implying smaller regional differences in $\beta^\text{z}$. 
The maximum and minimum temperature changes across all ESMs, along with the associated 'maximum minus minimum' range (displayed in the three rightmost columns), show that for most regions, this range is clearly smaller than 1 °C (indicated by blue and white colors). This range is, therefore, significantly smaller than the regional warming observed by any of the 41 ESMs.

\clearpage
\bibliography{main_R1_EJ_0625.bbl}

\end{document}

%% file: fig/capartment_model.tex
\begin{minipage}[t]{0.4\linewidth}%
\begin{tikzpicture}
	\begin{scope}[shift={(0,0)},scale=.7]
        \node  at (1.6,5.1) {\footnotesize{From}}; 	
		\node  at (0.4,4.4) {\small$\text{A}$};
		\node  at (1.2,4.4) {\small$\text{O}_1$}; 
		\node  at (2.0,4.4) {\small$\text{O}_2$};
		 
		\node  at (2.8,4.4) {\small$\text{L}$}; 	

        \node[rotate=90]  at (-1.3,2.4) {\footnotesize{To}}; 	
		\node  at (-0.5,3.6) {\small$\text{A}$}; 
		\node  at (-0.5,2.8) {\small$\text{O}_1$}; 
  	    \node  at (-0.5,2.0) {\small$\text{O}_2$};
  	     	
		\node  at (-0.5,1.2) {\small$\text{L}$};

		\node[fill=black,scale=1.5] at (0.4,3.6) (description) {};
		\node[fill=black,scale=1.5] at (1.2,3.6) (description) {};
		\node[fill=black,scale=1.5] at (2.8,3.6) (description) {};
  
		\node[fill=blue!50 ,scale=1.5] at (0.4,2.8) (description) {};
		\node[fill=black   ,scale=1.5] at (1.2,2.8) (description) {};
		\node[fill=black   ,scale=1.5] at (2.0,2.8) (description) {};
  
		\node[fill=blue!100 ,scale=1.5] at (1.2,2.0) (description) {};
		\node[fill=black    ,scale=1.5] at (2.0,2.0) (description) {};
  
		\node[fill=black     ,scale=1.5] at (2.0,2.0) (description) {};

		\node[fill=Green!70 ,scale=1.5] at (0.4,1.2) (description) {};
		\node[fill=black    ,scale=1.5] at (2.8,1.2) (description) {};
  

        \draw[step=.8, black] (0,.8) grid (3.2,4);  
	\end{scope}
\end{tikzpicture}
\end{minipage}
\noindent
\begin{minipage}[t]{0.4\linewidth}%
\centering
\begin{tikzpicture}[scale=0.6,->,shorten >=1pt,auto,
                        ultra thick,main node/.style={circle,draw,font=\sffamily\Large\bfseries}]
      \node[main node] (A) at (0,0) {A};
      \node[main node] (O1) at (-3,-3) {$\text{O}_1$};
      \node[main node] (O2) at (-7,-3) {$\text{O}_2$};
      \node[main node] (L1) at (3,-3) {$\text{L}$ };
\begin{scope}[ 
    shorten > = 1pt,
node distance = 3cm and 4cm,
    el/.style = {inner sep=2pt, align=left, sloped},
every label/.append style = {font=\tiny}      ]
      \path[->,ultra thick,blue!50] (A) edge [bend right] node [above left] {$\color{black}{\footnotesize{\too{A}{$\text{O}_1$}}}$} (O1);
      \path[->,very thick] (O1) edge [bend right] node [left] {} (A);      
    \path[->,ultra thick,Green!70] (A) edge [bend left] node [above right]  {$\color{black}{\footnotesize{\too{A}{$\text{L}$}}}$} (L1);
    \path[->,very thick] (L1) edge [bend left] node [right] {} (A);
    %

      \path[->,ultra thick,blue!100] (O1) edge [bend right] node [above]  {$\color{black}{\footnotesize{\too{$\text{O}_1$}{$\text{O}_2$}}}$} (O2);
      \path[->,very thick] (O2) edge [bend right] node [above] {} (O1);
    \end{scope}
\end{tikzpicture}
\end{minipage}